\documentclass[12pt,onecolumn]{IEEEtran}

\usepackage[T1]{fontenc}
\usepackage{url}
\usepackage{color} 
\usepackage{ifthen}
\usepackage{cite}
\usepackage{amssymb} 
\usepackage[cmex10]{amsmath} 
\usepackage{graphicx} 
\usepackage[pdfencoding=auto,psdextra]{hyperref} 
\usepackage{longtable}
\usepackage{cancel}
\usepackage{balance}
\usepackage{comment}

\newtheorem{theorem}{\textbf{Theorem}}

\newtheorem{definition}{\textbf{Definition}}
\newtheorem{algorithm}{\textbf{Algorithm}}
\newtheorem{example}{\textbf{Example}}

\newcommand{\el}
{\!\in\!}
\newcommand{\inat}
{\mathbb{N}}
\renewcommand{\iint}
{\mathbb{Z}}
\newcommand{\blackbox}[3]
{{\vrule height #1 width #2 depth #3}}
\def\topbotatom#1{\mathbin{\buildrel{\hbox{$#1.$}}\over{\smash{\hbox{$#1-$}}\vphantom{_t}}}}
\newcommand{\dotmin}{\mathrel{\mathchoice{\topbotatom\displaystyle}
{\topbotatom\textstyle}
{\topbotatom\scriptstyle}
{\topbotatom\scriptscriptstyle}}}
\newcommand{\css}[2]
{\buildrel{\rm #1}\over{#2}}
\newcommand{\iia}
{\left\lfloor}
\newcommand{\iic}
{\right\rfloor}
\newcommand{\isa}
{\left\lceil}
\newcommand{\isc}
{\right\rceil}

\newcommand{\sse}
{\subseteq}
\newcommand{\irea}
{\mathbb{R}}

\newcommand{\nel}
{\not\,\el}
\newcommand{\imp}
{\;\Rightarrow\;}

\definecolor{pinegreen}{rgb}{0,0.7,0.3}
\definecolor{paleblue}{rgb}{0.5,0.5,1}
\definecolor{orange}{rgb}{1,0.3,0}
\definecolor{gold}{rgb}{0.8,0.6,0}
\newcommand{\did}
{d_{0\mbox{\small\it-}\!D\!/\!I}}

\newcommand*{\QEDA}{\hfill\ensuremath{\blacktriangle}}%

\allowdisplaybreaks 

\begin{document}
%
\title{Efficient Systematic Deletions/Insertions of $0$'s Error Control Codes and the $L_{1}$ Metric\IEEEauthorrefmark{1}\vglue -5pt
{\normalsize (Extended version)}%
\thanks{\IEEEauthorrefmark{1}This paper was presented in part at the \emph{2022 IEEE Information Theory Workshop (ITW)}, Mumbai, India, November 1-9, 2022.
}}


\author{
\IEEEauthorblockN{Luca G. Tallini\IEEEauthorrefmark{2}, Nawaf Alqwaifly\IEEEauthorrefmark{3}$^{,}$\IEEEauthorrefmark{4} and Bella Bose\IEEEauthorrefmark{3}}\vglue 1mm
\IEEEauthorblockA{\small\IEEEauthorrefmark{2}Facolt\`{a} di Scienze della Comunicazione, Universit\`{a} degli Studi di Teramo, Teramo, Italy. E-mail: {\tt ltallini@unite.it}}\\
\IEEEauthorblockA{\small\IEEEauthorrefmark{3}School of EECS, Oregon State University, Corvallis, OR, USA. E-mail: {\tt alqwaifn@oregonstate.edu}, {\tt bose@eecs.orst.edu}}\\
\IEEEauthorblockA{\small\IEEEauthorrefmark{4}College of Engineering, Qassim University, Saudi Arabia.}
 }

\IEEEoverridecommandlockouts
\IEEEaftertitletext{\vglue -10mm}
\maketitle
\begin{abstract}
This paper gives some theory and efficient design of binary block systematic codes capable of controlling the deletions of the symbol ``$0$'' (referred to as $0$-deletions) and/or the insertions of the symbol ``$0$'' (referred to as $0$-insertions). The problem of controlling $0$-deletions and/or $0$-insertions (referred to as $0$-errors) is known to be equivalent to the efficient design of $L_{1}$ metric asymmetric error control codes over the natural alphabet, $\inat$. So, $t$ $0$-insertion correcting codes can actually correct $t$ $0$-errors, detect $(t+1)$ $0$-errors and, simultaneously, detect all occurrences of only $0$-deletions or only $0$-insertions in every received word (briefly, they are $t$-Symmetric $0$-Error Correcting/$(t+1)$-Symmetric $0$-Error Detecting/All Unidirectional $0$-Error Detecting ($t$-Sy$0$EC/$(t+1)$-Sy$0$ED/AU$0$ED) codes). From the relations with the $L_{1}$ distance, optimal systematic code designs are given. In general, \textbf{for all} $t,k\el\inat$, a recursive method is presented to encode $k$ information bits into efficient systematic $t$-Sy$0$EC/$(t+1)$-Sy$0$ED/AU$0$ED codes of length
$$
n\leq k+t\log_{2}k+o(t\log n)
$$
as $n\el\inat$ increases. Decoding can be efficiently performed by algebraic means using the Extended Euclidean Algorithm (EEA).
\end{abstract}
\begin{IEEEkeywords}
deletion/insertion of zero errors, repetition/sticky errors, $L_{1}$ distance, asymmetric distance, elementary symmetric functions, constant weight codes.
\end{IEEEkeywords}

\section{Introduction}
\label{secIntro}
In communication and magnetic recording systems, the channel may cause two types of synchronization errors. The first one is not receiving a transmitted symbol (a deletion error), and the second one is receiving a spurious symbol (an insertion error). The propagation of these errors will significantly reduce the performance of the systems.

The general problem of designing efficient codes capable of correcting $t$ insertion and deletion of symbols is still an open research problem even though some results have been reported in these research papers \cite{ABD12,FER97,GUR21,HEL02,HEL02,JAI17,KAU65,LEV65,LEV66,KUL13,PAL12,SIM19,SIMA20a,SIMA20b,SIMA20c,SLO00} (also please see the references in these papers). However, some efficient code designs for correcting insertion/deletion of some fixed symbol, say $0$, are given in \cite{DOL10,KOV18,KOV19,KUL14,LEV65,MAH17,TAL19,TAL10b,TAL23}. In the present paper, some efficient systematic codes capable of correcting $t$ insertion and deletion of the symbol $0$ are given which are superior to the codes given in \cite{MAH17} and \cite{TAL10b} in terms of redundancy and reliability.

Let $\iint_{2}^{*}$ be the set of all finite length binary sequences where $\iint_{2}\css{def}{=}\{0,1\}$. In this paper, we are interested in the efficient design of binary block codes capable of correcting $t\el\inat$ or less deletions and/or insertions of a fixed binary symbol, say, $0\el\iint_{2}$. In this error model, if
\begin{equation}
\label{eqexw1}
X=0100 1010 0010 1110 \el\iint_{2}^{16}
\end{equation}
is a transmitted binary sequence of length $n= 16$, then
\setlength{\jot}{0pt}
\begin{align}
Y&=0\mathbf{0}10\lambda 1\lambda10 0\mathbf{0}01\lambda 11\mathbf{0}\mathbf{0}1\mathbf{0}0 \label{eqexw2} \\
&=0010 1100 0011 1001 00\qquad\;\el\iint_{2}^{18}\notag
\end{align}
is the received word obtained from $X$ due to $3$ deletions ($\lambda$ represents the empty symbol) and $5$ insertions of the symbol $0$. The problem of designing efficient codes to control these types of $0$-deletion and/or insertion errors (briefly, $0$-errors) is an open research problem introduced by Levenshtein in \cite{LEV65} which is important for at least two reasons. From the application perspective, through the Gray mapping, correcting $t$ deletions or insertions of $0$'s is equivalent to correcting $t$ repetition (or, sticky) errors which occur in high speed communication and data storage systems due to synchronization loss \cite{DOL10,TAL10b,MAH17}. From the theoretical perspective, the design problem of $t$ deletion and/or insertion of $0$'s Error Correcting (i.~e., $t$-Symmetric $0$-Error Correcting ($t$-Sy$0$EC)) codes is important because it is a particular instance of the general problem also introduced by Levenshtein in \cite{LEV66}. Even though the general problem of designing asymptotically optimal codes capable of correcting at most $t$ deletions and/or insertions of a symbol appears to be quite difficult, some efficient solutions have been given recently for the particular problems of correcting the $0$-insertion errors (i.~e., the insertion of $0$'s only) \cite{DOL10,MAH17} and the $0$-errors (i.~e., the deletion and/or insertion of $0$'s) \cite{TAL10b,TAL23}.

With regard to the $0$-error problem, for all $X,Y\el\iint_{2}^{*}$, let
\begin{equation}
\label{LevendhteinEqu}
\did(X,Y)\css{def}{=}
\parbox[t]{0.55\linewidth}{
the minimum number of deletions and/or insertions of $0$'s needed to transform the binary word $X$ to $Y$.
}
\end{equation}
For example, if $X$ and $Y$ are the words given in (\ref{eqexw1}) and (\ref{eqexw2}) respectively, then $\did(X,Y)=8$. The above function introduced in \cite{LEV65} is a distance (called here the deletion/insertion of $0$'s distance). In fact, it is a graph distance defined in the graph $(N,E)$ where the set of nodes $N\css{def}{=}\iint_{2}^{*}$ and the set of edges $E\css{def}{=}\left\{(X,Y)\el N^{2}:\;\did(X,Y)=1\right\}$. Synchronization errors due to $0$-errors can be controlled by inserting a marker or synchronization sequence between consecutive codewords in the sequences that are sent \cite{LEV65,SEL62,FER97}. Thus, we assume no synchronization errors due to erroneous receptions of sequences of codewords (i.~e., we assume that the receiver knows the length of the received word). In this case, since $1$-errors are forbidden in our error model,
\begin{equation}
\label{eqwxeqwyiffdidinfty}
w_{H}(X)\neq w_{H}(Y)
\iff
\did(X,Y)=\infty;
\end{equation}
where $w_{H}(Z)$ denotes the Hamming weight of $Z\el\iint_{2}^{*}$. In this way, the metric space $(\iint_{2}^{*},\did)$ or its associated graph $(N,E)$ remains partitioned into many distinct connected components, one for each possible Hamming weight, $w=w_{H}(X)\el\inat$, of words $X\el\iint_{2}^{*}$.

Because of an isometry between the above metric (\ref{LevendhteinEqu}) and the $L_{1}$ metric, in \cite{TAL23}, it is shown how the problem of controlling $0$-errors is equivalent to the efficient design of $L_{1}$ metric error control codes over the natural alphabet, $\inat$ in such a way that $t$ $0$-insertion correcting codes can actually correct $t$ $0$-errors, detect $(t+1)$ $0$-errors and, simultaneously, detect all occurrences of only $0$-deletions or only $0$-insertions in every received word (briefly, they are $t$-Symmetric $0$-Error Correcting/$(t+1)$-Symmetric $0$-Error Detecting/All Unidirectional $0$-Error Detecting ($t$-Sy$0$EC/$(t+1)$-Sy$0$ED/AU$0$ED) codes; or, equivalently, just $t$-Sy$0$EC). In \cite{TAL23}, some non-asymptotic bounds are given, for all $n,t\el\inat$, on the cardinality, $D(n,t)\el\inat$, of the optimal $t$-Sy$0$EC codes of length $n$. Let us recall the following definition from \cite{TAL23} (see also \cite{MAH17}).
\begin{definition}[asymptotically optimal codes]
\label{defaoc}
A family of $t$-Sy$0$EC/$(t+1)$-Sy$0$ED/AU$0$ED binary codes of length $n$, $\mathcal{C}(n,t)\sse\iint_{2}^{n}$, $n,t\el\inat$, is asymptotically optimal if, and only if, the ratio between the redundancy of $\mathcal{C}(n,t)$ and the optimal redundancy, $D(n,t)$, approaches $1$ as $s\css{def}{=}n+t$ goes to infinity; i.~e., 
$$
\lim_{s\to\infty}\frac{n-\log_{2}|\mathcal{C}(n,t)|}{n-\log_{2}|D(n,t)|}=1.
$$
\end{definition}
From the non-asymptotic bounds, the following theorem was derived in \cite{TAL23}.
\begin{theorem}[on the optimal redundancy of $t$-Sy$0$EC codes]
\label{thoptrel}
Let $n,t\css{def}{=}t(n)\el\inat$. If
$$
\mbox{$\log_{2}t=o(\log n)$ $\iff$ $t=2^{o(\log n)}$}
$$
(for example, $t=2^{\sqrt{\log_{2}n}}=2^{o(\log n)}$) then, the optimal redundancy of the $t$-Sy$0$EC/$(t+1)$-Sy$0$ED/AU$0$ED binary codes is
\begin{equation}
\label{eqoptrel}
n-\log_{2}|D(n,t)|
=
t\log_{2}n+o(t\log n).
\end{equation}
So, any family of $t$-Sy$0$EC/$(t+1)$-Sy$0$ED/AU$0$ED binary codes whose redundancy is $t\log_{2}n+o(t\log n)$ and $t=2^{o(\log n)}$ is asymptotically optimal according to Definition \ref{defaoc}. On the other hand, if $t\geq n-1$ then the optimal redundancy is exactly,
\begin{equation}
\label{eqoptrel2_tgtenm1}
n-\log_{2}|D(n,t)|=n-\log_{2}(n+1).
\end{equation}
\end{theorem}
Based on Elementary Symmetric Functions, certain $\sigma$-codes \cite{TAL13} are used in \cite{TAL23} to design some optimal non-systematic code designs. Decoding can be efficiently performed by algebraic means using the Extended Euclidean Algorithm (EEA).

Here, the main contribution is the following. \textbf{For all} $t,k\el\inat$, the present paper gives a recursive method to encode $k$ information bits into efficient systematic $t$-Sy$0$EC/$(t+1)$-Sy$0$ED/AU$0$ED codes of length
$$
n\leq k+t\log_{2}k+o(t\log n)
$$
as $n\el\inat$ increases. These codes are obtained by applying the theory in \cite{TAL23}. From Theorem \ref{thoptrel} if $t=2^{o(\log n)}$ then such codes are asymptotically optimal according to Definition \ref{defaoc}. On the other hand, the codes in Subsection \ref{subsecdisti} are optimal systematic $t$-Sy$0$EC/$(t+1)$-Sy$0$ED/AU$0$ED codes for $t\geq n-1$.

The paper is organized as follows. In Section \ref{secbackg}, some necessary background from \cite{TAL23} is given on $t$-Sy$0$EC, $L_{1}$ metric EC codes and $\sigma$-codes. In Section \ref{secsimpl}, some simple systematic non-recursive code design are presented. Such designs are used as base codes in Table \ref{tabV} which summarizes the present systematic code design parameters for some values of $k$ and $t$. In Section \ref{secsyste_idea}, how to use $\sigma$-codes for the systematic recursive code design is discussed together with various synchronization problems among the various parts of a codeword (note that in a systematic code a codeword is the concatenation of at least two parts: the information part and the check part which must be separated by some synchronization marker). In Section \ref{secbasec}, some efficient base codes are shown. In Section \ref{secsyste_defi}, the $\sigma$-code based design is defined. In Section \ref{secredun}, the redundancy analysis is given. Finally, in Section \ref{secconcl} some concluding remarks are given.

\section{Background on the Theory of \texorpdfstring{$t$}{\textit{t}}-Sy\texorpdfstring{$0$}{0}EC Codes, \texorpdfstring{$L_{1}$}{\textit{L{1}}} Metric and \texorpdfstring{$\sigma$}{\textsigma}-codes}
\label{secbackg}
In \cite{TAL23}, it is shown that the design problem of $t$-Sy$0$EC codes is equivalent to the design problem of some $L_{1}$ metric asymmetric error control codes over the natural alphabet, $\inat$.

Before describing this result, some background materials are given.

For $m\el\inat\cup\{\infty\}$ let
$$
\iint_{m}\css{def}{=}\{0,1,\ldots,m-1\}\sse\inat\css{def}{=}\iint_{\infty}.
$$
Also, for $x,y\el\iint_{m}$, define the natural subtraction as $x \dotmin y=\max\{0,x-y\}$. For example, if $x = 2$ and $y = 0$ then $x \dotmin y$ = $2$ and $y \dotmin x$ = $0$. Given any two words $X,Y\el\iint_{m}^{n}$ of length $n\el\inat$, the operations $X\cap Y\el\iint_{m}^{n}$, $X\cup Y\el\iint_{m}^{n}$, $X+Y\el\inat^{n}$, and $X\dotmin Y\el\iint_{m}^{n}$ are defined as the digit by digit $\min$, $\max$, integer addition and $\dotmin$ operation between $X$ and $Y$, respectively. For example, if $m=3$, $n=9$, $X=012012012$ and $Y=000111222$ then $X\cap Y=000011012$, $X\cup Y=012112222$, $X+Y=012123234$, $X\dotmin Y=012001000$ and $Y\dotmin X=000100210$. In addition, the support of a word $X=x_{1}x_{2}\ldots x_{n}\el\iint_{m}^{n}$ is $\partial{X}=s_{1}s_{2}\ldots s_{n}\el\iint_{2}^{n}$ where $s_{i} =1$ if $x_{i}\neq0$ and $s_{i}=0$ otherwise. For example $\partial(42101) = (11101)$. Given a support $\partial{S}$ as an index set, say $\partial{S}=[1,n]$, every word in $X=x_{1}x_{2}\ldots x_{n}\el\iint_{m}^{n}$ can be regarded as a multiset over the index set $\partial{S}$ where each component, $x_{i}$ of $X$ defines the multiplicity of $i\el\partial{S}$ as an element of $X$. In this way, there is a one-to-one correspondence between $m$-ary words and multisets; and the above operations can be regarded as multisets operations too. So, in the following, we will identify $m$-ary words of length $n$ with multisets over an index set containing $n$ distinct elements (which, for code construction purposes, will be contained in a field). The cardinality of a word/multiset $X=x_{1}x_{2}\ldots x_{n}\el\iint_{m}^{n}$ is the $L_{1}$ weight of $X$ and is naturally defined as the real sum
$$
|X|\css{def}{=}w_{L_{1}}(X)\css{def}{=}\sum_{i\in\partial{S}}x_{i}.
$$
For example, $|01232|=w_{L_{1}}(01232)=8$. Note that for $m=2$ the $L_{1}$ weight and the Hamming weight coincide. So, when this creates no confusion we will indicate the weight of $X$ as $w(X)$.

To better describe the error control properties of codes for the $L_{1}$ metric, the following distances between $m$-ary words $X,Y\el\iint_{m}^{n}$ are considered in \cite{TAL11a,TAL13} (the ``$+$'' sign below indicates an integer sum).
\setlength{\jot}{2pt}
\begin{alignat}{1}
\mbox{symmetric $L_{1}$:}\; d_{L_{1}}^{sy}(X,Y)&\css{def}{=}|Y\dotmin X|+|X\dotmin Y|,\label{eqdistdef}\\
\mbox{asymmetric $L_{1}$:}\; d_{L_{1}}^{as}(X,Y)&\css{def}{=}\max\{|Y\dotmin X|,|X\dotmin Y|\},\notag\\
\mbox{Hamming:}\,\,\,d_{H}(X,Y)&\css{def}{=}|\partial(Y\dotmin X)|+|\partial(X\dotmin Y)|.\notag
\end{alignat}
For example, if $m=5$, $n=5$, $X=01423$, $Y=43213$ then $|X\dotmin Y|=3$, $|Y\dotmin X|=6$, $|\partial(X\dotmin Y)|=2$, $|\partial(Y\dotmin X)|=2$ and $d_{L_{1}}^{sy}(X,Y)=3+6=9$, $d_{L_{1}}^{as}(X,Y)=\max\{6,3\}=6$ and $d_{H}(X,Y)=2+2=4$. From the error control perspective, if $X$ is the transmitted word and $Y$ is the received word then $Y \dotmin X$ and $X \dotmin Y$ give the increasing and decreasing error vectors, respectively. Thus,
$$
X =Y-(Y \dotmin X)+(X \dotmin Y).
$$
Note that,
\begin{equation}
\label{eqHdisleqL1dis}
\mbox{for all $X,Y\el\iint_{m}^{n}$},
\quad
d_{H}(X,Y)\leq d_{L_{1}}^{sy}(X,Y)
\end{equation}
because $w_{H}(X)=|\partial{X}|\leq|X|\css{def}{=}w_{L_{1}}(X)$, for all $X\el\iint_{m}^{n}$.

Constant weight codes play an important role in what follows. Thus, given $n,w\el\inat$ and any numeric set $A\sse\inat$ as alphabet, let
\begin{equation}
\label{eqAnwdef}
\mathcal{S}(A,n,w)
\css{def}{=}
\left\{
X\el A^{n}:\;
w_{L_{1}}(X)=|X|=w
\right\}
\end{equation}
be the set of all words over $A$ of length $n$ and constant weight $w$. We readily note, from (\ref{eqAnwdef}), that
\begin{equation}
\label{UpperboundC}
\mathcal{S}(A,n,w)
=
\bigcup_{x\in A}
\mathcal{S}(A,n-1,w-x)x;
\end{equation} 
where the above union is a disjoint union of sets and $\mathcal{S}x\sse A^{n}$ indicates the set of words obtained concatenating every word in the set $\mathcal{S}\sse A^{n-1}$ with $x\el A$. Hence, the general recurring formula,
\begin{equation}
\label{eqAgenstiffel}
|\mathcal{S}(A,n,w)|
=
\sum_{x\in A}|\mathcal{S}(A,n-1,w-x)|,
\end{equation}
holds for, say, the ``$A$-nominal coefficient $n$ choose $w$'', $|\mathcal{S}(A,n,w)|$. If $A=\iint_{m}$ then the cardinality of the above set is the $m$-nominal coefficient $n$ choose $w$
\begin{equation}
\label{Mnominal}
|\mathcal{S}(\iint_{m},n,w)|
=
\binom{n}{w}_{m}
=
\sum_{v=0}^{m-1}\binom{n-1}{w-v}_{m},
\end{equation}
for all integers $m\el\inat$. The quantity $\binom{n}{w}_{m}$ is the coefficient of the monomial $z^{w}$ in the standard form of the polynomial $[1+z+...+z^{(m-1)}]^{n}$ which, for $m=2$, reduces to the usual binomial coefficient (i.~e., $\binom{n}{w}_{2}=\binom{n}{w}$). The $m$-nomial coefficient sequence has been studied in the ambit of $m$-ary unordered codes and share many properties with the binomial coefficient sequence obtained for $m=2$ \cite{PEZ12}. If instead, $A=\iint_{\infty}=\inat$ then we can define
$$
\binom{n}{w}_{\infty}\css{def}{=}|\mathcal{S}(\inat,n,w)|
$$
and note that the cardinality of $\mathcal{S}(\inat,n,w)$ is the composition of a natural number $w$ into $n$ natural numbers. In this way,
\begin{equation}
\label{eqmnomialcomp}
|\mathcal{S}(\inat,n,w)|\!\css{def}{=}\!\!\binom{n}{w}_{\infty}\!\!\!\!=\!\binom{n+w-1}{n-1}\!=\!\binom{n+w-1}{w}.
\end{equation}
In this case, the recursive formula (\ref{eqAgenstiffel}) becomes
\setlength{\jot}{8pt}
\begin{alignat}{2}
|\mathcal{S}(\inat,n,w)|=\,&\binom{n}{w}_{\infty}=\sum_{v=0}^{w}\binom{n-1}{v}_{\infty}=\label{eqAgenstiffelinf}\\
&\sum_{v=0}^{w}\binom{n+v-2}{n-2}=\binom{n+w-1}{n-1}\notag
\end{alignat}
because $x\geq0$ and $(w-x)\geq0$ ($\iff$ $x,(w-x)\el\inat$).

Now, if $X\el\iint_{2}^{*}$ then $X$ can be uniquely written as \cite{LEV65,LEV93},
\begin{equation}
\label{eqXwrasrl0}
X=0^{v_{1}}10^{v_{2}}10\ldots010^{v_{w}}10^{v_{w+1}}
\end{equation}
where, for all integers $i\el[1,w+1]$, $v_{i}\css{def}{=}v_{i}(X)\el\iint_{l-w+1}\sse\inat$ is the $i$-th run length of $0$'s in the word $X$, $l=l(X)\el\inat$ indicates the length of any $X\el A^{*}$ and $w=w_{H}(X)\el[0,l(X)]$ is the Hamming weight of $X$. Note that
\begin{equation}
\label{eqvwp1}
v_{w+1}=(l(X)-w(X))-\sum_{i=1}^{w}v_{i}.
\end{equation}

Given the above representation, consider the following bijective function (which we call here the bucket of $0$'s mapping)
\begin{equation}
\label{eqfunV}
V:\iint_{2}^{*}\to\iint_{\infty}^{*}=\inat^{*}
\end{equation}
which associates any $X\el\iint_{2}^{*}$ represented as in (\ref{eqXwrasrl0}) with
$$
V(X)
\css{def}{=}
(v_{1},v_{2},\ldots,v_{w},v_{w+1})\el\inat^{*}.
$$
For example, if
$$
X= 01\,001\,01\,0001\,01\,1\,1\,0000000\el\iint_{2}^{*}
$$
then
$$
V(X)=(1,2,1,3,1,0,0,7)\el\inat^{*}.
$$
The mapping $V$ in (\ref{eqfunV}), already considered by Levensthein in \cite{LEV65}, defines a bijection from the set of all binary words of any finite length $n\el\inat$ and Hamming weight $w$ ($=$ number of $1$'s of the binary words) into the words over $\inat$ of length $w+1$ ($=$ number of buckets defined by the $w$ $1$'s of the binary words) and $L_{1}$ weight $n-w$ ($=$ number of $0$'s of the binary words). Except for the rightmost ``$1$'' which is dropped, the function
$$
V^{-1}:\iint_{\infty}^{*}=\inat^{*}\to\iint_{2}^{*}
$$
is nothing but the prefix free unary representation of a sequence of integer numbers. Hence, both $V$ and $V^{-1}$ are one-to-one mappings such that
$$
V(\mathcal{S}(\iint_{2},n,w))
=
\mathcal{S}(\inat,w+1,n-w),
$$
and
$$
\mathcal{S}(\iint_{2},n,w)
=
V^{-1}(\mathcal{S}(\inat,w+1,n-w)).
$$
For example, for $n=4$, the mapping $V$ acts on $\iint_{2}^{4}$ is as reported in Table \ref{tab0}.
\begin{table*}[t]
\caption{The mapping $V$ acting on $\iint_{2}^{4}$.
\blackbox{0mm}{\textwidth}{0mm}
\vglue -1\baselineskip
{\rm In the table $v_{w(X)+1}$ is in boldface and $l(X)$ indicates the length of any $X\el A^{*}$.}}
\begin{center}
\renewcommand{\tabcolsep}{2pt}
\renewcommand{\arraystretch}{1.1}
\begin{tabular}{|c|c|c||l|c|c|}
\hline
\blackbox{3.5mm}{0mm}{2mm}
$l(X)=n$ & $w(X)$ & $X$ & $V(X)=\hat{V}(X)\boldsymbol{v}_{\boldsymbol{w}+1}$ & $l(V(X))$ & $w(V(X))$\\\hline\hline
$4$ & $0$ & $0000$ & $\mathbf{4}$ & $1$ & $4$\\\hline
      && $0001$ & $3\mathbf{0}$ &&       \\
$4$ & $1$ & $0010$ & $2\mathbf{1}$ & $2$ & $3$\\
       && $0100$ & $1\mathbf{2}$ &&       \\
       && $1000$ & $0\mathbf{3}$ &&       \\\hline
       && $0011$ & $20\mathbf{0}$ &&       \\
       && $0101$ & $11\mathbf{0}$ &&       \\
$4$ & $2$ & $0110$ & $10\mathbf{1}$ & $3$ & $2$\\
      && $1001$ & $02\mathbf{0}$ &&       \\
      && $1010$ & $01\mathbf{1}$ &&       \\
      && $1100$ & $00\mathbf{2}$ &&       \\\hline
      && $0111$ & $100\mathbf{0}$ &&       \\
$4$ & $3$ & $1011$ & $010\mathbf{0}$ & $4$ & $1$\\
      && $1101$ & $001\mathbf{0}$ &&       \\
      && $1110$ & $000\mathbf{1}$ &&       \\\hline
$4$ & $4$ & $1111$ & $0000\mathbf{0}$ & $5$ & $0$\\\hline
\end{tabular}
\end{center}
\label{tab0}
\end{table*}
Let
\begin{equation}
\label{eqfunVhat}
\hat{V}:\iint_{2}^{*}\to\inat^{*}
\end{equation}
be the function obtained from $V$ by dropping the last component; $\hat{V}$ associates any $X\el\iint_{2}^{*}$ represented as in (\ref{eqXwrasrl0}) with
$$
\hat{V}(X)
\css{def}{=}
(v_{1},v_{2},\ldots,v_{w})\el\inat^{*}.
$$
Obviously, since $V$ is a one-to-one function, it is possible to reconstruct $X$ from $V(X)$; likewise, even though $\hat{V}$ is not one-to-one (for example, $\hat{V}(0110)=\hat{V}(011000)=(1,0)$), it is possible to reconstruct $X$ from $\hat{V}(X)$ and $n=l(X)$ because of (\ref{eqvwp1}). In this case, $v_{w+1}$ can be considered as a parity digit which makes the $L_{1}$ weight $w_{L_{1}}(V(X))=n-w$. Both functions $V$ and $\hat{V}$ play important roles in our code designs and analysis. Consider the following example words
\setlength{\jot}{0pt}
\begin{alignat*}{3}
X&= 01\,001\,01\,0001\,01\,1\,1\,0\qquad&\el\iint_{2}^{16},\\
Y&=001\,001\,1\,00001\,1\,1\,001\,00&\el\iint_{2}^{19},\\
Y'&=001\,001\,01\,0001\,01\,00&\el\iint_{2}^{16}.
\end{alignat*}
Then their associated $V$ values are
\setlength{\jot}{0pt}
\begin{alignat*}{3}
V(X)&=(1,2,1,3,1,0,0,\mathbf{1})\;&\el\inat^{8},\\
V(Y)&=(2,2,0,4,0,0,2,\mathbf{2})&\el\inat^{8},\\
V(Y')&=(2,2,1,3,1,\mathbf{2})&\el\inat^{6}.
\end{alignat*}
Note that if $X$ is sent, $Y'$ can never be received because $7=w(X)\neq w(Y')=5$ and $1$-errors are forbidden in our channel model; whereas, $Y$ can erroneously be received and the number of $0$-deletions ($=2$) plus the number of $0$-insertions ($=5$) from $X$ to $Y$ is equal to the $L_{1}$ distance between $V(X)$ and $V(Y)$, $d_{L_{1}}^{sy}(V(X),V(Y))=2+5=7$. In fact, in general, a sequence $Y\el\iint_{2}^{*}$ is obtained from the sequence $X\el\iint_{2}^{*}$ due to $t_{-}$ deletions and $t_{+}$ insertions of the symbol $0$ if, and only if, $w(Y)=w(X)$ and $d_{L_{1}}^{sy}(V(Y),V(X))=t_{-}+t_{+}$; that is, $V(Y)$ is obtained from $V(X)$ due to a negative error pattern of magnitude $t_{-}$ and a positive error pattern of magnitude $t_{+}$. Hence, the bucket of $0$'s mapping $X\to V(X)$ reduces the $t_{-}$ $0$-deletion and $t_{+}$ $0$-insertion error correction problem into the $t_{-}$ negative and $t_{+}$ positive error correction problem for the $L_{1}$ distance over $\inat$. The following theorem is proved in \cite{TAL23}.
\begin{theorem}[isometry between $(\iint_{2}^{*},\!\did\!)$ and $(\inat^{*}\!,\!d_{L_{1}}^{sy}\!)$]
\label{thmisodidl1}
For all $X,Y\el\iint_{2}^{*}$,
\begin{equation}
\label{eqdidugdl0}
\!\!\!\!
\did(X,Y)
=
\!
\left\{
\renewcommand{\arraystretch}{1.1}
\renewcommand{\arraycolsep}{1mm}
\!\!
\begin{array}{ll}
d_{L_{1}}^{sy}(V(X),V(Y)) &\mbox{if $w(X)\!=\!w(Y)$},\\
\infty &\mbox{if $w(X)\!\neq\!w(Y)$}.\\
\end{array}
\right.
\end{equation}
Relation (\ref{eqdidugdl0}) implies that $\did(X,Y)<\infty$ if, and only if, $w(X)=w(Y)$. So, if we extend the domain of $d_{L_{1}}^{sy}$ from $\inat^{l}\times\inat^{l}$, $l\el\inat$, to $\inat^{*}\times\inat^{*}$ by letting $d_{L_{1}}^{sy}(U,V)=\infty$ whenever $l(U)\neq l(V)$ then,
$$
\mbox{for all $X,Y\el\iint_{2}^{*}$, $\quad\did(X,Y)=d_{L_{1}}^{sy}(V(X),V(Y))$}.
$$
This implies that the mapping $V$ in (\ref{eqfunV}) is an isometry between the metric spaces $(\iint_{2}^{*},\did)$ and $(\inat^{*},d_{L_{1}}^{sy})$.
\end{theorem}

In general, the isometry $V$ in (\ref{eqfunV}) reduces the design problem of error control codes for the insertion/deletion of $0$'s problem to the design problem of error control codes under the $L_{1}$ metric. In particular, for all $w\el[0,n]$, the one-to-one function $V$ transforms any word $X\el\mathcal{S}(\iint_{2},n,w)\sse\iint_{2}^{n}$ into a word $V(X)=(v_{1},v_{2},\ldots,v_{w+1})\el\mathcal{S}(\inat,w+1,n-w)\sse\iint_{n-w+1}^{w+1}$. Furthermore, any fixed length $n\el\inat$ binary code, $\mathcal{C}\sse\iint_{2}^{n}$, is union of block (i.~e., constant) length $n\el\inat$ constant weight $w\el[0, n]$ codes, where the union is over $w$; say, $\mathcal{C}=\bigcup_{w\in[0,n]}\mathcal{C}_{w}$, with $\mathcal{C}_{w}\sse\mathcal{S}(\iint_{2},n,w)$. So, the image of $\mathcal{C}$ through the isometry $V$ is equal to
$$
V\left(\mathcal{C}\right)=V\left(\bigcup_{w\in[0,n]}\mathcal{C}_{w}\right)=\bigcup_{w\in[0,n]}V\left(\mathcal{C}_{w}\right)\sse\inat^{n}
$$
with $\mathcal{A}_{w}\css{def}{=}V\left(\mathcal{C}_{w}\right)\sse\mathcal{S}(\inat,w+1,n-w)$, for all $w\el[0, n]$. Since the $\did$ distance between binary words of distinct weight is $\infty$, the insertion/deletion of $0$'s code design problem is reduced to the proper design of the $L_{1}$ metric constant weight error control codes $\mathcal{A}_{w}$, for all $w\el[0, n]$. Thus, in general, any $L_{1}$ distance error control property of codes over $\inat$ reflects into the analogous $\did$ distance error control property of codes over $\iint_{2}$ because of Theorem~\ref{thmisodidl1}. So, from the $L_{1}$ metric asymmetric/unidirectional coding theory \cite{BOS82,TAL11a,TAL13,WEB92,BLA93} and Theorem~\ref{thmisodidl1}, the following theorem holds which gives only some (maximal) error correction capabilities of $t$-Sy$0$EC codes. Following the classical asymmetric/unidirectional coding theory notation \cite{BLA93}, in the theorem below, $t$-Sy$X$C/$d$-Sy$X$D/AU$X$D indicates the class of codes capable of correcting $t$ symmetric errors, detecting $d$ symmetric errors and, simultaneously detecting all unidirectional errors; where the errors are of type $X$ defined as follows. If $X=$``$0$E" then the codes are in the binary sequences domain of the function $V$ in (\ref{eqfunV}) and the errors are $0$-errors, if instead $X=$``E" then then the codes are in the integer sequences codomain of the function $V$ and the errors are $L_{1}$ distance errors. Analogously, $(t_{-},t_{+})$-$0$EC indicates the class of codes capable of correcting $t_{-}$ deletions of $0$'s and, simultaneously, $t_{+}$ insertions of $0$'s; and $(t_{-},t_{+})$-EC indicates the class of codes capable of correcting $t_{-}$ negative errors and, simultaneously, $t_{+}$ positive errors in the $L_{1}$ metric \cite{TAL08,TAL10,TAL10b,TAL11a,TAL12a,TAL13,TAL18b}. The next theorem is proved in \cite{TAL23}.
\begin{theorem} [Error control capabilities and combinatorial characterizations of $t$-Sy$0$EC]
\label{thtdi0ECCequidec2}
Let $t,t_{-},t_{+}$, $\tau\el\inat$ be any numbers such that $t_{-}+t_{+}=t$ and $\tau\el[0,t]$. If
$$
\mathcal{C}=\bigcup_{w\in[0,n]}\mathcal{C}_{w}\sse\iint_{2}^{n}
$$
is a binary code of length $n\el\inat$ and $\mathcal{C}_{w}\css{def}{=}\mathcal{C}\cap\mathcal{S}(\iint_{2},n,w)$, for all integer $w\el[0,n]$, then $V\left(\mathcal{C}_{w}\right)\sse\mathcal{S}(\inat,w+1,n-w)$, for all $w\el[0, n]$; and the following statements are equivalent:
\begin{itemize}
\item[1)] $\mathcal{C}$ is a $t$-Sy$0$EC code (i.~e., $\mathcal{C}$ is a $t$-Symmetric $0$-error Correcting Code);
\item[2)] $\mathcal{C}$ is a $(t=t_{-},0)$-$0$EC code (i.~e., $\mathcal{C}$ is a $t$ deletion of $0$'s error correcting code);
\item[3)] $\mathcal{C}$ is a $(0,t=t_{+})$-$0$EC code (i.~e., $\mathcal{C}$ is a $t$ insertion of $0$'s error correcting code);
\item[4)] $\did(\mathcal{C})>2t$;
\item[5)] for all $w\el[0,n]$, $\did(\mathcal{C}_{w})>2t$ ($\!\iff$ $\mathcal{C}_{w}$ is a $t$-Sy$0$EC);
\item[6)] for all $w\el[0,n]$, $d_{L_{1}}^{sy}(V(\mathcal{C}_{w}))\geq2(t+1)$ ($\iff$ $V(\mathcal{C}_{w})$ is a $\tau$-SyEC/$(2t-\tau+1)$-SyED/AUED code over $\inat$);
\item[7)] for all $w\el[0,n]$, $d_{L_{1}}^{as}(\hat{V}(\mathcal{C}_{w}))\geq t+1$ ($\iff$ $\hat{V}(\mathcal{C}_{w})$ is a $(t_{-},t_{+})$-EC code over $\inat$);
\item[9)] $\did(\mathcal{C})>2t+1$;
\item[10)] $\mathcal{C}$ is a $\tau$-Sy$0$EC/$(2t-\tau+1)$-Sy$0$ED/AU$0$ED code.
\end{itemize}
\end{theorem}

\subsection{\texorpdfstring{$t$-Sy$0$EC/$(t+1)$-Sy$0$ED/AU$0$ED Decoding Algorithm for $t$-Sy$0$EC codes}{\textit{t}-Sy\textit{0}EC/(\textit{t}+1)-Sy\textit{0}ED/AU\textit{0}EC Decoding Algorithm}}
\label{subsectsy0e}
Let $\mathcal{C}=\bigcup_{w\in[0,n]}\mathcal{C}_{w}$ be a $t$-Sy$0$EC code of length $n$, where $\mathcal{C}_{w}\css{def}{=}\mathcal{C}\cap\mathcal{S}(\iint_{2},n,w)$. From Theorem~\ref{thtdi0ECCequidec2} with $\tau=t$, $\mathcal{C}$ is actually a $t$-Sy$0$EC/$(t+1)$-Sy$0$ED/AU$0$ED code and here we recall the efficient $t$-Sy$0$EC/$(t+1)$-Sy$0$ED/AU$0$ED decoding algorithm for $\mathcal{C}$ which exploits the maximum error control capabilities of $\mathcal{C}$ (see Subsection III.B in \cite{TAL23}). Such algorithm is as follows. If $C\el\mathcal{C}\sse\iint_{2}^{n}$ is sent and $R\el\iint_{2}^{*}$ is received, the decoder computes $w\css{def}{=}w(R)\el[0,n]$ and applies Algorithm \ref{alggentsyECdUEDcwc} below with input 1) the constant $L_{1}$ weight $\omega\css{def}{=}n-w$ code of length $\nu\css{def}{=}w+1$ over the alphabet $\inat$,
\begin{equation}
\label{eqAnuomegadef}
\mathcal{A}\css{def}{=}\mathcal{A}(\inat,\nu,\omega)\css{def}{=}
V(\mathcal{C}_{w})\sse\mathcal{S}(\inat,w+1,n-w)
\end{equation}
and 2) the word $Y=V(R)\el\inat^{\nu}$. On getting as output the word $X'\el\inat^{\nu}$ the decoder computes $C'\css{def}{=}V^{-1}(X')$ as the estimate of the sent codeword $C$. The output signal $cor\el\{0,1\}$ is such that if $cor=1$ then $0$-errors are corrected.

The following Algorithm \ref{alggentsyECdUEDcwc} is a general efficient error control algorithm for any $m$-ary constant weight $w$ code, $\mathcal{A}$, of length $n$ with minimum $L_{1}$ distance $d_{L_{1}}^{sy}(\mathcal{A})\geq2(t+1)$. Note that Algorithm \ref{alggentsyECdUEDcwc} efficiently reduces the $t$-SyEC/$(t+1)$-SyED/AUED decoding design problem for constant weight codes to the less powerful $(\tau_{-},\tau_{+})$-EC decoding design problem; proving that the two problems are indeed equivalent.
\begin{algorithm}[General $t$-SyEC/$(t+1)$-SyED/AUED decoding algorithm for Constant Weight codes \protect{\cite[Algorithm 3.1]{TAL23}}]
\label{alggentsyECdUEDcwc}\rm
\\
{\bf Input}:
\begin{itemize}
\item[1)] The constant weight code $\mathcal{A}\css{def}{=}\hat{\mathcal{A}}\,x_{n}\sse\mathcal{S}(\iint_{m},n,w)$, where
$$
x_{n}\css{def}{=}w-w_{L_{1}}(\hat{X})\el\iint_{m},\quad\hat{X}\el\hat{\mathcal{A}},
$$
is the parity digit; together with a set, $\mathcal{D}ec(\hat{\mathcal{A}})$, of \textbf{any} (possibly efficient) $(\tau_{-},\tau_{+})$-EC decoding algorithm, $\mathcal{D}ec(\hat{\mathcal{A}},\tau_{-},\tau_{+})$, for the punctured code $\hat{\mathcal{A}}$, for all $\tau_{-},\tau_{+}\el\inat$ such that $\tau_{-}+\tau_{+}=t<d_{L_{1}}^{as}(\hat{\mathcal{A}})$; and,
\item[2)] the (received) word $Y=\hat{Y}\,y_{n}\el\iint_{m}^{n}$ with $\hat{Y}\el\iint_{m}^{n-1}$ and $y_{n}\el\iint_{m}$.
\end{itemize}
{\bf Output}:
\begin{itemize}
\item[1)] A word $X'=\hat{X}'\,x_{n}'\el\iint_{m}^{n}$, where $\hat{X}'\el\iint_{m}^{n-1}$ and $x_{n}'\el\iint_{m}$ (the word $X'$ represents the estimate of the sent codeword $X\css{def}{=}\hat{X}\,x_{n}\el\mathcal{A}$); and,
\item[2)] a signal $cor\el\{0,1\}$ such that if $cor=1$ then errors are corrected; i.~e., $X'=X$.
\end{itemize}
Execute the following steps.\\
{\bf S1}: Compute
\begin{equation}
\label{eqconddeltatilde}
\mbox{$\Delta(X,Y)\css{def}{=}|Y|-w=|Y\dotmin X|-|X\dotmin Y|$}.
\end{equation}
\noindent
{\bf S2}: If $|\Delta(X,Y)|\geq t+1$ then set $cor=0$, set $X'$ to be any word, output $cor$, output $X'$ and \textbf{exit}.\\
{\bf S3}: Otherwise, if $|\Delta(X,Y)|\leq t$ then execute the following steps.\\
{\bf S3.1}: Compute
\begin{equation}
\label{eqtauepsdef}
\tau_{-}
\css{def}{=}
\iia\frac{t-\Delta(X,Y)}{2}\iic
\;
\mbox{and}
\;
\tau_{+}
\css{def}{=}
\iia\frac{t+\Delta(X,Y)}{2}\iic.
\end{equation}
\noindent
Note that $0\leq\tau_{-},\tau_{+}\leq t$ (because $|\Delta(X,Y)|\leq t$) and
\begin{equation}
\label{eqt-+t+leqt}
\tau_{-}+\tau_{+}
\leq
\frac{t-\Delta(X,Y)}{2}+\frac{t+\Delta(X,Y)}{2}
=
t.
\end{equation}
{\bf S3.2}: With the word $\hat{Y}\el\iint_{m}^{n-1}$ as input, execute the algorithm $\mathcal{D}ec(\hat{\mathcal{A}},\tau_{-},t-\tau_{-})$ for $\hat{\mathcal{A}}$. Let $\hat{X}'\el\iint_{m}^{n-1}$ be its output word.\\
{\bf S3.3}: Set $X'=\hat{X}'\,x_{n}'\el\mathcal{A}$ if $\hat{X}'\el\hat{\mathcal{A}}$, and $X'=$ any word
if $\hat{X}'\nel\hat{\mathcal{A}}$; where
\begin{equation}
\label{eqpardigit}
x_{n}'=w-w_{L_{1}}(\hat{X}')
\end{equation}
is the parity digit of $\hat{X}'$.\\
{\bf S3.4}: Set
\begin{equation}
\label{eqcordefalg1}
cor
=
\left\{
\renewcommand{\arraystretch}{1.1}
\renewcommand{\arraycolsep}{0pt}
\begin{array}{ll}
1\quad&\mbox{if $X'\el\mathcal{A}$ and $d_{L_{1}}^{sy}(X',Y)\leq t$},\\
0&\mbox{otherwise}
\end{array}
\right.
\end{equation}
{\bf S3.5}: Output $X'$, output $cor$ and \textbf{exit}.
\end{algorithm}
For completeness we recall
\begin{theorem}[Correctness of Algorithm \ref{alggentsyECdUEDcwc} \protect{\cite[Algorithm 3.1]{TAL23}}]
Given $m\el\inat\cup\{\infty\}$ and $n,w,t\el\inat$, let $\mathcal{A}$ be any $m$-ary constant weight $w$ code of length $n$ with minimum $L_{1}$ distance
\begin{equation}
\label{eqhypdistthmbase}
d_{L_{1}}^{sy}(\mathcal{A})\geq2t+2
\quad\Longleftrightarrow\quad
d_{L_{1}}^{as}(\hat{\mathcal{A}})\geq t+1.
\end{equation}
If for all (sent codeword) $X\el\mathcal{A}$ and (received word) $Y\el\iint_{m}^{n}$,
\begin{equation}
\label{eqcondtSydUEDAlgcwc}
\left\{
\renewcommand{\arraystretch}{1.1}
\renewcommand{\arraycolsep}{0pt}
\begin{array}{ll}
\mbox{either }&\mbox{$\delta(X,Y)\css{def}{=}\min\{|Y\dotmin X|,|X\dotmin Y|\}=0$},\\
\mbox{or}&\mbox{$d_{L_{1}}^{sy}(X,Y)\leq t+1$},
\end{array}
\right.
\end{equation}
then Algorithm \ref{alggentsyECdUEDcwc} gives the correct output as a $t$-SyEC/$(t+1)$-SyED/AUED decoding algorithm for $\mathcal{A}$; that is, by definition of $t$-SyEC/$(t+1)$-SyED/AUED decoding,
\begin{itemize}
\item[C1)]
if (\ref{eqcondtSydUEDAlgcwc}) holds and $cor=1$ then $X'=X$; and,
\item[C2)]
if $d_{L_{1}}^{sy}(X,Y)\leq t$ then $cor=1$ (and hence, $X'=X$).
\end{itemize}
\end{theorem}
\begin{IEEEproof}
Let $X\el\mathcal{A}$, $Y\el\iint_{m}^{n}$ and assume (\ref{eqcondtSydUEDAlgcwc}) holds. First, let us prove that if $cor=1$ then $X'=X$. Note that $cor=1$ if, and only if, step S3.4 is executed and (\ref{eqcordefalg1}) evaluates to $1$. In particular, if $cor=1$ then $X\el\mathcal{A}$, $|\Delta(X,Y)|\leq t$, $X'\el\mathcal{A}$ and $d_{L_{1}}^{sy}(X',Y)\leq t$. And so, $X\el\mathcal{A}$, $d_{L_{1}}^{sy}(X,Y)\leq t+1$, $X'\el\mathcal{A}$ and $d_{L_{1}}^{sy}(X',Y)\leq t$ because of (\ref{eqcondtSydUEDAlgcwc}) and
$
d_{L_{1}}^{sy}(X,Y)=|\Delta(X,Y)|+2\delta(X,Y).
$
Hence, $X\el\mathcal{A}$, $X'\el\mathcal{A}$ and
$
d_{L_{1}}^{sy}(X,X')
\leq
d_{L_{1}}^{sy}(X,Y)+d_{L_{1}}^{sy}(Y,X')
\leq
t+1+t<2t+2.
$
This implies $X'=X$ because $d_{L_{1}}^{sy}(\mathcal{A})\geq2(t+1)$. So, condition C1) of the theorem is satisfied. Now we prove that if $d_{L_{1}}^{sy}(X,Y)\leq t$ then $cor=1$. First note that, from (\ref{eqdistdef}) and (\ref{eqconddeltatilde}), the following relations hold for any $X\css{def}{=}\hat{X}x_{n},Y\css{def}{=}\hat{Y}y_{n}\el\inat_{m}^{n-1}\times\inat_{m}$:
\setlength{\jot}{-4pt}
\begin{alignat}{1}
|\hat{X}\dotmin\hat{Y}|
\leq
|X\dotmin Y|
=
\frac{d_{L_{1}}^{sy}(X,Y)-\Delta(X,Y)}{2},\notag\\
\label{eqrelxmenoy}\\
|\hat{Y}\dotmin\hat{X}|
\leq
|Y\dotmin X|
=
\frac{d_{L_{1}}^{sy}(X,Y)+\Delta(X,Y)}{2}.\notag
\end{alignat}
Now, if $d_{L_{1}}^{sy}(X,Y)\leq t$ then $|\Delta(X,Y)|\leq d_{L_{1}}^{sy}(X,Y)\leq t$ and so, step S3 is executed. In this case, from the relations in (\ref{eqrelxmenoy}), $d_{L_{1}}^{sy}(X,Y)\leq t$, (\ref{eqtauepsdef}) and (\ref{eqt-+t+leqt}), it follows,
\begin{equation}
\label{eqrelxmenoyhat}
|\hat{X}\dotmin\hat{Y}|
\leq
\tau_{-},
\quad\mbox{and}\quad
|\hat{Y}\dotmin\hat{X}|
\leq
\tau_{+}\leq t-\tau_{-}.
\end{equation}
From the hypothesis (\ref{eqhypdistthmbase}), $d_{L_{1}}^{as}(\hat{\mathcal{A}})\geq t+1$, and so, from (\ref{eqrelxmenoyhat}), decoding algorithm $\mathcal{D}ec(\hat{\mathcal{A}},\tau_{-},t-\tau_{-})$ will give the correct output in step S3.2. Hence, $\hat{X}'=\hat{X}\el\hat{\mathcal{A}}$, and so, from (\ref{eqpardigit}), $X'=X\el\mathcal{A}$ and $d_{L_{1}}^{sy}(X',Y)=d_{L_{1}}^{sy}(X,Y)\leq t$. This implies that $cor=1$ is set in (\ref{eqcordefalg1}). In this way, also condition C2) of the theorem is satisfied.
\end{IEEEproof}

\subsection{Non Systematic Code Design based on \texorpdfstring{$\sigma$}{\textsigma}-codes}
\label{subsecnonsy}
The $L_{1}$ metric $t$-SyEC codes over $\iint_{m}$ are designed based on the $\sigma$-codes defined in \cite{TAL08,TAL10,TAL10b,TAL11a,TAL12a,TAL13,TAL18,TAL19}. The $\sigma$-code theory is based on the sigma polynomials of a word defined below. Let $m\el\inat\cup\{\infty\}$, $\mathbb{F}$ be {\bf any} field and $\partial{S}\sse\mathbb{F}$ be a set of $n\el\inat$ distinct elements in $\mathbb{F}$. The $\sigma$-polynomial associated with a word $X\css{def}{=}(x_{a})_{a\in\partial{S}}\el\iint_{m}^{n}$ is defined as \cite{TAL11a},
\setlength{\jot}{1pt}
\begin{align}
\sigma_{X}(z)\css{def}{=}\,&z^{x_{0}}\prod_{a\in\partial{S}-\{0\}}(1-az)^{x_{a}}=\label{sigmadefm}\\
&z^{x_{0}}\left(1+\sigma_{1}(X)z+\sigma_{2}(X)z^{2}+\ldots\right)\el\mathbb{F}[z].\notag
\end{align}
For example, if $n=7$, $\partial{S}=\{a_{0},a_{1},a_{2},a_{3},a_{4},a_{5},a_{6}\}\sse\mathbb{F}-\{0\}$ and $X=(3021000)=\{a_{0},a_{0},a_{0},a_{2},$ $a_{2},a_{3}\}$
then
\setlength{\jot}{1pt}
\begin{alignat*}{2}
\sigma_{X}(z)=\,&(1-a_{0}z)^{3}(1-a_{2}z)^{2}(1-a_{3}z)=1-\\
&(3a_{0}+2a_{2}+a_{3})z+(3a_{0}^{2}\!+6a_{0}a_{2}+\\
&\!3a_{0}a_{3}+\!a_{2}^{2}+2a_{2}a_{3})z^{2}+\ldots-(a_{0}^{3}a_{2}^{2}a_{3})z^{7}.
\end{alignat*}
Note that $\sigma_{X}(z)$ is a polynomial of degree $\deg(\sigma_{X})=w_{L_{1}}(X)=|X|$ having $w_{H}(X)=|\partial{X}|$ distinct roots in $\mathbb{F}$, each with multiplicity $x_{a}$, for $a\el\partial{S}\sse\mathbb{F}$. In particular, $X$ coincides with the multiset of all the inverses of the roots of $\sigma_{X}(z)$, where we let $1/0\css{def}{=}0$. Hence, its coefficient sequence is given by the elementary symmetric functions, $1$, $\sigma_{1}(X)$, $\sigma_{2}(X)$, $\ldots\el\mathbb{F}$, of the elements in the multiset $X-\{0\}$ ordered in increasing order of their degrees, and eventually right shifted by $x_{0}\el\iint_{m}\sse\inat$ if $0\el\partial{S}\sse\mathbb{F}$. At this point the general definition of $\sigma$-code is the following. For all polynomials $g(z),\sigma(z)\el\mathbb{F}[z]$, the $m$-ary $\sigma$-code of length $n$ associated with $g$ and $\sigma$ is defined as
\setlength{\jot}{0pt}
\begin{alignat}{1}
\mathcal{C}_{g,\sigma}(\iint_{m},n)\css{def}{=}
\left\{
X\el\iint_{m}^{n}\left|\;
\parbox[c]{0.442\linewidth}{
$\sigma_{X}(z)=c_{X}\sigma(z)\bmod{g(z)}$, with $c_{X}\el \mathbb{F}-\{0\}$
}
\right.
\right\}.\label{eqsigmagencodes}
\end{alignat}
For simplicity, here we can choose $g(z) = z^{t+1}$.

To define a $t$-Sy$0$EC code $\mathcal{C}\sse\iint_{2}^{n}$, the $\sigma$-codes are used in the function $\hat{V}$ codomain; where $\hat{V}$ is given in (\ref{eqfunVhat}). So, $X\el\mathcal{C}$ if, and only if $\sigma_{\hat{V}(X)}(z)=\sigma(z)\bmod{z^{t+1}}$, where $\sigma(z)$ is a monic polynomial of degree $t$. Note that under the mapping $X\to\sigma_{\hat{V}(X)}(z)\bmod{z^{t+1}}$, the set of constant weight $w$ vectors of length $n$ over $\iint_{2}$ (and in fact, the set $\mathcal{S}(\inat,w+1,n-w)$) is partitioned into $|\mathbb{F}|^{t}$ classes, $\mathcal{D}_{1}, \mathcal{D}_{2}, \ldots, \mathcal{D}_{\mathbb{F}^{t}}$, where, $X$ and $Y$ are in $\mathcal{D}_{i}$ if, and only if, $\sigma_{\hat{V}(X)}(z)=\sigma_{\hat{V}(Y)}(z)\bmod z^{t+1}$. Now, we prove that each of the $\hat{V}(\mathcal{D}_{i})$'s is an asymmetric $L_1$ distance $t+1$ code. Suppose $X,Y\el\mathcal{D}_{i}$, let $\hat{V}\css{def}{=}\hat{V}(X)$ and $\hat{U}\css{def}{=}\hat{V}(Y)$. Then, $\sigma_{\hat{V}}(z)=\sigma_{\hat{U}}(z) \bmod z^{t+1}$ and this implies $\sigma_{\hat{V}\dotmin\hat{U}}(z)=\sigma_{\hat{U}\dotmin\hat{V}}(z)\bmod z^{t+1}$ because
\begin{equation}
\label{eqkey}
\!\mbox{for all $A,B\el\inat^{n}$,}
\;
\sigma_{A}(z)\sigma_{B\dotmin A}(z)
=
\sigma_{B}(z)\sigma_{A\dotmin B}(z).
\end{equation}
Now, if the asymmetric $L_{1}$ distance between $\hat{V}$ and $\hat{U}$ is $s<t+1$ then the degrees of $\sigma_{\hat{V}\dotmin\hat{U}}(z)$ and $\sigma_{\hat{U} \dotmin\hat{V}}(z)$ are $s< t+1$ and so, $\sigma_{\hat{V}\dotmin\hat{U}}(z)=\sigma_{\hat{U}\dotmin\hat{V}}(z)$. This means, $\sigma_{\hat{V}\dotmin\hat{U}}(z)$ has $2s$ roots (i.~e., the $s$ roots of $\sigma_{\hat{V}\dotmin\hat{U}}(z)$ and the $s$ roots of $\sigma_{\hat{U}\dotmin\hat{V}}(z)$), which gives a contradiction. Therefore, the minimum asymmetric $L_1$ distance of the code is at least $t+1$. So, under the $X\to\sigma_{\hat{V}(X)}(z)\bmod{z^{t+1}}$ mapping the set $\mathcal{S}(\iint_{2},n,w)$ is partitioned into the $|\mathbb{F}|^{t}$ classes $\mathcal{D}_{i}$'s. Thus, by pigeon-hole principle, one of the classes, say $\tilde{\mathcal{D}}(\mathbb{F};n,w)$ should have at least $\binom{n}{w}/|\mathbb{F}|^t$ codewords. From equivalence 7) of Theorem~\ref{thtdi0ECCequidec2}, the $t$-Sy$0$EC code, $\mathcal{C}$, can be simply defined by letting for all $w\el[0,w]$, $\mathcal{C}_{w}\css{def}{=}\tilde{\mathcal{D}}(\mathbb{F};n,w)\sse\mathcal{S}(\iint_{2},n,w)$; where, to maximize $|\mathcal{C}|$, the algebraic structure $\mathbb{F}$ is chosen to be the smallest possible field if $t > 1$ or the smallest group if $t=1$. In this way, the number of codewords is 
\begin{equation}
\label{eqlowerbound}
|\mathcal{C}|\geq\sum_{w=0}^{n}
\isa\left.\binom{n}{w}\right/|\mathbb{F}_{w}|^{t}\isc.
\end{equation}
where $\mathbb{F}_{w}$ is the smallest field, $\mathbb{F}$, whose cardinality is $|\mathbb{F}|>w$, when $t>1$ and $\mathbb{F}_{w}=(\iint_{w+1},+\bmod{(w+1)})$ when $t=1$. Note that if $t=1$, then $|\mathbb{F}_{w}|=w+1$.

In the case of the $\sigma$-codes in (\ref{eqsigmagencodes}), the efficient $(\tau_{-},\tau_{+})$-EC decoding algorithm, $\mathcal{D}ec(\hat{\mathcal{A}},\tau_{-},\tau_{+})$, for the code $\hat{\mathcal{A}}$ is based on the key equation \cite{TAL08,TAL10,TAL10b,TAL11a,TAL12a,TAL13,TAL18b},
\begin{equation}
\label{eqkey2}
\!\!\!
\sigma_{X}(z)\sigma_{Y\dotmin X}(z)
=
\sigma_{Y}(z)\sigma_{X\dotmin Y}(z),
\;
\mbox{for all $X,Y\el\iint_{m}^{n}$},
\end{equation}
relating the $\sigma$-polynomials (\ref{sigmadefm}). Again, for simplicity, assume $\partial{S}\sse\mathbb{F}-\{0\}$ with $|\partial{S}|=n-1$, $g(z)=z^{t+1}$ with $\gcd\{z^{t+1},\tilde{\sigma}(z)\}=1$, so that
\setlength{\jot}{4pt}
\begin{alignat}{1}
\hat{\mathcal{A}}\css{def}{=}&\,\mathcal{C}_{z^{t+1},\tilde{\sigma}}(\iint_{m},n-1)=\label{eqsimpsigmacodehat}\\
&\left\{
\hat{X}\el\iint_{m}^{n-1}\left|\;
\parbox[c]{7em}{
$\sigma_{1}(\hat{X})=\tilde{\sigma}_{1}$,\\
$\sigma_{2}(\hat{X})=\tilde{\sigma}_{2}$, $\ldots$,\\
$\sigma_{t}(\hat{X})=\tilde{\sigma}_{t}$}
\right.
\right\}
\notag
\end{alignat}
and, hence,
\setlength{\jot}{4pt}
\begin{alignat}{1}
\mathcal{A}\css{def}{=}&\,\mathcal{A}_{z^{t+1},\tilde{\sigma}}(\iint_{m},n,w)\css{def}{=}\label{eqsimpsigmacode}\\
&\left\{
X\el\iint_{m}^{n}\left|\;
\parbox[c]{8em}{
$X=\hat{X}\,x_{n}$ with\\
$x_{n}=w-w_{L_{1}}(\hat{X})$,\\
$\sigma_{1}(\hat{X})=\tilde{\sigma}_{1}$,\\
$\sigma_{2}(\hat{X})=\tilde{\sigma}_{2}$, $\ldots$,\\
$\sigma_{t}(\hat{X})=\tilde{\sigma}_{t}$}
\right.
\right\}.
\notag
\end{alignat}
If $X=\hat{X}\,x_{n}\el\mathcal{A}=\hat{\mathcal{A}}\,x_{n}$ is sent and $Y=\hat{Y}\,y_{n}\el\inat^{n}$ is received then, from (\ref{eqkey2}),
\setlength{\jot}{1pt}
\begin{alignat}{1}
\mbox{for all}&\;\mbox{$\hat{X}\el\hat{\mathcal{A}}\sse\iint_{m}^{n-1}$ and $\hat{Y}\el\inat^{n-1}$},\label{eqkeyeqmmodzt}\\
&\sigma_{\hat{Y}\dotmin\hat{X}}(z)=[\sigma_{\hat{Y}}(z)/\sigma(z)]\sigma_{\hat{X}\dotmin\hat{Y}}(z)\bmod{z^{t+1}}\notag
\end{alignat}
where $\sigma_{\hat{Y}\dotmin \hat{X}}(z)$ and $\sigma_{\hat{X}\dotmin\hat{Y}}(z)$ are unknown and $[\sigma_{\hat{Y}}(z)/\sigma(z)]$ is known to the receiver. In this way, algorithm $\mathcal{D}ec(\hat{\mathcal{A}},\tau_{-},\tau_{+})$ consists in solving the equation (\ref{eqkeyeqmmodzt}) with the constrains $\deg(\sigma_{\hat{Y}\dotmin\hat{X}})\leq t_{+}$ and $\deg(\sigma_{\hat{X}\dotmin\hat{Y}})\leq t_{-}$ required by $(\tau_{-},\tau_{+})$-EC decoding. This can be efficiently performed with the Extended Euclidean Algorithm. Note, however, that Algorithm \ref{alggentsyECdUEDcwc} is of general type and can be efficiently applied to any constant weight code, $\mathcal{A}$, with minimum distance $2t+2$ having efficient $(\tau_{-},\tau_{+})$-EC decoding algorithms for its punctured code, $\hat{\mathcal{A}}$. So, in general, efficiently decodable $t$-Sy$0$EC based $\sigma$-codes can be defined by choosing as $\hat{\mathcal{A}}$ the general codes in (\ref{eqsigmagencodes}). In this case, the $t$-Sy$0$EC codes are, $\mathcal{C}=$
$$
\bigcup_{w\in[0,n]}\!\left\{X\el\iint_{2}^{n}:\,V(X)\el\mathcal{A}_{g_{w,t},\tilde{\sigma}_{w,t}}(\iint_{m},w+1,n-w)\right\}\!;
$$
where, for all $w\in[0,n]$, the field $\mathbb{F}_{w,t}$, the code support $\partial S_{w,t}\sse\mathbb{F}_{w}$ and the polynomial $g_{w,t}(z)\el\mathbb{F}_{w}[z]$ define a given triplet, $(\mathbb{F}_{w},\partial S_{w},g_{w}(z))$, of set $\Gamma(w,t)$ defined in \cite[relation (50)]{TAL23}; and $\tilde{\sigma}_{w,t}(z)\el\mathbb{F}_{w,t}[z]$ is a given polynomial such that $\gcd(\tilde{\sigma}_{w,t},g_{w,t})=1$.

Note that, if the receiver knows the check information, $C\css{def}{=}\sigma_{V(X)}(z)\bmod{g(z)}$, of any sent word $X\el\iint_{2}^{n}$ then it is capable of decoding the corresponding received word, $Y\el\iint_{2}^{*}$. Following and improving the fixed length recursive code design idea in \cite{TAL10b}, here, we have give systematic code designs whose strategy is to recursively send a $(t-1)$-Sy$0$EC encoding of $C$ to the receiver; strangely enough, the $(t-1)$-Sy$0$EC capability is enough for the recursive $t$-Sy$0$EC design to be well defined. This comes from the combinatorial properties of the constant weight $\sigma$-codes. So, for these recursive codes, the challenging problem is to give a well defined $t$-Sy$0$EC/$(t+1)$-Sy$0$ED/AU$0$ED error control algorithm by keeping the redundancy below the optimal value of $t\log_{2}k+o(t\log n)$ given by Theorem \ref{thoptrel}. In fact, here we give fixed length $n\el\inat$ systematic recursive $\sigma$-code based asymptotically optimal codes to efficiently encode $k$ information bits. These codes have efficient $t$-Sy$0$EC/$(t+1)$-Sy$0$ED/AU$0$ED error control algorithms and redundancy $n-k\leq t\log_{2}k+o(t\log n)$ bits, for all $k,t\el\inat$.

We conclude this subsection with the following theorem which explicitely gives the lower bounds on the cardinality of non systematic $t$-Sy$0$EC CW codes. For small length values these codes can be efficiently implemented with a table look-up and efficiently decoded by means of Algorithm \ref{alggentsyECdUEDcwc}, and will be used in Subsection \ref{subsecreeds}.
\begin{theorem}[lower bound on $t$-Sy$0$EC CW codes]
\label{thLBtdi0codesCW}
For all $t,n,w\el\inat$ there exists a binary $t$-Sy$0$EC extended $\sigma$-code with constant weight $w$
\begin{align*}
\mathcal{C}(\iint_{2},n,w,t)=\,&V^{-1}\left(\mathcal{A}_{g(z)=z^{t+1},\tilde{\sigma}}(\inat,w+1,n-w)\right)\sse\\
&\mathcal{S}(\iint_{2},n,w)\sse\iint_{2}^{n};
\end{align*}
of length $n$ whose cardinality is
\begin{equation}
\label{eqlowerboundCW}
|\mathcal{C}(\iint_{2},n,w,t)|\geq\isa\left.\binom{n}{w}\right/|\mathbb{F}_{w}|^{t}\isc;
\end{equation}
where $\mathbb{F}_{w}$ is the smallest field, $\mathbb{F}$, whose cardinality is $|\mathbb{F}|>w$, when $t>1$; and $\mathbb{F}_{w}=(\iint_{w+1},+\bmod{(w+1)})$ when $t=1$. In particular, if $t>0$ and $w=\isa n/2\isc$ then the binary balanced $t$-Sy$0$EC extended $\sigma$-code $\mathcal{C}_{t}(n)\css{def}{=}\mathcal{C}(\iint_{2},n,\isa n/2\isc,t-1)$ has cardinality
\begin{equation}
\label{eqlowerboundbalcode}
|\mathcal{C}_{t}(n)|\geq\isa\left.\binom{n}{\isa n/2\isc}\right/n^{t-1}\isc.
\end{equation}
Thus, the length of $\mathcal{C}_{t}(n)$ is
\setlength{\jot}{1pt}
\begin{alignat}{1}
n\css{def}{=}n(t)\leq\,&\log_{2}|\mathcal{C}_{t}(n)|+t\log_{2}n-\frac{1}{2}\log_{2}n+1.17657.\label{eqlowerboundbalcodered0}
\end{alignat}
Given $k\el\irea$, let $n\css{def}{=}n(t,k)$ be such that
\begin{equation}
\label{eqCWtm1sm}
|\mathcal{C}_{t}(n-1)|<2^{k}\leq|\mathcal{C}_{t}(n)|,
\end{equation}
and $r\css{def}{=}r(t,k)\css{def}{=}n(t,k)-k$. If $c\el\irea$ is any real constant such that
\begin{equation}
\label{eqCWtm1hp}
n\geq2^{c}\qquad\mbox{and}\qquad\frac{t\log_{2}e}{k}\leq\frac{1}{2}\;\;\mbox{(i. e., $t\leq0.34657\cdot k$)},
\end{equation}
then
\begin{equation}
\label{eqlowerboundbalcodered}
n(t,k)<k+2t\log_{2}k+c.
\end{equation}
\end{theorem}
\begin{IEEEproof}
Relation (\ref{eqlowerboundCW}) comes from the arguments used to prove (\ref{eqlowerbound}). Relation (\ref{eqlowerboundbalcode}) comes from (\ref{eqlowerboundCW}) for $w=\isa n/2\isc$ and Bertrand's Postulate \cite{BER45}. With regard to relation (\ref{eqlowerboundbalcodered0}), note that if $n\el\inat$ then
\begin{equation}
\label{eqstirling}
\sqrt{2\pi n}n^{n}e^{-n}e^{1/(12n+1)}
<
n!
<
\sqrt{2\pi n}n^{n}e^{-n}e^{1/(12n)};
\end{equation}
and so, for all $n,w\el\inat$ with $w\el[0,n]$,
\begin{alignat}{1}
\log_{2}\!\binom{n}{w}&>nh\left(\frac{w}{n}\right)+\frac{1}{2}\log_{2}n+\frac{\log_{2}e}{12n+1}-\label{eqbetterboundncw}\\
&\frac{1}{2}\left[
\log_{2}[2\pi w(n-w)]+\frac{\log_{2}e}{6w}+\frac{\log_{2}e}{6(n-w)}
\right];\notag
\end{alignat}
where
$$
h(x)\css{def}{=}-[x\log_{2}x+(1-x)\log_{2}(1-x)],\;
\mbox{with $x\el[0,1]$}.
$$
Now, (\ref{eqlowerboundbalcodered0}) comes from (\ref{eqlowerboundbalcode}) and (\ref{eqbetterboundncw}). Let $c'\css{def}{=}c/2\el\irea$. As regard (\ref{eqlowerboundbalcodered}), from (\ref{eqlowerboundbalcodered0}), the leftmost relations in (\ref{eqCWtm1sm}) and (\ref{eqCWtm1hp}), respectively, $n=k+r$ and $\log_{2}(1+x)\leq(\log_{2}e)x$, it follows,
\setlength{\jot}{4pt}
\begin{align}
r\leq\,&c'+t\log_{2}n=\label{eqlowerboundbalcodered1}\\
&c'+t\log_{2}(k+r)=c'+t\log_{2}\left[k\left(1+\frac{r}{k}\right)\right]=\notag\\
&c'+t\log_{2}k+t\log_{2}\left(1+\frac{r}{k}\right)\leq c'+t\log_{2}k+\frac{t\log_{2}e}{k}r.\notag
\end{align}
Let $s\el\inat$ be any natural. By applying (\ref{eqlowerboundbalcodered1}) $s$ times, it follows,
$$
r\leq \left[\sum_{i=0}^{s}\left(\frac{t\log_{2}e}{k}\right)^{i}\right](c'+t\log_{2}k)+t\left(\frac{t\log_{2}e}{k}\right)^{s}\log_{2}\left(1+\frac{r}{k}\right).
$$
Thus, from the second relation in (\ref{eqCWtm1hp}),
\begin{equation}
\label{eqlowerboundbalcodered2}
r(t,k)<2(t\log_{2}k+c')+\left(\frac{1}{2}\right)^{s}t\log_{2}\left(1+\frac{r}{k}\right),\quad\mbox{for all $s\el\inat$}.
\end{equation}
From (\ref{eqlowerboundbalcodered2}), relation (\ref{eqlowerboundbalcodered}) follows because $s$ is independent from $t$ and $k$.
\end{IEEEproof}

\section{Some Simple Systematic Non-recursive Code Designs}
\label{secsimpl}
In this Section some simple systematic code designs which are non-recursive are given. When $t$ is very big, these codes are less redundant than the systematic code construction in Section \ref{secsyste_idea} (see Table \ref{tabV}).

\subsection{Repetition Codes}
\label{subsecrepet}
A $(t+1)$ repetition code is a code where each information symbol is repeated $(t+1)$ times. Such codes fall into this class of codes in Subsection \ref{subseclimit} and so are efficient $t$-Sy$0$EC/$(t+1)$-Sy$0$ED/AU$0$ED codes. A $(t+1)$ repetition code with $k\el\inat$ information symbols has length $n=(t+1)k$ and $r=tk$ check bits.

\subsection{Distinct Weight Codes}
\label{subsecdisti}
In a distinct weight code no two codewords have the same Hamming weight. From (\ref{eqdidugdl0}), the minimum distance of any distinct weight code is $\infty$ and so can be used to correct any number of $0$-errors; in particular, $t\el\inat$ $0$-errors, for any $t$. Just by counting the number of the received $1$'s (which are received error free) the receiver can correct all the $0$-errors and so, a fortiori, they are $t$-Sy$0$EC/$(t+1)$-Sy$0$ED/AU$0$ED codes. An optimal length $n\el\inat$ distinct weight code contains $n+1$ distinct codewords. Hence, an optimal distinct weight code with $k\el\inat$ information bits has length $n=2^{k}-1$. Let $d:\iint_{2}^{k}\to[0,2^{k}-1]$ be the ono-to-one function which associates and $X\el\iint_{2}^{k}$ with the natural number $d(X)$ whose binary representation is $X$. An efficient optimal distinct weight systematic encoding, $\mathcal{E}:\iint_{2}^{k}\to\iint_{2}^{n}$ for $k$ information bits can be defined as
$$
\mathcal{E}(X)
\css{def}{=}
X\;0^{n-[d(X)-w_{H}(X)]}1^{d(X)-w_{H}(X)}.
$$
Table \ref{tabdwcex} gives the systematic distinct weight code with $k=4$ information bits. 
\begin{table*}
\caption{The $16$ codewords for the systematic distinct weight code with $k=4$ information bits and length $n=2^{4}-1=15$.
}
\begin{center}
\renewcommand{\tabcolsep}{2pt}
\renewcommand{\arraystretch}{1.1}
\begin{tabular}{|c||c|c||c|}
\hline
\blackbox{3.5mm}{0mm}{2mm}
$d(X)$ & $X$ & Check of $X$ & $w_{H}(\mathcal{E}(X))$\\\hline\hline
$0$ & $0000$ & $00000000000$ & $0$ \\\hline
$1$ & $0001$ & $00000000000$ & $1$\\\hline
$2$ & $0010$ & $00000000001$ & $2$ \\\hline
$3$ & $0011$ & $00000000001$ & $3$ \\\hline
$4$ & $0100$ & $00000000111$ & $4$ \\\hline
$5$ & $0101$ & $00000000111$ & $5$ \\\hline
$6$ & $0110$ & $00000001111$ & $6$ \\\hline
$7$ & $0111$ & $00000001111$ & $7$ \\\hline
$8$ & $1000$ & $00001111111$ & $8$ \\\hline
$9$ & $1001$ & $00001111111$ & $9$ \\\hline
$10$ & $1010$ & $00011111111$ & $10$ \\\hline
$11$ & $1011$ & $00011111111$ & $11$ \\\hline
$12$ & $1100$ & $01111111111$ & $12$ \\\hline
$13$ & $1101$ & $01111111111$ & $13$ \\\hline
$14$ & $1110$ & $11111111111$ & $14$ \\\hline
$15$ & $1111$ & $11111111111$ & $15$ \\\hline
\end{tabular}
\end{center}
\label{tabdwcex}
\end{table*}

\section{Systematic Recursive \texorpdfstring{$\sigma$}{\textsigma}-code Based Code Design}
\label{secsyste_idea}
\begin{figure*}[t]
\centering
\includegraphics[width=\linewidth]{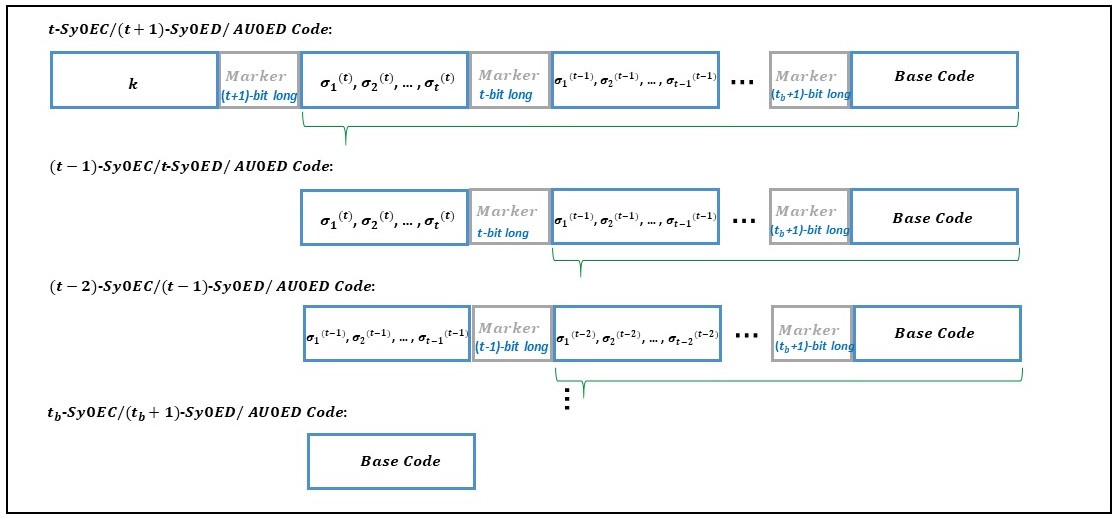}
\vglue -2mm
\caption{Proposed recursive code construction.}
\label{fig1}
\end{figure*}

In the proposed non-systematic $t$-Sy$0$EC code design in Subsection \ref{subsecnonsy}, any given word $X\el\iint_{2}^{k}$ is mapped to
$$
\sigma_{\hat{V}(X)0^{*}}(z)=1+\sigma_{1}\!\left(\hat{V}_{X}\right)z+\ldots+\sigma_{t}\!\left(\hat{V}_{X}\right)z^{t}\bmod{z^{t+1}};
$$
where
\begin{equation}
\label{eqVx}
\hat{V}_{X}
\css{def}{=}
(v_{1},v_{2},\ldots,v_{w},0,0,\ldots,0)
\el
\iint_{k}^{k}\equiv\mathbb{F}^{k}.
\end{equation}
Now, all input words mapping into the same $\sigma_{1},\sigma_{2},\ldots,\sigma_{t}\el\mathbb{F}_{w(X)}$ form a $t$-Sy$0$EC. To design the code, if, for simplicity, we use the same field $\mathbb{F}\css{def}{=}\mathbb{F}_{k}$, for all possible weights $w=w(X)\el[0,k]$, then the set of input vectors is partitioned into $|\mathbb{F}|^{t}$ classes $\mathcal{C}_{1}, \mathcal{C}_{2}, \ldots, \mathcal{C}_{\mathbb{F}^{t}}$, and each of these classes is a $t$-Sy$0$EC (or, equivalently, a $t$-Sy$0$EC/$(t+1)$-Sy$0$ED/AU$0$ED) code. In the proposed systematic $t$-Sy$0$EC recursive code design given in this section, for the given information word $X\el\iint_{2}^{k}$, we first find the values of its $\sigma_{i}\css{def}{=}\sigma_{i}\left(\hat{V}_{X}\right)$'s, $i=1,2,\ldots,t$, and append them as check. Then, assuming these $\sigma_{i}$'s as an information word, interestingly, we encode them with a $(t-1)$-Sy$0$EC (or, $(t-1)$-Sy$0$EC/$t$-Sy$0$ED/AU$0$ED) code. This process continues until a base code is used. However, we need to insert a marker between the successive words generated in these recursive iterations. This code design is shown in Figure~\ref{fig1}.

We now explain why this code gives distance $\did(C)\geq2t+2$. If two information words $X$ and $Y$ map to the same value then the $\did(X, Y)\geq2t+2$. On the other hand if they map to different values, by our construction, the checks will have $\did\geq2t$. Since $X$ and $Y$ are constant weight words,  $\did(X, Y)\geq2$ and so the distance between these two codewords is at least $2t + 2$.

For decoding, note that if we know the check symbols (i.~e., the $\sigma_{i}$'s) then the sent information word can be recovered by first applying the $\hat{V}$ mapping to the information part and then applying the $L_{1}$ metric $t$-Sy$0$EC/$(t+1)$-Sy$0$ED/AU$0$ED error control Algorithm \ref{alggentsyECdUEDcwc} in Subsection \ref{subsectsy0e}. Thus, in the entire decoding process, once the correct parsing of the received information word is done, we sequentially decode the remaining received check part starting from the base code, all the way up to the first iteration. Theorem \ref{thmcorralg0eccrec} shows that this code design can correct $t$ $0$-errors (and actually, it has even more error control capabilities).

Since the proposed efficient code designs rely on the concatenation of some codewords, we need to be aware of the following unexpected behavior of the $\did$ distance. Unlike usual metrics, the metric function $\did$ is not additive with respect to concatenation. 
In fact, for example, if $X_{1}=010$, $X_{2}=010$, $Y_{1}=0001$, $Y_{2}=001$ then
\setlength{\jot}{2pt}
\begin{align*}
\did(X_{1}&X_{2},Y_{1}Y_{2})=\did(010\,010,0001\,001)=3\neq\\
&\did(010,0001)+\did(010,001)=\\
&\did(X_{1},Y_{1})+\did(X_{2},Y_{2})=3+2=5.
\end{align*}
In general, the $\did$ metric is not additive with respect to the concatenation. However, the following theorems hold.
\begin{theorem}[concatenation subadditivity]
\label{thmconcatprop}
\setlength{\jot}{2pt}
\begin{alignat}{1}
\mbox{For all $X_{1},X_{2},Y_{1},Y_{2}\el\iint_{2}^{*}$},\label{concsubadd}\\
\did(X_{1}X_{2},Y_{1}Y_{2})\leq\;&\did(X_{1},Y_{1})+\did(X_{2},Y_{2}).\notag
\end{alignat}
In particular,
\setlength{\jot}{2pt}
\begin{alignat}{1}
&\mbox{for all $X_{1},X_{2},Y_{1},Y_{2}\el\iint_{2}^{*}$},\quad w(X_{1})=w(Y_{1})\;\imp\label{concadd}\\
&\did(X_{1}X_{2},Y_{1}Y_{2})=\did(X_{1},Y_{1})+\did(X_{2},Y_{2})-Q;\notag
\end{alignat}
where the quantity $Q\el\inat$ is defined as,
\setlength{\jot}{2pt}
\begin{alignat*}{1}
Q\css{def}{=}\;&Q(X_{1},X_{2},Y_{1},Y_{2})\css{def}{=}\\
&|v_{w_{1}+1}(X_{1})-v_{w_{1}+1}(Y_{1})|+|v_{1}(X_{2})-v_{1}(Y_{2})|-\\
&|(v_{w_{1}+1}(X_{1}) + v_{1}(X_{2}))- (v_{w_{1}+1}(Y_{1}) + v_{1}(Y_{2}))|\geq0.
\end{alignat*}
\end{theorem}
\begin{IEEEproof}
Please, see the appendix section.
\end{IEEEproof}
\begin{theorem}[concatenation additivity]
\label{thmconcatprop1}
In general, the following relation holds.
\setlength{\jot}{2pt}
\begin{alignat}{1}
&\mbox{for all $X_{1},X_{2},Y_{1},Y_{2}\el\iint_{2}^{*}$},\quad w(X_{1})=w(Y_{1})\;\imp\label{concadd1}\\
&\quad\did(X_{1}1X_{2},Y_{1}1Y_{2})=\notag\\
&\qquad\qquad\did(X_{1}1,Y_{1}1)+\did(X_{2},Y_{2})=\notag\\
&\qquad\qquad\did(X_{1},Y_{1})+\did(X_{2},Y_{2}).\notag
\end{alignat}
\end{theorem}
\begin{IEEEproof}
From Theorem \ref{thmconcatprop}, the first equality follows by replacing $X_{1}$ with $X_{1}1$ and $Y_{1}$ with $Y_{1}1$, respectively. In this case, in (\ref{concadd}), $w(X_{1})=w(Y_{1})$ and $Q=0$. Analogously, the second equality follows by letting $X_{2}=Y_{2}=1$.
\end{IEEEproof}

Synchronization errors due to $0$-errors can be controlled by inserting a marker or synchronization sequence between consecutive codewords in the sequences of codewords that are sent \cite{SEL62,LEV65,FER97}. These markers do not depend on the code used, convey no information, and the receiver can recover the exact alignment of transmitted codeword sequence from the received bit sequence. For this reason, in this recursive systematic code design, we may need to use some $0$-error markers in the code definition. In this case, let $n\el\inat$ be the code length and assume that the received bit sequence is parsed from left to right. If synchronization should be immune by at most $t\el\inat$ $0$-errors in the codeword concatenated with the marker, then an optimal (i.~e., shortest) marker has length $2t+1$ \cite{SEL62,LEV65,FER97}, and can be chosen to be the binary word
$$
M^{sy}_{t}\css{def}{=}\overbrace{00\ldots0}^{\mbox{\scriptsize $2t$ zeros}}1=0^{2t}1\el\iint_{2}^{2t+1}.
$$
In this case, the receiver recovers synchronization by simply parsing at the first $1$ to the right of the $(n+t+1)$-th received bit. Afterwards, the receiver can even correct the received synchronization sequence by eventually adding the missing $0$'s on the left of that first $1$. If, instead, synchronization should be immune by at most $t\el\inat$ $0$-insertion (or $0$-deletion) errors (i.~e., in a totally asymmetric case), then an optimal marker has length $t+1$ and can be chosen to be
\begin{equation}
\label{eqMtas}
M^{tas}_{t}\css{def}{=}\overbrace{00\ldots0}^{\mbox{\scriptsize $t$ zeros}}1=0^{t}1\el\iint_{2}^{t+1}.
\end{equation}
The receiver can recover synchronization by parsing at the first $1$ to the right of the $(n+t+1)$-th (or $(n+1)$-th, respectively) received bit. Here, even if the $0$-deletions (or $0$-insertions) are no more than the $0$-insertions (or $0$-deletions) and all at most $t$ $0$-errors occur before the above rightmost $1$ (or leftmost $0$), then the receiver recovers synchronization by parsing at the first $1$ to the right of the $(n+t+1)$-th (or $(n+1)$-th, respectively) received bit. Then the receiver eventually corrects the received synchronization sequence by adding the missing $0$'s on the left of that $1$. Clearly, if synchronization should be immune by at most $t_{-}\el\inat$ $0$-deletion errors and at most $t_{+}\el\inat$ $0$-insertion errors (i.~e., in the general asymmetric case), then a marker choice is
$$
M^{gas}_{t_{-},t_{+}}\css{def}{=}\overbrace{00\ldots0}^{\mbox{\scriptsize ($t_{+}+t_{-}$) zeros}}\!\!\!\!\!1
=
0^{t_{+}+t_{-}}1\el\iint_{2}^{t_{+}+t_{-}+1}
$$
of length $t_{-}+t_{+}+1$. As above, synchronization is recovered by looking at the $(n+t_{+}+1)$-th received bit and then parsing at the first $1$ to the right of that $(n+t_{+}+1)$-th received bit. Here also, the receiver can correct the received synchronization sequence by adding the missing $0$'s on the left of that $1$ in such a way that the overall length of the received sequence becomes $(n+t_{+}+t_{-}+1)$. In all cases, other choices of markers are possible. We also note that if the information needs not to be systematic then other synchronizing techniques are possible to recover synchronization being immune by at most $t$ $0$-errors of any of the above types, for example, by employing codes of length $n$ which are subsets of Borden codes \cite{BOR82} followed by a $1$. This is because of the Hamming weight (i.~e., the number of $1$'s) of a binary word of length $n$ sent through a $0$-insertion/deletion channel can be modeled as a limited magnitude error channel \cite{TAL18}. When $t=\infty$, the codes are the so called balanced or constant weight codes \cite{KNU86,TAL98,TAL99,PEL15}. Note that the marker technique has the advantage of recovering synchronization both at the beginning and at the end of the marker. Thus, since the marker is unique, it can be corrected and then deleted from the received bit sequence. We will make a profitable use of this in the proposed code design.

Before defining the recursive code design and the decoding algorithm, some base code designs are described in Section \ref{secbasec}.

\section{Base Code Designs}
\label{secbasec}
In this section, two base code designs are given, one in Subsection \ref{subseclimit} and  the other in Subsection \ref{subsecreeds}. Depending on the value of $t/k$, one design gives better information rate than the other as shown in Table \ref{tabV}.

\subsection{Limited Magnitude Based Codes for Small \texorpdfstring{$t/k$}{{\textit t/k}}}
\label{subseclimit}
In \cite{TAL22}, some theory and design of $(t_{d},t_{i})$-A$0$EC codes capable of correcting all zero deletion errors up to $t_{d}$ and all zero insertion errors up to $t_{i}$ in \textbf{each} $0$-run is proposed for any given $t_{d},t_{i}\el\inat$. It is shown that this problem is equivalent to the design problem of All Error Correcting Codes in limited magnitude error channels \cite{TAL18} where the the following max $L_{1}$ distance, $D_{max}^{(L_{1})}(X,Y)$, between $X,Y\el\inat^{n}$,
\begin{equation}
\label{D_max}
{D_{max}^{(L_{1})}}(X,Y)\css{def}{=}\max_{i\in[1, n]}
\left\{
D^{(L_{1})}(x_{i}, y_{i})=|x_{i}-y_{i}|
\right\}
\end{equation}
plays the same important role as the symmetric $L_{1}$ distance in (\ref{eqdistdef}) through the $V$ mapping (\ref{eqfunV}). In fact, a theorem analogous to Theorem \ref{thmisodidl1} holds for the A$0$EC code design problem in \cite{TAL22}, and a code is $(t_{d},t_{i})$-A$0$EC if, and only if, its minimum max $L_{1}$ distance is greater than $D-1\css{def}{=}t_{d}+t_{i}$ (see Theorem 1 in \cite{TAL22}). In particular, efficient $(t_{d},t_{i})$-A$0$EC binary codes, $\mathcal{C}$, of length $n\el\inat$ containing (see (9) in \cite{TAL22})
$$
|\mathcal{C}|
=
\sum_{w=0}^{n}\binom{w+\iia(n-w)/D\iic}{w}
$$
codewords are proposed with their encoding and decoding algorithms. Here, we note that, a fortiori, $(t_{d},t_{i})$-A$0$EC codes are also capable of correcting all zero deletion errors up to $t_{d}$ and all zero insertion errors up to $t_{i}$ in \textbf{the entire} received word (i.~e., they are $(t_{d},t_{i})$-$0$EC). Hence, letting $t=D-1$, from Theorem \ref{thtdi0ECCequidec2} they are $t$-Sy$0$EC/$(t+1)$-Sy$0$ED/AU$0$ED codes and Algorithm \ref{alggentsyECdUEDcwc} defines an efficient $t$-Sy$0$EC/$(t+1)$-Sy$0$ED/AU$0$ED error control algorithm for these codes. Also note that the $(t+1)$ repetition codes in Subsection \ref{subsecrepet} fall into this class of codes and so they can be efficiently decoded as $t$-Sy$0$EC/$(t+1)$-Sy$0$ED/AU$0$ED in the same way.

\subsection{Reed-Solomon Based Codes for Large \texorpdfstring{$t/k$}{{\textit t/k}}}
\label{subsecreeds}
Given $t,k\el\inat$, the basic idea of this code construction is a generalization of the following. Divide $k$ information bits into $\isa k/b\isc$ bytes of $b$ bits. Each of these $b$-bit bytes can be considered as an element in a field $\mathbb{F}$, where $\max\{2^{b},\isa k/b\isc+t\}\leq|\mathbb{F}|$. Design a distance $t+1$ Reed-Solomon code with these bytes as the information digits. Note that this RS code generates $t$ check digits. The next step is to map each codeword digit of the generated RS code to a balanced code. In general, we use a $(\tau-1)$-Sy0EC constant weight codes, for some integer $\tau\geq1$. In this way, the RS code distance can be reduced up to $\iia t/\tau\iic+1$. Finally, to separate the bytes, insert a $1$ after each byte for synchronization. The following example explains this base code design.
\begin{example}
\label{ex1}
Suppose we are given $k=9$ information bits and we want to design a $4$-Sy0EC code. Choose $b=3$ and so the field $\mathbb{F}=GF(2^{3})$ can be used for the code design because $\max\{2^{b},\isa k/b\isc+t\}=\max\{2^{3},9/3+4\}=2^{3}\leq|\mathbb{F}|$. Assume the given information word is
$$
x_{1}x_{2}x_{3}\,x_{4}x_{5}x_{6}\,x_{7}x_{8}x_{9}\css{def}{\equiv}X_{1}\,X_{2}\,X_{3}\el\left(\iint_{2}^{3}\right)^{3}\equiv\iint_{2}^{9}.
$$
Each of the byte, $X_{i}\el\iint_{2}^{3}$, $i=1,2,3$, can be considered as an element in the field $\mathbb{F}=GF(2^{3})$. The $\left(\mathbb{F};7,3,5\right)$ RS code of length $7$ and minimum distance $t+1=5$ can be designed by taking the generator polynomial $g(z)=(z-\alpha^{0})(z-\alpha^{1})(z-\alpha^{2})(z-\alpha^{3})$ where $\alpha$ is a root of the primitive polynomial $z^{3}+z+1$. Thus, $\alpha^{0}\equiv001$, $\alpha^{1}\equiv010$, $\alpha^{2}\equiv100$, $\alpha^{3}\equiv011$, $\alpha^{4}\equiv110$, $\alpha^{5}\equiv111$, $\alpha^{6}\equiv101$ and $\mathbf{0}\equiv000$. For simplicity, assume the given information word is $000\,000\,000\equiv\mathbf{0}\,\mathbf{0}\,\mathbf{0}$ so that its associated RS codeword is
\begin{equation}
\label{eq0excodeword}
(\mathbf{0},\mathbf{0},\mathbf{0},\mathbf{0},\mathbf{0},\mathbf{0},\mathbf{0})\el\mathbb{F}^{7}.
\end{equation}
Now we need to design a one-to-one mapping of the symbols in $\mathbb{F}=GF(2^{3})$ to the codewords of a $(\tau-1)$-Sy0EC constant weight code. For this example, assume $\tau=1$ and so we can use $2$-out-of-$5$ words for this mapping since $\binom{5}{2}=10\geq8=|\mathbb{F}|$. One of these mappings is $\mathbf{0}\equiv000\to00011$, $\alpha^{0}\equiv001\to00101$, $\alpha^{1}\equiv010\to00110$, $\alpha^{2}\equiv100\to01001$, $\alpha^{3}\equiv011\to01010$, $\alpha^{4}\equiv110\to01100$, $\alpha^{5}\equiv111\to10001$, $\alpha^{6}\equiv101\to10010$. Thus, for the all $0$ RS codeword, after this mapping and also adding an additional $1$ at the end of each byte, the codeword is
$$
000111\, 000111\, 000111\, 000111\, 000111\, 000111\, 000111.
$$
Suppose $A$ and $B$ are two codewords. Since the Hamming distance between them is at least five and each symbol is mapped into a balanced $2$-out-$5$ codeword, the $D_{0-D/I}\geq 2\cdot 5=10$. Thus the code can correct $4$-Sy$0$EC/$5$-Sy$0$ED/AU0ED code. 

Now we explain how the $t=4$ $0$-error correction is done. This is based on, as explained later, $e\el[0,t]$ erasures error correction, $\theta\css{def}{=}\iia(t-e)/2\iic$ error correction and $\delta\css{def}{=}\isa(t-e)/2\isc$ error detection ($e$-EEC/$\theta$-EC/$\delta$-ED) for this code. In particular, the $e$-EEC/$\theta$-EC/$\delta$-ED error control algorithm for Reed-Solomon code is used to simulate the $(e+2\theta)$-Sy$0$EC/$(e+2\theta+1)$-Sy$0$ED part of the control algorithm for this code. Since a $1$ is inserted at the end of each byte, by reading from left to right of the received word, the bytes can be parsed correctly even with some $0$-errors. In general, if a single or an odd number of $0$-error occurs in a byte, then this byte can be identified as erroneous and, hence, set equal to an erasure byte. In the case that two or an even number of $0$-errors occur in a byte, then that byte is an erroneous byte which, a priori, can not be identified as erroneous. For example, suppose the received word is
$$
001011\, 0001011 \,0000111\,000111\, 000111\, 000111\, 000111.
$$
By counting the number of $1$'s from left to right, it can be noticed that the balanced encoding of the second and third bytes are $6$ bit long (excluding the synchronizing bit $1$) and so these can be set as erasure bytes by the receiver. After inverse mapping from $2$-out-of-$5$ codewords to $GF(2^{3})$ the received word is
$$
(\alpha^{0}\equiv001,*,*, \mathbf{0},\mathbf{0},\mathbf{0},\mathbf{0})\el\left\{\mathbb{F}\cup\{*\}\right\}^{7};
$$
where ``$*$'' stands for an erasure symbol. Using $2$-EEC/$1$-EC decoding algorithm, the receiver can correct and obtain the sent codeword in (\ref{eq0excodeword}).
\QEDA
\end{example}

Now, generalizing Example \ref{ex1}, for all $t,k\el\inat-\{0\}$, with $t\lesssim2\left(\log_{e}\sqrt{2}\right)^{2}k\log^{(2)}_{2}(k+t)$, efficient designs are given for $t$-Sy$0$EC/$(t+1)$-Sy$0$ED/AU$0$ED codes with $k$ information bits and length
\begin{equation}
\label{eqredBCD_A}
n(t,k)\lesssim k+\sqrt{8tk\log^{(2)}_{2}(k+t)}+2t\log^{(2)}_{2}(k+t);
\end{equation}
where $\log^{(i)}(x)\css{def}{=}\log(\log^{(i-1)}(x))$ and $\log^{(0)}(x)\css{def}{=}x$; for all $x\el\irea$ and $i\el\inat$.

Given $t,k,\tau,b\el\inat-\{0\}$, let $k_{b}\css{def}{=}k\bmod{b}\el[0,b)\cap\inat$ and consider any information word $X\el\iint_{2}^{k}$ as a sequence of $\isa k/b\isc$ binary words $X_{1},X_{2},\ldots,X_{\iia k/b\iic}\el\iint_{2}^{b}$ and $X_{\iia k/b\iic+1}\el\iint_{2}^{k_{b}}$. In this way,
$$
X
\equiv
X_{1}\,X_{2}\,\ldots\,X_{\isa k/b\isc}
\el
\left(\iint_{2}^{b}\right)^{\iia k/b\iic}\times\iint_{2}^{k_{b}}
\equiv
\iint_{2}^{k}.
$$
Now, let $\mathbb{F}$ be any field whose cardinality is
\setlength{\jot}{2pt}
\begin{alignat}{1}
\nu\css{def}{=}&\,\nu(t,k,\tau,b)\css{def}{=}\label{eqnuA}\\
&\max\left\{2^{b},\isa\frac{k}{b}\isc+\isa\frac{t-(\tau-1)}{\tau}\isc\right\}\leq|\mathbb{F}|\notag
\end{alignat}
and note that if the field $\mathbb{F}\css{def}{=}\mathbb{F}_{\nu-1}$ then
\begin{equation}
\label{eqlBERA}
\nu\leq|\mathbb{F}|<2\nu-1.
\end{equation}
Relation (\ref{eqlBERA}) is a consequence of Bertrand's postulate \cite{BER45}. Assume (\ref{eqlBERA}) holds, let
\setlength{\jot}{4pt}
\begin{alignat}{1}
&k_{\scriptscriptstyle RS}\css{def}{=}k_{\scriptscriptstyle RS}(k,b)\css{def}{=}\isa\frac{k}{b}\isc,\label{eqRSparam}\\
&t_{\scriptscriptstyle RS}\css{def}{=}t_{\scriptscriptstyle RS}(t,\tau)\css{def}{=}\isa\frac{t-(\tau-1)}{\tau}\isc=\iia\frac{t}{\tau}\iic,\notag\\
&n_{\scriptscriptstyle RS}\css{def}{=}n_{\scriptscriptstyle RS}(t,k,\tau,b)\css{def}{=}k_{\scriptscriptstyle RS}+t_{\scriptscriptstyle RS},\notag
\end{alignat}
and consider the (eventually extended) Reed-Solomon code over the alphabet $\mathbb{F}$,
$$
\mathcal{C}_{\scriptscriptstyle RS}=\mathcal{C}_{\scriptscriptstyle RS}\left(\mathbb{F};n_{\scriptscriptstyle RS},k_{\scriptscriptstyle RS},t_{\scriptscriptstyle RS}+1\right),
$$
of length
\begin{equation}
\label{eql_{RS}A}
n_{\scriptscriptstyle RS}=k_{\scriptscriptstyle RS}+t_{\scriptscriptstyle RS}\leq\nu\leq|\mathbb{F}|
\end{equation}
with $k_{\scriptscriptstyle RS}=\isa k/b\isc$ information digits,
\begin{equation}
\label{eqtrsbound}
t_{\scriptscriptstyle RS}=\isa\frac{(t-\tau+1)}{\tau}\isc\leq\frac{t}{\tau}
\end{equation}
check digits and minimum Hamming distance $t_{\scriptscriptstyle RS}+1$. From (\ref{eqnuA}), it follows
$$
\left|\iint_{2}^{k_{b}}\right|<\left|\iint_{2}^{b}\right|=2^{b}\leq|\mathbb{F}|,
$$
and so, $\iint_{2}^{b}\hookrightarrow\mathbb{F}$ (or $\iint_{2}^{k_{b}}\hookrightarrow\mathbb{F}$), that is, any $X\el\iint_{2}^{b}$ (or $X\el\iint_{2}^{k_{b}}$) can be considered as an element in the field $\mathbb{F}$ through a fixed one-to-one correspondance from $\iint_{2}^{b}$ (or $\iint_{2}^{k_{b}}$, respectively) to $\mathbb{F}$. In this way, we can assume that any information word $X\el\iint_{2}^{k}$ is really a $\isa k/b\isc$-tuple of field elements. Under this representation, for any
$X\el\mathbb{F}^{k_{\scriptscriptstyle RS}}\equiv\iint_{2}^{k}$, let
\setlength{\jot}{2pt}
\begin{alignat}{1}
&\mathcal{E}_{\scriptscriptstyle RS}(X)\css{def}{=}R=\left(R_{1},R_{2},\ldots,R_{n_{\scriptscriptstyle RS}}\right)\css{def}{=}\label{eqE_{RS}A}\\
&\;\left(X_{1},X_{2},\ldots,X_{k_{\scriptscriptstyle RS}},C_{1},C_{2},\ldots,C_{t_{\scriptscriptstyle RS}}\right)\el\mathcal{C}_{\scriptscriptstyle RS}\sse\mathbb{F}^{n_{\scriptscriptstyle RS}}\notag
\end{alignat}
be the RS codeword associated with $X$. At this point, the binary codeword to be sent through the insertion/deletion channel is obtained by concatenation of the encodings of each $\mathbb{F}$-digit of the above RS codeword to any binary $(\tau-1)$-Sy$0$EC constant weight code followed by a single bit equal to $1$ (in this way, the rightmost bit of any digit is $1$). For example, the $\sigma$-code based constant weight codes given in Subsection \ref{subsecnonsy} can be used efficiently for small values of $\log_{2}|\mathbb{F}|$. With this constant weight encoding ending with a $1$, some $0$-errors in the codeword associated with a digit affect at most that digit. In fact, the propagation of synchronization errors due to the occurrence of $0$-errors in the transmission of a constant weight codeword is prevented by simply counting the number of $1$'s of the received word, if parsing is done from left to right. Note that, from (\ref{eqlowerboundCW}), there exist such $(\tau-1)$-Sy$0$EC constant weight $w$ $\sigma$-codes of length $n$, say $\mathcal{C}_{\tau}(n,w)$, with exactly
$$
|\mathcal{C}_{\tau}(n,w)|=\isa\left.\binom{n}{w}\right/|\mathbb{F}_{w}|^{\tau-1}\isc
$$
codewords. In particular, let $\mathcal{C}_{\tau}(n)\css{def}{=}\mathcal{C}_{\tau}(n,\iia n/2\iic)$ be the balanced $(\tau-1)$-Sy$0$EC $\sigma$-codes. Also, for any set $\mathcal{I}$, let
$
\tilde{n}\css{def}{=}\tilde{n}(\mathcal{I},\tau)
$
be the smallest $n\el\inat$, such that $|\mathcal{I}|\leq|\mathcal{C}_{\tau}(n)|$ and
\begin{equation}
\label{eqbetadef}
\beta_{\mathcal{I},\tau}1:\mathcal{I}\to\mathcal{C}_{\tau}(\tilde{n})1\sse\iint_{2}^{\tilde{n}+1}
\end{equation}
be any encoding for the elements in $\mathcal{I}$ to the elements in $\mathcal{C}_{\tau}(\tilde{n})$ followed by the symbol $1$. Now note that, from the following well known approximation
$$
\sqrt{2\pi n}n^{n}e^{-n}e^{1/(12n+1)}<n!<\sqrt{2\pi n}n^{n}e^{-n}e^{1/(12n)};
$$
the following lower bound holds
\setlength{\jot}{4pt}
\begin{alignat}{1}
\binom{n}{w}\geq\,&\frac{2^{h(w/n)\cdot n}}{\sqrt{8n(w/n)(1-w/n)}}=\label{eqwkbinLB}\\
&2^{h(w/n)\cdot n-(1/2)[\log_{2}w+3+\log_{2}(1-w/n)]}\notag\\
&\qquad\qquad\mbox{for $n,w\el\inat$ and $w\el[1,n-1]$}.\notag
\end{alignat}
So, from the minimality of $\tilde{n}$, Bertrand's postulate \cite{BER45} and (\ref{eqwkbinLB}),
$$
|\mathcal{I}|>|\mathcal{C}_{\tau}(\tilde{n}-1)|\geq
$$
$$
\isa\left.\binom{\tilde{n}-1}{\iia (\tilde{n}-1)/2\iic}\right/|\mathbb{F}_{\iia (\tilde{n}-1)/2\iic}|^{\tau-1}\isc\geq
$$
$$
\left.\binom{\tilde{n}-1}{\iia (\tilde{n}-1)/2\iic}\right/(\tilde{n}-1)^{\tau-1}\gtrsim
$$
$$
\frac{2^{h(1/2)\cdot(\tilde{n}-1)}}{\sqrt{2(\tilde{n}-1)}(\tilde{n}-1)^{\tau-1}}=\frac{2^{\tilde{n}-1}}{\sqrt{2}(\tilde{n}-1)^{(\tau-1/2)}},
$$
and so,
$$
\kappa\css{def}{=}\log_{2}|\mathcal{I}|\gtrsim\tilde{n}-1-1/2-(\tau-1/2)\log_{2}(\tilde{n}-1)\imp
$$
\begin{equation}
\label{eqtildenub}
\tilde{n}\lesssim\kappa+3/2+(\tau-1/2)\log_{2}(\tilde{n}-1)\leq\kappa+\tau\log_{2}\tilde{n}-1.
\end{equation}
Hence, if $\tau\leq\kappa\log_{e}2$ then
$$
\tilde{r}\css{def}{=}\tilde{n}-\kappa\lesssim\tau\log_{2}\tilde{n}=\tau\log_{2}(\kappa+\tilde{r})=
$$
$$
\tau\log_{2}\kappa+\tau\log_{2}\left(1+\frac{\tilde{r}}{\kappa}\right)
\leq
\tau\log_{2}\kappa+\frac{\tau\tilde{r}}{\kappa\log_{e}2}
\lesssim
$$
$$
\tau\log_{2}\kappa+\frac{\tau^{2}\log_{2}(\kappa+\tilde{r})}{\kappa\log_{e}2}=
\tau\log_{2}\kappa+\frac{\tau^{2}\log_{2}\kappa}{\kappa\log_{e}2}+\frac{\tau^{2}}{\kappa\log_{e}2}\log_{2}\left(1+\frac{\tilde{r}}{\kappa}\right)
\leq
$$
$$
\tau\log_{2}\kappa+\frac{\tau^{2}\log_{2}\kappa}{\kappa\log_{e}2}+\frac{\tau^{2}\tilde{r}}{(\kappa\log_{e}2)^{2}}
\lesssim
\tau\log_{2}\kappa+\frac{\tau^{2}\log_{2}\kappa}{\kappa\log_{e}2}+\frac{\tau^{3}\log_{2}(\kappa+\tilde{r})}{(\kappa\log_{e}2)^{2}}\leq
$$
$$
\tau\log_{2}\kappa+\frac{\tau^{2}\log_{2}\kappa}{\kappa\log_{e}2}+\frac{\tau^{3}\log_{2}\kappa}{(\kappa\log_{e}2)^{2}}+\frac{\tau^{3}\tilde{r}}{(\kappa\log_{e}2)^{3}}
\lesssim\ldots\lesssim
$$
$$
\tau\log_{2}\kappa\sum_{i=0}^{+\infty}\left(\frac{\tau}{\kappa\log_{e}2}\right)^{i}=\tau\log_{2}\kappa\cdot\frac{\kappa\log_{e}2}{\kappa\log_{e}2-\tau}.
$$
At this point, from (\ref{eqtildenub}) and the above, if, in particular,
$$
\frac{\tau}{\kappa}=\frac{\tau}{\log_{2}|\mathcal{I}|}\leq\frac{\log_{e}2}{2}=\log_{e}\sqrt{2}=0.34657
$$
then $\kappa\log_{e}2/(\kappa\log_{e}2-\tau)\leq2$ and then the length, $n_{bal}\css{def}{=}\tilde{n}$, of the encoding $\beta_{\mathcal{I},\tau}1$ is,
\setlength{\jot}{1pt}
\begin{alignat}{1}
n_{bal}+1\lesssim\log_{2}|\mathcal{I}|+2\tau\log_{2}\log_{2}|\mathcal{I}|;\label{eqnbalA}
\end{alignat}
as $|\mathcal{I}|$ (and hence, $n_{bal}$) increases. Now, given (\ref{eqE_{RS}A}), the encoding of any information word $X\el\iint_{2}^{k}$ to a $t$-Sy$0$EC codes of length $n$ can be simply defined by applying $\beta_{\iint_{2}^{b},\tau}1$ (or $\beta_{\iint_{2}^{k_{b}},\tau}1$) to any information digit and $\beta_{\mathbb{F},\tau}1$ to any check digit of the RS codeword associated with $X$, as follows,
\setlength{\jot}{2pt}
\begin{alignat}{1}
&\mathcal{E}(X)\css{def}{=}\mathcal{E}_{\mathbb{F},t,\tau,k,b}(X)\css{def}{=}\beta1(\mathcal{E}_{\scriptscriptstyle RS}(X))=\beta1(R)=\label{eqcorEbaseA}\\
&\;\beta1(\left(R_{1},R_{2},\ldots,R_{n_{\scriptscriptstyle RS}}\right))\css{def}{=}\beta_{\iint_{2}^{b},\tau}(X_{1})1\beta_{\iint_{2}^{b},\tau}(X_{2})1\ldots\notag\\
&\;\;\beta_{\iint_{2}^{b},\tau}\left(X_{\iia k/b\iic}\right)\!1\beta_{\iint_{2}^{k_{b}}\!,\tau}\left(X_{k_{b}}\right)\!1\beta_{\mathbb{F},\tau}(C_{1})1\beta_{\mathbb{F},\tau}(C_{2})1\ldots\notag\\
&\quad\;\beta_{\mathbb{F},\tau}(C_{t_{\scriptscriptstyle RS}})1\el\iint_{2}^{n};\notag
\end{alignat}
where
$$
\beta1:\left(\iint_{2}^{b}\right)^{\iia k/b\iic}\times\iint_{2}^{k_{b}}\times\mathbb{F}^{t_{\scriptscriptstyle RS}}
\to
\mathcal{B}_{\tau}1\sse\mathcal{S}(\iint_{2},n,w),
$$
is a constant weight
\setlength{\jot}{4pt}
\begin{alignat*}{1}
w\css{def}{=}\,&w(t,k,\tau,b)=\iia\frac{k}{b}\iic\isa\frac{n_{bal}\left(\iint_{2}^{b},\tau\right)}{2}\isc+\\
&\isa\frac{n_{bal}\left(\iint_{2}^{k_{b}},\tau\right)}{2}\isc+t_{\scriptscriptstyle RS}\isa\frac{n_{bal}(\mathbb{F},\tau)}{2}\isc+n_{\scriptscriptstyle RS}
\end{alignat*}
concatenation encoding of length (below, $\rho(a)\css{def}{=}0$ if $a=0$ and $\rho(a)\css{def}{=}1$ if $a\neq0$, for $a\el\inat$)
\setlength{\jot}{6pt}
\begin{alignat}{1}
n\css{def}{=}\,&n(t,k,\tau,b)=\iia\frac{k}{b}\iic\left(n_{bal}\left(\iint_{2}^{b},\tau\right)+1\right)+\label{eqexactlengEbaseA}\\
&\rho(k_{b})\left(n_{bal}\left(\iint_{2}^{k_{b}},\tau\right)+1\right)+t_{\scriptscriptstyle RS}(n_{bal}(\mathbb{F},\tau)+1);\notag
\end{alignat}
and,
$$
\mathcal{B}_{\tau}1\css{def}{=}[\mathcal{C}_{\tau}(n_{1})1]^{\iia k/b\iic}\times[\mathcal{C}_{\tau}(n_{2})1]^{k_{b}}\times[\mathcal{C}_{\tau}(n_{3})1]^{t_{\scriptscriptstyle RS}}
$$
with
\setlength{\jot}{2pt}
\begin{alignat*}{1}
&n_{1}\css{def}{=}n_{bal}\left(\iint_{2}^{b},\tau\right),\;n_{2}\css{def}{=}n_{bal}\left(\iint_{2}^{k_{b}},\tau\right),\;\mbox{and}\\
&n_{3}\css{def}{=}n_{bal}(\mathbb{F},\tau).
\end{alignat*}
Now, assume the code word $E\css{def}{=}\mathcal{E}(X)$ in (\ref{eqcorEbaseA}) is sent. On receiving a word
\begin{equation}
\label{eqrecwordbase}
E'=e'_{1}e'_{2}\ldots e'_{n'}\el\iint_{2}^{n'},
\end{equation}
the receiver can correct the errors by first parsing $E'$ from left to right aiming to recover the received version,
\setlength{\jot}{2pt}
\begin{alignat}{1}
R'\css{def}{=}&\,(\beta1)^{-1}\left(E'\right)\css{def}{=}\left(R'_{1},R'_{2},\ldots,R'_{n_{\scriptscriptstyle RS}}\right)\css{def}{=}\label{eqrecwordbase2}\\
&\left(X'_{1},X'_{2},\ldots,X'_{\isa k/b\isc},C'_{1},C'_{2},\ldots,C'_{t_{\scriptscriptstyle RS}}\right)\el\left(\mathbb{F}\cup\{*\}\right)^{n_{\scriptscriptstyle RS}}\notag
\end{alignat}
of the RS codeword, $R$, associated with $X$, where ``$*$'' stands for an erasure symbol. Note that if $E'\el\iint_{2}^{n'}$ is parsed correctly, the situation is as if the Reed-Solomon codeword $R\el\mathbb{F}^{n_{\scriptscriptstyle RS}}$ as in (\ref{eqE_{RS}A}) was sent on a $|\mathbb{F}|$-ary symmetric channel and some $R'\el(\mathbb{F}\cup\{*\})^{n_{\scriptscriptstyle RS}}$ as in (\ref{eqrecwordbase2}) was received. Also, we readily note that the minimum $0$-deletion/insertion distance of our code
$$
\mathcal{C}\css{def}{=}\left\{\mathcal{E}(X):\;X\el\iint_{2}^{k}\right\}
$$
defined in (\ref{eqcorEbaseA}) is no less than $2(t+1)$. In fact, let
$$
\mathcal{C}_{\tau}(\mathcal{I})1
=
\{\beta_{\mathcal{I},\tau}1(R):\;R\el\mathcal{I}\}\sse\iint_{2}^{n_{bal}(\mathcal{I},\tau)+1}
$$
be the binary $(\tau-1)$-Sy$0$EC constant weight code followed by a single synchronizing bit equal to $1$, for $\mathcal{I}=\iint_{2}^{b},\iint_{2}^{k_{b}}$ and $\mathbb{F}$. In general, note that for all $R,R'\el\mathcal{I}\hookrightarrow\mathbb{F}$,
\setlength{\jot}{2pt}
\begin{alignat*}{1}
R\neq R' \imp&\beta_{\mathcal{I},\tau}1(R),\beta_{\mathcal{I},\tau}1(R')\el\mathcal{C}_{\tau}(\mathcal{I})1\;\mbox{and}\\
&\qquad\beta_{\mathcal{I},\tau}1(R)\neq\beta_{\mathcal{I},\tau}1(R');
\end{alignat*}
Hence, from Theorem \ref{thtdi0ECCequidec2} and $\did(\mathcal{C}_{\tau}(\mathcal{I})1)>2(\tau-1)$, for all $R,R'\el\mathcal{I}$,
\setlength{\jot}{2pt}
\begin{alignat}{1}
R\neq R'\imp\did\left(\beta_{\mathcal{I},\tau}1(R),\beta_{\mathcal{I},\tau}1(R')\right)\geq2\tau.\label{eqRneqRpimpdBB2tau}
\end{alignat}
Now, if $E'=\beta1(R')$, for some $R'\el\left(\iint_{2}^{b}\right)^{\iia k/b\iic}\times\iint_{2}^{k_{b}}\times\mathbb{F}^{t_{\scriptscriptstyle RS}}$, then, from the constant weight property of the codes $\mathcal{C}_{\tau}(\mathcal{I})1$, from (\ref{eqcorEbaseA}), (\ref{eqrecwordbase2}), Theorem \ref{thmconcatprop1} and (\ref{eqRneqRpimpdBB2tau}), it follows,
\setlength{\jot}{2pt}
\begin{alignat*}{1}
\did(E,E')=\;&\sum_{i=1:\,R_{i}\neq R'_{i}}^{n_{\scriptscriptstyle RS}}\did(\beta_{\mathcal{I},\tau}1(R_{i}),\beta_{\mathcal{I},\tau}1(R'_{i}))\geq\notag\\
&2\tau\sum_{i=1:\,R_{i}\neq R'_{i}}^{n_{\scriptscriptstyle RS}}1=2\tau d_{H}(R,R');\notag
\end{alignat*}
that is,
\begin{equation}
\label{eqdHm2dddiBBp}
d_{H}(R,R')
\leq
\frac{\did(E,E')}{2\tau}.
\end{equation}
So, if $E'=\beta1(R')$, with $R'\el\left(\iint_{2}^{b}\right)^{\iia k/b\iic}\times\iint_{2}^{k_{b}}\times\mathbb{F}^{t_{\scriptscriptstyle RS}}$, then no more than $t$ 0-errors in $E'$ cause no more than $\iia t/(2\tau)\iic$ Hamming errors in $R'$. In particular, if $E$ and $E'$ are both codewords of our code $\mathcal{C}$ then $R$ and $R'$ are codewords of $\mathcal{C}_{\scriptscriptstyle RS}$, and so, from (\ref{eqdHm2dddiBBp}) and (\ref{eqRSparam}), the minimum distance of $\mathcal{C}$ is
\setlength{\jot}{6pt}
\begin{alignat}{1}
\did(\mathcal{C})\geq\,&2\tau d_{H}(\mathcal{C}_{\scriptscriptstyle RS})\geq2\tau(t_{RS}+1)=\label{eqmindisRSbc}\\
&2\tau\left(\isa\frac{t-(\tau-1)}{\tau}\isc+1\right)\geq\\
&2\tau\left(\frac{t-\tau+1}{\tau}+1\right)=2\tau\left(\frac{t+1}{\tau}\right)=\\
&2(t+1);
\end{alignat}
that is, from Theorem \ref{thtdi0ECCequidec2}, there exists a $t$-Sy$0$EC/$(t+1)$-Sy$0$ED/AU$0$ED decoding algorithm for $\mathcal{C}$. An efficient design of such an algorithm may be given as follows. On receiving the binary sequence $E'\el\iint_{2}^{n'}$ as in (\ref{eqrecwordbase}), the receiver starts to recover the rightmost $\iia k/b\iic$ received information digits $X'_{1},X'_{2},\ldots,X'_{\iia k/b\iic}$ by letting $j_{0}=0$ and counting the number of $1$'s from left to right so that to identify the positions $j_{i}\el[1,n']$ of the $[i*(\isa n_{1}/2\isc+1)]$-th $1$ in $E'$, for all integers $i\el[1,\iia k/b\iic]$. While doing so, it considers
\begin{equation}
\label{eqBi}
B'_{i}\css{def}{=}e'_{(j_{i-1})+1}e'_{(j_{i-1})+2}\ldots e'_{(j_{i})-1}\el\iint_{2}^{j_{i}-(j_{i-1})-1}
\end{equation}
and applies, with the word $B'_{i}$ as input, any $(\tau-1)$-Sy$0$EC/$\tau$-Sy$0$ED/AU$0$ED algorithm for the binary $(\tau-1)$-Sy$0$EC balanced code $\mathcal{C}_{\tau}(n_{1})$. Such an algorithm exists because of Theorem \ref{thtdi0ECCequidec2}. Let $\tilde{B'}_{i}$ be the output from this algorithm. If errors are corrected and $(\beta_{\iint_{2}^{b},\tau}1)^{-1}(\tilde{B'}_{i}1)$ is well defined then the receiver sets
$$
X'_{i}=\left(\beta_{\iint_{2}^{b},\tau}1\right)^{-1}\left(\tilde{B'}_{i}1\right)\el\iint_{2}^{b}\hookrightarrow\mathbb{F}.
$$
Otherwise, it sets $X'_{i}=*$. Afterwards, only if $k_{b}=k\bmod{b}\neq0$, the receiver recovers the received information digit $X'_{\iia k/b\iic+1}$ by counting the number of $1$'s from left to right so that to identify the position $j_{\iia k/b\iic+1}\el[1,n']$ of the $[i\cdot(\isa n_{2}/2\isc+1)]$-th $1$ in $E'$. In this case, it considers
$$
B'_{\iia k/b\iic+1}\css{def}{=}e'_{j_{\iia k/b\iic}+1}e'_{j_{\iia k/b\iic}+2}\ldots e'_{(j_{\iia k/b\iic+1})-1}
$$
and applies, with the word $B'_{\iia k/b\iic+1}$ as input, any $(\tau-1)$-Sy$0$EC/$\tau$-Sy$0$ED/AU$0$ED algorithm for the binary balanced code $\mathcal{C}_{\tau}(n_{2})$. Let $\tilde{B'}_{\iia k/b\iic+1}$ be the output from this algorithm. If errors are corrected and $(\beta_{\iint_{2}^{k_{b}},\tau}1)^{-1}(\tilde{B'}_{\iia k/b\iic+1}1)$ is well defined then the receiver sets
$$
X'_{\iia k/b\iic+1}=\left(\beta_{\iint_{2}^{k_{b}},\tau}1\right)^{-1}\left(\tilde{B'}_{\iia k/b\iic+1}1\right)\el\iint_{2}^{k_{b}}\hookrightarrow\mathbb{F}.
$$
Otherwise, it sets $X'_{\iia k/b\iic+1}=*$. Finally, the receiver recovers the received check digits $C'_{1},C'_{2},\ldots,C'_{t_{\scriptscriptstyle RS}}$ by counting the number of $1$'s from left to right so that to identify the positions $j_{i}\el[1,n']$ of the $[i\cdot(\isa n_{3}/2\isc+1)]$-th $1$ in $E'$, for all integers $i\el[1,t_{\scriptscriptstyle RS}]$. Again, while doing so, it considers $B'_{i}$ as in (\ref{eqBi}) and applies, with the word $B'_{i}$ as input, any $(\tau-1)$-Sy$0$EC/$\tau$-Sy$0$ED/AU$0$ED algorithm for the binary balanced code $\mathcal{C}_{\tau}(n_{3})$. Let $\tilde{B'}_{i}$ be the output from this algorithm. If errors are corrected and $(\beta_{\mathbb{F},\tau}1)^{-1}(\tilde{B'}_{i}1)$ is well defined then the receiver sets
$$
C'_{i}=\left(\beta_{\mathbb{F},\tau}1\right)^{-1}\left(\tilde{B'}_{i}1\right)\el\mathbb{F}.
$$
Otherwise, it sets $X'_{i}=*$. In this way, the decoder computes the $n_{\scriptscriptstyle RS}=k_{\scriptscriptstyle RS}+t_{\scriptscriptstyle RS}$ received digits
$$
R'_{i}\el\mathbb{F}\cup\{*\},\;i\el[1,n_{\scriptscriptstyle RS}].
$$
Note that $E'\el\iint_{2}^{n'}$ is parsed correctly because the $1$-errors are forbidden. For the time being assume $\tau=1$. In this case, from (\ref{eqRSparam}), $t_{\scriptscriptstyle RS}=t$ and the code $V(\beta_{\mathcal{I}}1(\mathcal{I}))$ can detect $1$ (or, an odd number of) $L_{1}$ error(s) because $d_{L_{1}}^{sy}\left(V(\beta_{\mathcal{I}}1(\mathcal{I}))\right)\geq2$. For this reason, given any $i\el[1,n_{\scriptscriptstyle RS}]$, if a $B'_{i}$ is affected by $1$ (or, an odd number of) $0$-error(s) then this will cause an erasure in the above computation of $R'_{i}$; otherwise (i.~e., if the number of $0$-errors is $2$ or, more generally, even), the computation will miscorrect and cause $R'_{i}\neq R_{i}$, that is, an error. So, since $t_{\scriptscriptstyle RS}=t$, the receiver can successfully apply $e\el[0,t]$ erasure error correction combined with $\iia(t-e)/2\iic$ error correction/$\isa(t-e)/2\isc$ error detection for the code $\mathcal{C}_{\scriptscriptstyle RS}$ with input word $R'$, and either recover $R$, and hence, $\mathcal{E}(X)$, or detect more than $t$ $0$-errors. Indeed, from Theorem \ref{thtdi0ECCequidec2}, the code $\mathcal{C}$ can be used to control more than $t$ $0$-errors, for example, it can be used as a $t$-Sy$0$EC/$d$-Sy$0$ED/AU$0$ED code with $d=t+1$. A $t$-Sy$0$EC/$(t+1)$-Sy$0$ED/AU$0$ED algorithm is given below and is based on $e\el[0,t_{\scriptscriptstyle RS}]$ erasure error corrections, $\theta\css{def}{=}\iia(t_{\scriptscriptstyle RS}-e)/2\iic$ error corrections and $\delta\css{def}{=}\isa(t_{\scriptscriptstyle RS}-e)/2\isc$ error detections ($e$-EEC/$\theta$-EC/$\delta$-ED) for $\mathcal{C}_{\scriptscriptstyle RS}$, where the input is $R'$. In particular, the $e$-EEC/$\theta$-EC/$\delta$-ED error control algorithm for $\mathcal{C}_{\scriptscriptstyle RS}$ is used to simulate the $(e+2\theta)$-Sy$0$EC/$(e+2\theta+1)$-Sy$0$ED part of the control algorithm for $\mathcal{C}$ and the fixed length property of $\mathcal{C}$ (which is equivalent to the constant weight property of $V(\mathcal{C})$) is used to obtain the AU$0$ED part. The $t$-Sy$0$EC/$(t+1)$-Sy$0$ED/AU$0$ED decoding algorithm for this base code design is given in Algorithm \ref{algbasedect0ecc} below. First the following definition is needed.
\begin{definition}
\label{deftSy0EC/t+1Sy0ED/AU0ED}
Let $n,t,d\el\inat$ be given such that $t\leq d$. A binary code $\mathcal{C}\sse\iint_{2}^{n}$ of length $n$ can correct up to $t$ $0$-errors, detect $d$ $0$-errors and, simultaneously, detect any occurrence of only $0$-insertion errors or only $0$-deletion errors (i.~e., $\mathcal{C}$ is a $t$-Sy$0$EC/$d$-Sy$0$ED/AU$0$ED code) if, and only if, there exists an (error control) Algorithm for $\mathcal{C}$, say $\mathcal{D}ec(\mathcal{C},t ,d)$, which, for all sent codeword $X\el\mathcal{C}$ takes the corresponding received word $Y\el\iint_{m}^{n}$ and gives a word $\tilde{X}\el\iint_{2}^{n}$ (estimate of the sent word $X$) as the output with a signal $cor\el\{0,1\}$ ($cor=1$ means the received word is corrected) such that
\begin{itemize}
\item[C1)$\!$]
if $X$ and $Y$ differ by only $0$-deletion errors and $cor=1$ then $\tilde{X}=X$;
\item[C2)$\!$]
if $X$ and $Y$ differ by only $0$-insertion errors and $cor=1$ then $\tilde{X}=X$;
\item[C3)$\!$]
if $\did(X,Y)\leq d$ and $cor=1$ then $\tilde{X}=X$; and,
\item[C4)$\!$]
if $\did(X,Y)\leq t$ then $cor=1$.
\end{itemize}
Note that if $cor=1$ then $\tilde{X}=X$ (i.~e., errors are corrected) because of condition C3).
\end{definition}
\begin{algorithm}[$t$-Sy$0$EC/$(t+1)$-Sy$0$ED/AU$0$ED error control algorithm for the codes in (\ref{eqcorEbaseA})]
\label{algbasedect0ecc}\rm
\\
{\bf Input}:
\begin{itemize}
\item[1)]
The (received) word $E'\el\iint_{2}^{n'}$ in (\ref{eqrecwordbase}).
\end{itemize}
{\bf Output}:
\begin{itemize}
\item[1)] A word $\tilde{E}=\mathcal{E}(\tilde{X})\el\iint_{2}^{n}$, for some $\tilde{X}\el\iint_{2}^{k}$ (the word $\tilde{E}$ represents the estimate of the sent codeword $E=\mathcal{E}(X)\el\iint_{2}^{n}$ associated with the information word $X\el\iint_{2}^{k}$); and,
\item[2)] a signal $cor\el\{0,1\}$. The signal $cor=1$ means that errors are corrected. 
\end{itemize}
Execute the following steps.\\
{\bf S1} Check for more than $t$ unidirectional $0$-errors in $E'$. Execute the following steps.\\
\blackbox{0mm}{2mm}{0mm}
{\bf S1.1}: Compute $\delta\css{def}{=}n'-n\el\iint$ and then\\
\blackbox{0mm}{2mm}{0mm}
{\bf S1.2}: If $\delta\nel[-t,t]$ then set $cor=0$ (i.~e., errors are only detected), set $\tilde{E}=\mbox{any codeword}$, outuput the couple $(\tilde{E},cor)$ and \textbf{exit}.
\\
{\bf S2}: Otherwise, if $\delta\el[-t,t]$ then for all integer $\xi\el[1,\tau]$, execute step S2.1-S2.2.\\
\blackbox{0mm}{2mm}{0mm}
{\bf S2.1}: Parse $E'$ so that to compute the RS codeword $R'_{\xi}\el\mathbb{F}^{n_{\scriptscriptstyle RS}}\cup\{*\}$ in (\ref{eqrecwordbase2}) as explained previously by appling an efficient $(\tau-\xi)$-Sy$0$EC/$(\tau+\xi-1)$-Sy$0$ED/AU$0$ED error control algorithm for $\mathcal{C}_{\tau}(\mathcal{I})$ to each $B'_{i}\el\mathcal{I}$ parsed for all $i\el[1,n_{\scriptscriptstyle RS}]$.\\
\blackbox{0mm}{2mm}{0mm}
{\bf S2.2}: Perform $e_{\xi}$-EEC/$\theta_{\xi}$-EC/$\delta_{\xi}$-ED error control for $\mathcal{C}_{\scriptscriptstyle RS}$. Execute the following steps.\\
\blackbox{0mm}{2mm}{0mm}
\blackbox{0mm}{2mm}{0mm}
{\bf S2.2.1}: Compute
\setlength{\jot}{4pt}
\begin{alignat*}{1}
&e_{\xi}=\,\mbox{number of erasures in $R'_{\xi}\el\mathbb{F}^{n_{\scriptscriptstyle RS}}\cup\{*\}$},\\
&\theta_{\xi}=\iia\frac{t_{\scriptscriptstyle RS}-e_{\xi}}{2}\iic,\\
&\delta_{\xi}=\isa\frac{t_{\scriptscriptstyle RS}-e_{\xi}}{2}\isc;
\end{alignat*}
and then\\
\blackbox{0mm}{2mm}{0mm}
\blackbox{0mm}{2mm}{0mm}
{\bf S2.2.2}: If $e_{\xi}\el[0,t]$ then run any efficient $e_{\xi}$-EEC/$\theta_{\xi}$-EC/$\delta_{\xi}$-ED error control algorithm for $\mathcal{C}_{\scriptscriptstyle RS}$ with input word $R'_{\xi}$ and get the estimate $\tilde{R}_{\xi}\el\mathbb{F}^{n_{\scriptscriptstyle RS}}$.\\
\blackbox{0mm}{2mm}{0mm}
\blackbox{0mm}{2mm}{0mm}
{\bf S2.2.3}: Compute $\tilde{E}_{\xi}=\beta1(\tilde{R}_{\xi})\el\iint_{2}^{n}$.\\
\blackbox{0mm}{2mm}{0mm}
\blackbox{0mm}{2mm}{0mm}
{\bf S2.2.4}: if $\tilde{E}_{\xi}=\mathcal{E}(\tilde{X}_{\xi})\el\mathcal{C}$ for some $\tilde{X}_{\xi}\el\iint_{2}^{k}$ and $\did(\tilde{E}_{\xi},E')\leq t$ then set $cor=1$ (i.~e., errors are corrected), set $\tilde{E}=\tilde{E}_{\xi}$, outuput the couple $(\tilde{E},cor)$ and \textbf{exit}.\\
{\bf S3}: Set $cor=0$ (i.~e., errors are only detected), set $\tilde{E}=\mbox{any codeword}$, outuput the couple $(\tilde{E},cor)$ and \textbf{exit}.
\end{algorithm}

As we explained before, if $\tau=1$ then Algorithm \ref{algbasedect0ecc} gives the correct output. For $\tau>1$, if the errors in each $B'_{i}\el\mathcal{I}$ parsed in step {\bf S2.1} are of unidirectional type (i.~e., only $0$-deletions or only $0$-insertions) then Algorithm \ref{algbasedect0ecc} gives the correct output because the $(\tau-1)$-Sy$0$EC/$\tau$-Sy$0$ED/AU$0$ED error control algorithm alone becomes a $(\tau-1)$-U$0$EC/AU$0$ED error control algorithm and, as such, it does not give mis-corrections; but, eventually, only at most $\iia t/\tau\iic=t_{RS}$ error detections (which account for $e_{1}\leq t_{RS}$ correctable erasures in step {\bf S2.2}). So, the list decoding based Algorithm \ref{algbasedect0ecc} may fail in the general case of symmetric errors. However, note that if in step {\bf S2}, there exists $\xi_{1},\xi_{2}\el[1,\tau]$ such that $cor=1$, then, from step {\bf S2.2.4}, $\tilde{E}_{\xi_{1}},\tilde{E}_{\xi_{2}}\el\mathcal{C}$ and, from (\ref{eqmindisRSbc}),
\setlength{\jot}{6pt}
\begin{alignat*}{1}
\did(\tilde{E}_{\xi_{1}},\tilde{E}_{\xi_{2}})\leq\;&\did(\tilde{E}_{\xi_{1}},E')+\did(E',\tilde{E}_{\xi_{2}})\leq\\
&2t<2(t+1)\leq\did(\mathcal{C});
\end{alignat*}
and so, $\tilde{E}_{\xi_{1}}=\tilde{E}_{\xi_{2}}$. This means that if Algorithm \ref{algbasedect0ecc} fails then it is because it is too weak and, in step {\bf S2} does not find a codeword $\tilde{E}_{\xi}$ at a distance less than $t+1$ from the received word $E'$. In this case, however, it just detects the symmetric errors where, instead, it should correct them. We readily say that we were unable to prove the existence of such $\tilde{E}_{\xi}\el\mathcal{C}$; and hence, the correctness of Algorithm \ref{algbasedect0ecc} for $\tau>1$. In any case, assume we want to correct $t$ $0$-errors and let $e_{1}$ and $f_{1}$ be the number of erasure errors and errors, respectively, in $R'_{1}$ (the input to step {\bf S2.2.2} of Algorithm \ref{algbasedect0ecc} for $\xi=1$). Since $(\tau-1)$-Sy$0$EC/$\tau$-Sy$0$ED/AU$0$ED is applied in step {\bf S2.1}, the number of $0$-errors in $E'$ is at least
$$
\tau e_{1}+(\tau+1)f_{1}=\tau(e_{1}+f_{1})+f_{1}\leq t;
$$
which, implies,
$$
\tau(e_{1}+f_{1})\leq t-f_{1}\leq t;
$$
and so, the number of errors in $R'_{1}$ is
$$
e_{1}+f_{1}\leq\iia\frac{t}{\tau}\iic=t_{\scriptscriptstyle RS}.
$$
This means that, in our construction for the general case of symmetric errors and $\tau>1$, if we choose a minimum Hamming distance
$$
d_{H}(\mathcal{C}_{\scriptscriptstyle RS})\geq t'_{\scriptscriptstyle RS}+1\css{def}{=}2\iia\frac{t}{\tau}\iic+1
$$
for the RS code then Algorithm \ref{algbasedect0ecc} will certainly give the correct output. Even though, in this way, we may double the redundancy of $\mathcal{C}_{\scriptscriptstyle RS}$, this will be enough for our theoretical purposes in Subsection \ref{secredun} where, really, only $d_{H}(\mathcal{C}_{\scriptscriptstyle RS})\geq t_{\scriptscriptstyle RS}=\Theta(t/\tau)$ can be assumed. Note that proving the correctness of Algorithm \ref{algbasedect0ecc} only assuming $d_{H}(\mathcal{C}_{\scriptscriptstyle RS})\geq t_{\scriptscriptstyle RS}$, it is a matter of taking into account how the at most $t$ $0$-errors are partitioned among the $B'_{i}$'s, $i\el[1,n_{\scriptscriptstyle RS}]$. For $\tau=2,3$, we have tried all the partitions of $t\leq16$ and Algorithm \ref{algbasedect0ecc} always finds a codeword $\tilde{E}_{\xi}$ in step {\bf S2}. So, our conjecture is that even if $d_{H}(\mathcal{C}_{\scriptscriptstyle RS})\geq t_{\scriptscriptstyle RS}$ then Algorithm \ref{algbasedect0ecc} for the minimum $\did$ distance $2(t+1)$ code $\mathcal{C}$ is a correct $t$-Sy$0$EC/$(t+1)$-Sy$0$ED/AU$0$ED error control algorithm. For these reasons and for simplicity, in the following we assume the analysis using $t_{\scriptscriptstyle RS}$ as in (\ref{eqRSparam}).

Let us analyze the redundancy of this base code design. Readily note that, to minimize redundancy, it is more convenient to use values for $\tau,b\el\inat-\{0\}$, say $\tau_{min},b_{min}$, which minimize the code length $n(t,k,\tau,b)$ in (\ref{eqexactlengEbaseA}), for any given $t,k\el\inat-\{0\}$. However, finding a simple explicit expression for the minimum $n_{min}(t,k)\css{def}{=}n(t,k,\tau_{min},b_{min})$ might be difficult, but a useful asymptotic upper bound on this minimum can be derived as in the following theorem where particular values of $b$ and $\tau$ are chosen which give efficient code designs both in terms of redundancy and complexity. Such an upper bound is used in the redundancy analysis of Ssection \ref{secredun}.
\begin{theorem}[Asymptotic upper bound on $n_{min}(t,k)$]
\label{thmubnsmin}
Given $t,k\el\inat-\{0\}$, let $\tau,b\el\inat-\{0\}$ be
\setlength{\jot}{4pt}
\begin{alignat}{1}
&b\css{def}{=}b(t,k)\css{def}{=}\isa\log_{2}(k+t)\isc\simeq\log_{2}(k+t),\label{eqbtkdef}\\
&\tau\css{def}{=}\tau(t,k)\css{def}{=}\isa\sqrt{\frac{tb}{2\isa k/b\isc\log_{2}b}}\blackbox{7mm}{0mm}{0mm}\isc\leq\sqrt{\frac{t}{2k\log_{2}b}}\,b+1\label{eqtautkdef}.
\end{alignat}
If $\tau/b\leq\log_{e}\sqrt{2}$ then
\setlength{\jot}{4pt}
\begin{alignat*}{1}
n(t,k)\css{def}{=}&\min_{\theta,\beta\,\el\,\inat-\{0\}}n(t,k,\theta,\beta)=n(t,k,\tau_{min},b_{min})\lesssim\\
&k+\sqrt{8t\isa\frac{k}{b}\isc b\log_{2}b}+2t\log_{2}b\simeq\\&k+\sqrt{8tk\log_{2}b}+2t\log_{2}b,
\end{alignat*}
as $\max\{t,k\}\el\inat$ increases.
\end{theorem}
\begin{IEEEproof}
From (\ref{eqnuA}) and $k_{b}=k\bmod{b}\el[0,b)$, it follows
$$
\left|\iint_{2}^{k_{b}}\right|<\left|\iint_{2}^{b}\right|=2^{b}\leq|\mathbb{F}|.
$$
So, from (\ref{eqbtkdef}), if $\max\{t,k\}$ is big then $b=\left|\iint_{2}^{b}\right|>0$ is big and $|\mathbb{F}|\geq\nu\geq 2^{b}$ is big. Let $\rho(a)\css{def}{=}0$ if $a=0$ and $\rho(a)\css{def}{=}1$ if $a\neq0$, where $a\el\inat$. From (\ref{eqexactlengEbaseA}), the relations $\tau/\log_{2}|\mathbb{F}|\leq\tau/\left|\iint_{2}^{b}\right|\leq\log_{e}\sqrt{2}=0.34657$, (\ref{eqnbalA}), and (\ref{eqtrsbound}), for any value of $k_{b}\el[0,b)$, it follows,
\setlength{\jot}{4pt}
\begin{alignat}{1}
n\simeq\,&\iia\frac{k}{b}\iic\left(\log_{2}\left|\iint_{2}^{b}\right|+2\tau\log_{2}\log_{2}\left|\iint_{2}^{b}\right|\right)+\label{eqnsimder}\\
&\qquad\left(\log_{2}\left|\iint_{2}^{k_{b}}\right|+\rho(k_{b})2\tau\log_{2}\log_{2}\left|\iint_{2}^{b}\right|\right)+\notag\\
&\qquad\quad t_{\scriptscriptstyle RS}\left(\log_{2}|\mathbb{F}|+2\tau\log_{2}\log_{2}|\mathbb{F}|\right)\lesssim\notag\\
&\iia\frac{k}{b}\iic\left(b+2\tau\log_{2}b\right)+(k_{b}+2\rho(k_{b})\tau\log_{2}b)+\notag\\
&\qquad\qquad t_{\scriptscriptstyle RS}\left(\log_{2}\nu+2\tau\log_{2}\log_{2}\nu\right)\leq\notag\\
&\left(\iia\frac{k}{b}\iic b+k_{b}\right)+2\left(\iia\frac{k}{b}\iic+\rho(k_{b})\right)\tau\log_{2}b+\notag\\
&\qquad\qquad \frac{t}{\tau}\left(\log_{2}\nu+2\tau\log_{2}\log_{2}\nu\right)=\notag\\
&\;k+2\isa\frac{k}{b}\isc\tau\log_{2}b+\frac{t}{\tau}\left(\log_{2}\nu+2\tau\log_{2}\log_{2}\nu\right)=\notag\\
&\;k+2t\log_{2}\log_{2}\nu+2\isa\frac{k}{b}\isc\tau\log_{2}b+\frac{t}{\tau}\log_{2}\nu.\notag
\end{alignat}
With the above choice of $b$ in (\ref{eqbtkdef}),
$$
b=\isa\log_{2}(k+t)\isc\geq\log_{2}\left(\isa\frac{k}{b}\isc+t_{\scriptscriptstyle RS}\right)
$$
because $k\geq\isa k/b\isc$ and, form (\ref{eqtrsbound}) and $\tau>0$, $t\geq t/\tau\geq t_{\scriptscriptstyle RS}$. Thus, from (\ref{eqnuA}), $\log_{2}\nu=b$ and so, from (\ref{eqnsimder}),
\begin{equation}
\label{eqrne}
n\lesssim k+2t\log_{2}b+2\isa\frac{k}{b}\isc\tau\log_{2}b+\frac{tb}{\tau}.
\end{equation}
At this point note that, given $t,k$ and $b$, the above bound is minimized for
$$
\hat{\tau}_{min}(b)\css{def}{=}\sqrt{\frac{tb}{2\isa k/b\isc\log_{2}b}}\el\irea^{+}.
$$
So, letting
$$
\tau\css{def}{=}\tau_{min}(b)\css{def}{=}\isa\hat{\tau}_{min}(b)\isc\simeq\sqrt{\frac{t}{2k\log_{2}b}}\,b
$$
in (\ref{eqrne}), we have $\tau\el\inat-\{0\}$,
$$
n\lesssim k+2t\log_{2}b+2\sqrt{2t\isa\frac{k}{b}\isc b\log_{2}b};
$$
and hence, the theorem.
\end{IEEEproof}

The upper bound given in (\ref{eqredBCD_A}) comes from Theorem \ref{thmubnsmin}. We note that, if $\tau=1$ then the balanced encoding can be implemented with the methods in \cite{KNU86,PEL15,TAL98,TAL99}. If $\tau>1$ the encoding can be implemented with a table look-up to store $k+t$ codewords of length $n_{bal}=b+\tau\log_{2}b+O(1)$ (see (\ref{eqnbalA})). Also, for our purposes of obtaining a good (i.~e., small enough) upper bound in (\ref{eqredBCD_A}), other choices of $b$ and $\tau$ (such as, $b=\isa\log_{2}k\isc$ and $\tau=\isa\sqrt{t\log_{2}k/(2\isa k/b\isc\log_{2}b)}\isc$) are possible in (\ref{eqbtkdef}) and (\ref{eqtautkdef}), respectively.

In the next section, we give a systematic recursive code design which lowers the redundancy and uses the codes defined in this and the previous subsection as base codes.

\section{Systematic Recursive Code Definition}
\label{secsyste_defi}
The systematic recursive code design idea is as explained at the beginning of Section \ref{secsyste_idea} and shown in Figure~\ref{fig1}. Now, the code design is formally described in this section. For all $t,k\el\inat$, the length, $n_{t}(k)\el\inat$, of the codes is
\setlength{\jot}{4pt}
\begin{alignat}{1}
n_{t}(k)\leq k+\min&\left\{t+1+n_{t-1}(\isa t\log_{2}(2k)\isc),\blackbox{4mm}{0mm}{0mm}\right.\label{eqsysreccdlenbound}\\
&\left.\;\;t\log_{2}k+O\left(t\log^{(2)}(t\log k)\sqrt{\log k}\right)\right\}=\notag\\
&\blackbox{4mm}{-12mm}{0mm} k + t\log_{2}k+o(t\log_{2}n_{t}(k)).\notag
\end{alignat}

Again by using the functions $V$ in (\ref{eqfunV}) and $\hat{V}$ in (\ref{eqfunVhat}), the code design is based on the $\sigma$-codes described in Subection \ref{subsecnonsy}. In particular, for simplicity, we consider designs based on constant weight codes in (\ref{eqAnuomegadef}) which are parity extended $\sigma$-codes; i.~e., in (\ref{eqAnuomegadef}) we let
$$
\mathcal{A}(\inat,\nu,\omega)=
\mathcal{A}_{z^{t+1},\tilde{\sigma}}(\inat,\nu,\omega)=
\mathcal{C}_{z^{t+1},\tilde{\sigma}}(\inat,\nu-1)x_{\nu-1}\sse\mathcal{S}(\inat,\nu,\omega);
$$
where $\mathcal{C}_{z^{t+1},\tilde{\sigma}}(\inat,\nu-1)$ is defined in (\ref{eqsigmagencodes}) or (\ref{eqsimpsigmacodehat}) and the parity digit is defined as $x_{\nu-1}\css{def}{=}\omega-w(X)$, for all $X\el\mathcal{C}_{z^{t+1},\tilde{\sigma}}(\inat,\nu-1)$. Indeed, these codes are $t$-SyEC/$(t+1)$-ED/AUED codes on the $L_{1}$ metric in (\ref{eqsimpsigmacode}). Now, the code design idea is the following. Given $t,k\el\inat$, let
$$
X\el\iint_{2}^{k}=\bigcup_{w\in[0,k]}\mathcal{S}(\iint_{2},k,w)
$$
be an information binary word, $\nu=w+1\el[1,k+1]$, $\omega=k-w\el[0,k]$ with $w=w(X)$ and
\begin{equation}
\label{eqVhatofX}
\hat{V}_{X}\css{def}{=}(v_{1},v_{2},\ldots,v_{w},0,0,\ldots,0)\el\iint_{k+1}^{k}\sse\inat^{k}
\end{equation}
be the sequence $\hat{V}(X)$ defined in (\ref{eqfunVhat}) followed by $(k-w)$ number of $0$'s to make a length $k$ vector. Note that $v_{w+1}=v_{w+1}(X)$ can be computed from $\hat{V}_{X}$ using (\ref{eqvwp1}). In this way, let $\mathbb{F}$ be any field and $\partial{S}\sse\mathbb{F}-\{0\}$ be a subset of $k=|\partial{S}|$ distinct non-zero elements in $\mathbb{F}$. Assume $X$ is sent through a $0$-error channel and $Z\el\iint_{2}^{*}$ is received. This is as if $V(X)\el\iint_{k+1}^{k}\sse\inat^{k}$ is sent through an $L_{1}$ error channel and $V(Z)\el\inat^{*}$ is received, because of the isometry stated in Theorem \ref{thmisodidl1}. Now, if the receiver knows the $\sigma$-polynomial in (\ref{sigmadefm}) associated with the vector $\hat{V}_{X}\el\inat^{k}$,
\begin{equation}
\label{eqsigmatildeVX}
\tilde{\sigma}(z)=\sigma_{\hat{V}_{X}}(z)\bmod{z^{t+1}}\el\left\{1+\frac{\mathbb{F}[z]}{(z^{t+1})}\right\},
\end{equation}
then it can correct the $L_{1}$ errors in $V(Z)$ (and, consequently, the associated $0$-errors in $Z$) by applying any $t$-SyEC/$(t+1)$-ED/AUED error control algorithm for the code $\mathcal{A}_{z^{t+1},\tilde{\sigma}}(\inat,\nu,\omega)$ (which contains the vector $V(X)\equiv(\hat{V}_{X},v_{w+1}(X))$) with input the (received) word $V(Z)$. Once $V(Z)$ has been corrected, $V(X)$ and, hence, $X=V^{-1}(V(X))$ are known to the receiver if $Z$ contains less than $t$ $0$-errors. We can let the receiver know $\tilde{\sigma}$ in (\ref{eqsigmatildeVX}) by appending to $X$ any (the shortest, if we aim reducing redundancy) binary encoding, $\gamma_{2}(\tilde{\sigma})$, of $\tilde{\sigma}$; encoded so that to protect $\gamma_{2}(\tilde{\sigma})$ against an appropriate number of $0$-errors. In fact, this encoding strategy can be implemented in a recursive manner if $\gamma_{2}(\tilde{\sigma})$ is encoded using the same scheme with $k$ being replaced by the length of $\gamma_{2}(\tilde{\sigma})$ and, surprisingly, $t$ being replaced with $t-1$. Note that, in order for all this to work some synchronization issues need to be addressed. First, we need to encode the set of all possible polynomials $\tilde{\sigma}$'s in (\ref{eqsigmatildeVX}) that, really, is
$$
\mathcal{I}\css{def}{=}\left\{\mbox{$\sigma_{\hat{V}_{X}}(z)\bmod{z^{t+1}}$: $X\el\iint_{2}^{k}$}\right\}\sse\left\{1+\frac{\mathbb{F}[z]}{(z^{t+1})}\right\}
$$
with a cardinality of
$$
|\mathcal{I}|\leq\min\left\{\left|\iint_{2}^{k}\right|,\left|\left\{1+\frac{\mathbb{F}[z]}{(z^{t+1})}\right\}\right|\right\}.
$$
Since we are interested in fixed length encoding schemes, and $\tilde{\sigma}\el\mathcal{I}$, the length of any binary encoding $\gamma_{2}(\tilde{\sigma})$ is
\setlength{\jot}{4pt}
\begin{alignat}{1}
l(\gamma_{2}(\tilde{\sigma}))=&\,\isa\log_{2}|\mathcal{I}|\isc\leq\label{eqlengthgamma}\\
&\min\left\{k,\isa\log_{2}\left|\left\{1+\frac{\mathbb{F}[z]}{(z^{t+1})}\right\}\right|\isc\right\}.\notag
\end{alignat}
If $\max\{t,k\}$ is large, a simple function $\gamma_{2}$ whose length is the above upper bound can be defined as follows. First, given any field $\mathbb{F}$ and $t\el\inat$, let
$$
[\cdot]_{2}:\left\{1+\frac{\mathbb{F}[z]}{(z^{t+1})}\right\}\to\iint_{2}^{l_{t}},\qquad\mbox{$l_{t}\css{def}{=}l_{t}(\mathbb{F})\el\inat$},
$$
be any (concise) finite fixed length binary encoding of the monic polynomials over $\mathbb{F}$ with degree less than $t+1$ which are coprime with $z^{t+1}$. They are of the form $\sigma(z)=1+\sigma_{1}z+\sigma_{2}z^{2}+\ldots+\sigma_{t}z^{t}$, and so, if $\mathbb{F}$ is a finite field then a simple encoding can be chosen as
\setlength{\jot}{4pt}
\begin{alignat*}{1}
[\sigma]_{2}\css{def}{=}&\,\frac{\sigma(|\mathbb{F}|)-1}{|\mathbb{F}|}=\\
&\sigma_{1}+\sigma_{2}|\mathbb{F}|^{1}+\ldots+\sigma_{t}|\mathbb{F}|^{t-1}\el[0,|\mathbb{F}|^{t}-1];
\end{alignat*}
where the operations are over the integers. In this way, the length of $[\sigma]_{2}$ is
\begin{equation}
\label{eqlengthtlog2F}
l_{t}=l([\cdot]_{2})=\isa t\log_{2}|\mathbb{F}|\isc
\end{equation}
bits. From this, if $\max\{t,k\}$ is large, a simple fixed length encoding
\begin{equation}
\label{eqfflbgamma}
\gamma_{2}:\mathcal{I}\to\iint_{2}^{k_{t}},\qquad\mbox{$k_{t}\css{def}{=}k_{t}(\mathbb{F})\el\inat$},
\end{equation}
which obtains the upper bound in (\ref{eqlengthgamma}) can be defined, for all $\sigma_{\hat{V}_{X}}\el\mathcal{I}$, as
\begin{equation}
\label{eqgammadef}
\gamma_{2}\left(\sigma_{\hat{V}_{X}}\right)\css{def}{=}
\renewcommand{\arraystretch}{1.1}
\left\{\begin{array}{ll}
\left[\sigma_{\hat{V}_{X}}\right]_{2}&\mbox{if $l_{t}=\isa t\log_{2}|\mathbb{F}|\isc<l(X)=k$}, \\
X&\mbox{if $l_{t}=\isa t\log_{2}|\mathbb{F}|\isc\geq l(X)=k$}.
\end{array}
\right.
\end{equation}
Note that from any $X\el\iint_{2}^{k}$, it is always possible to compute $\tilde{\sigma}=\sigma_{\hat{V}_{X}}(z)$; that is, $X$ itself (as any other $X\el\mathcal{A}_{z^{t+1},\tilde{\sigma}}(\inat,\nu,\omega)$) encodes $\sigma_{\hat{V}_{X}}(z)$. In this way, from (\ref{eqlengthgamma}) and (\ref{eqlengthtlog2F}), the length of the fixed length binary encoding $\gamma_{2}$ defined in (\ref{eqfflbgamma}) is
$$
k_{t}=l(\gamma_{2})=\min\{l(X),l([\cdot]_{2})\}=\min\{k,\isa t\log_{2}|\mathbb{F}|\isc\}
$$
bits. At this point, the encoding function to a $t$ $0$-error correcting code can be recursively defined as follows. Let $\mathcal{E}_{t_{b}}^{Base}(X)$, with $t_{b},k_{t_{b}}\el\inat$ and $X\el\iint_{2}^{k_{t_{b}}}$, be any encoding function to a $t_{b}$ $0$-error correcting code to be used as base code for the code design (for example, $\mathcal{E}_{t_{b}}^{Base}(X)$ can be a code defined in the previous section). Now, given $t,k\el\inat$, for the time being let $m=\infty$ (otherwise, an $m\el[0,k)\sse\inat\cup\{\infty\}$ may be considered as a further constraint in the code design). Let $\mathbb{F}\css{def}{=}\mathbb{F}(k,t)$ be any field with finite length binary encoding (\ref{eqfflbgamma}) and $\partial{S}\sse\mathbb{F}-\{0\}$ be a subset of $k_{t+1}\css{def}{=}k=|\partial{S}|$ distinct non-zero elements in $\mathbb{F}$. Assuming that the receiver parses the received word from left to right, the codeword associated with the information word $X\css{def}{=}X_{t}\el\iint_{2}^{k}$ is defined as
\begin{equation}
\label{eqEtX}
\mathcal{E}_{t}(X)\css{def}{=}
\renewcommand{\arraystretch}{1.1}
\left\{\begin{array}{ll}
X_{t}\mu_{t}\;\mathcal{E}_{t-1}(X_{t-1})&\mbox{if $t>t_{b}$},\\
\mathcal{E}_{t_{b}}^{Base}(X)&\mbox{if $t=t_{b}$}
\end{array}
\right.
\el\iint_{2}^{n_{t}};
\end{equation}
where, in (\ref{eqEtX}),
\setlength{\jot}{4pt}
\begin{alignat*}{1}
\mu_{t}\el\;&\,
\parbox[t]{0.85\linewidth}{
\baselineskip 16pt
$\overbrace{00\ldots0}^{\mbox{\scriptsize $t$ zeros}}1=0^{t}1\el\iint_{2}^{t+1}$ is the marker in (\ref{eqMtas}) for $t$ asymmetric $0$-error immunity (in general, $\mu_{t}$ can be any synchronization sequence which allows the receiver to identify the last bit of the received version of $X_{t}$ when no more than $t_{-}\el\inat$ $0$-deletions and no more than $t_{+}\el\inat$ $0$-insertions, with $t_{-}+t_{+}=t$, have occurred during the transmission of $X_{t}\mu_{t}$),}\\\\
X_{t-1}\css{def}{=}&\,\parbox[t]{0.85\linewidth}{$\gamma_{2}\left(\sigma_{\hat{V}_{X_{t}}}(z)\bmod{z^{t+1}}\right)\el\iint_{2}^{k_{t}}$, where $\gamma_{2}$ is the fixed length
\begin{equation}
\label{eqktdef}
k_{t}=\min\{k_{t+1},\isa t\log_{2}|\mathbb{F}|\isc\}
\end{equation}
binary encoding defined in (\ref{eqgammadef}),}
\\\\
\sigma_{V}(z)\el\;&\,\parbox[t]{0.85\linewidth}{$\left\{1+\mathbb{F}[z]/(z^{t+1})\right\}$ is the $\sigma$-polynomial associated with $V\el\inat^{k}$ defined in (\ref{sigmadefm}), and}\\\\
\hat{V}_{X_{t}}\css{def}{=}&\,\parbox[t]{0.85\linewidth}{$(v_{1},v_{2},\ldots,v_{w},0,0,\ldots,0)\el\iint_{k}^{k}\sse\inat^{k}$ is the vector in (\ref{eqVhatofX}) associated with $X=X_{t}$.}
\end{alignat*}
From (\ref{eqEtX}), it follows
$$
\mathcal{E}_{t}(X)=X\mu_{t}\,X_{t-1}\mu_{t-1}\,X_{t-2}\mu_{t-2}\,\ldots\,\mathcal{E}_{t_{b}}^{Base}(X_{t_{b}})\el\iint_{2}^{n_{t}}.
$$
Now, the $t$-Sy$0$EC/$(t+1)$-Sy$0$ED/AU$0$ED decoder design idea is the following. Assume $E_{t}\css{def}=\mathcal{E}_{t}(X_{t})\el\iint_{2}^{n_{t}}$ is sent and $Y_{t}\el\iint_{2}^{*}$ is received. Also, assume either no more than $(t+1)$ $0$-errors occurred during the transmission or all the errors are of the same type: namely, either $0$-deletions or $0$-insertions. If $t=t_{b}$ then the receiver computes the codeword estimate $E_{t}'\el\iint_{2}^{n_{t}}$ and signal $cor_{t}\el\{0,1\}$ by executing any $t$-Sy$0$EC/$(t+1)$-Sy$0$ED/AU$0$ED algorithm for the code defined by $\mathcal{E}_{t_{b}}^{Base}$ with input word $Y_{t}$. If $t>t_{b}$ then
$$
E_{t}=\mathcal{E}_{t}(X_{t})=X\mu_{t}\,E_{t-1}\el\iint_{2}^{n_{t}}
\;\;
\mbox{and}
\;\;
Y_{t}\css{def}{=}Z\nu_{t}\,Y_{t-1}\el\iint_{2}^{*};
$$
with $Z\nu_{t}\el\iint_{2}^{*}1$ being the erroneous versions (whose length is unknown to the receiver) of $X\mu_{t}\el\iint_{2}^{k+t}1\sse\iint_{2}^{k+t+1}$. In this case, the receiver
\\
{\bf S1}: checks for more than $t$ unidirectional $0$-errors in $Y_{t}$;
\\
{\bf S2}: extracts $Z$ and $Y_{t-1}$ from the received sequence $Y_{t}$;
\\
{\bf S3}: recursively decodes $Y_{t-1}\el\iint_{2}^{*}$ and computes $E_{t-1}'=\mathcal{E}_{t-1}(X_{t-1}')=X_{t-1}'\mu_{t-1}\ldots\el\iint_{2}^{n_{t-1}}$ and $cor_{t-1}\el\{0,1\}$;
\\
{\bf S4}: computes $X_{t}'$, estimate of $X_{t}$, as follows. Let $w\css{def}{=}w(Z)$ and note that $w=w(Z)=w(X)$. If $cor_{t-1}=0$ then it\\
$\bullet$ sets $X_{t}'=Z$;\\
otherwise, if $cor_{t-1}=1$,\\
$\bullet$ sets $X_{t}'=V^{-1}(W)\el\iint_{2}^{k}$, where $W\el\iint_{m}^{w+1}$ is the output word from the general $t$-SyEC/$(t+1)$-SyED/AUED decoding Algorithm \ref{alggentsyECdUEDcwc} with input word $Y\css{def}{=}V(Z)\el\iint_{m}^{w+1}$ and input code
\setlength{\jot}{2pt}
\begin{alignat*}{1}
\mathcal{A}&\css{def}{=}\mathcal{A}_{z^{t+1},\tilde{\sigma}}(\iint_{m},w+1,k-w)\sse\\
&\sse\mathcal{S}(\iint_{m},w+1,k-w),\quad\mbox{with $w=w(Z)=w(X)$};
\end{alignat*} 
where, $\tilde{\sigma}(z)\el\left\{1+\mathbb{F}[z]/(z^{t+1})\right\}$ is computed from $X_{t-1}'$ (see the definition of $X_{t-1}$ in (\ref{eqEtX}));
\\
{\bf S5}: computes $E_{t}'=\mathcal{E}_{t}(X_{t}')$ as the estimate of $E_{t}$.
\\
{\bf S6}: finally, the receiver checks that no mis-correction occurred in step {\bf S5}, by letting $cor_{t}=1$ if, and only if, $\did(E_{t}',Y_{t})\leq t$.
\\
We will show that the code is indeed a $t$-Sy$0$EC/$(t+1)$-Sy$0$ED/AU$0$ED code even though the marker is $\mu_{t}=M^{tas}_{t}$ as in (\ref{eqMtas}) (i.~e., the length of $\mu_{t}$ is $t+1$ and no more). A key point is step $\bf S2$ whose implementation needs the function $Extract(Y,\mu_{t},t,i)$. $Extract(Y,\mu_{t},t,i)$ is used to extract $Z$ and $Y'=Y_{t-1}$ from the received sequence $Y=Y_{t}=Z\nu_{t}\,Y_{t-1}$. In general, we assume $\mu_{t}\css{def}{=}0^{s}1\el\iint_{2}^{*}$.
\begin{algorithm}[$Extract$]
\label{algextract}\rm
\\
{\bf Input}:
\begin{itemize}
\item[1)]
A sequence $Y\el\iint_{2}^{*}$.
\item[2)]
The marker $\mu_{t}\css{def}{=}0^{s}1\el\iint_{2}^{s+1}$, with $s\el\inat$.
\item[3)]
The error control capability $t\el\inat$.
\item[4)]
An index $i\el\inat$; the key index for parsing.
\end{itemize}
{\bf Output}:
\begin{itemize}
\item[1)] A sequence $Z\el\iint_{2}^{*}$ which is a prefix of $Y\el\iint_{2}^{*}$.
\item[2)] A sequence $Y'\el\iint_{2}^{*}$ which is a suffix of $Y\el\iint_{2}^{*}$.
\end{itemize}
It is assumed that $Y$ is read from left to right order.\\
Execute the following steps.\\
{\bf S1}: Parse (i.~e., put a comma in) $Y$ right after the first $1$ to the right of the bit whose index is $i$ (i.~e., the $i$-th bit). In this way, $Y$ remains splited into a prefix $Z'=W0^{v}1\el\iint_{2}^{*}$ ending with a $1$ and a suffix $Y'\el\iint_{2}^{*}$, where $v\el\inat$ and $W\el\iint_{2}^{*}$ either ends with a $1$ or is the empty string $\Lambda$.\\
{\bf S2}: Correct the suffix $0^{v}1$ of $Z'$ and extract the marker $\mu_{t}=0^{s}1$ from $Z'$ to get $Z$ as follows. If $v< s$ then let $Z\css{def}{=}W\el\iint_{2}^{*}$ else let $Z\css{def}{=}W0^{v-s}$ (succintely, $Z\css{def}{=}W0^{v\dotmin s}\el\iint_{2}^{*}$).\\
{\bf S3}: Output the couple $(Z,Y')$ and {\bf exit}.
\end{algorithm}
The following definition clarifies the concept of correct parsing for the above $Extract(B,\mu_{t},t,i)$ function.
\begin{definition}[successful key indices and marker immunity]
\label{defsucckeybit}
Given a $0$-error channel model and a marker $\mu_{t}\css{def}{=}0^{s}1\el\iint_{2}^{s}1$, $s\el\inat$, assume $A\css{def}{=}X\mu_{t}\,A'\el\iint_{2}^{*}$ is sent through the $0$-error channel, $B\css{def}{=}Z\nu_{t}\,B'\el\iint_{2}^{*}$ is received and at most $t\el\inat$ $0$-errors occurred during the transmission of $X\mu_{t}$ (i.~e., by definition, $\did(X\mu_{t},Z\nu_{t})\leq t$); where, $X\el\iint_{2}^{k}$, $k\el\inat$, $A',B'\el\iint_{2}^{*}$, $\nu_{t}\el0^{*}1$ and $Z\el\iint_{2}^{*}$ is the erroneous versions of $X\el\iint_{2}^{*}$ (i.~e., by definition, $w(X)=w(Z)$). We say that the function $Extract(B,\mu_{t},t,i)$ given in Algorithm \ref{algextract} is {\bf successful} if, and only if, it returns the ordered couple $Z$ and $Y'=B'$. An index $i\el\inat$ is called {\bf successful} key index for parsing if, and only if, the function $Extract(B,\mu_{t},t,i)$ is successful. Finally, the marker $\mu_{t}$ is {\bf immune} by at most $t$ $0$-errors in the channel model if, and only if, for any couple of sent and received words $A$ and $B$ as above there exists a {\bf successful} key index.
\end{definition}
For example, as we mentioned at the beginning of this section, if $s\geq t\css{def}{=}t_{-}+t_{+}$, with $t_{-},t_{+}\el\inat$, then $Extract(Y,\mu_{t}=0^{t},t,n+t_{+}+1)$ defined in Algorithm~\ref{algextract} is successful (i.~e., gives the correct output) for $(t_{-},t_{+})$-$0$EC immunity.

Taking into account these considerations and Definition \ref{deftSy0EC/t+1Sy0ED/AU0ED}, the decoding algorithm is given below. 
\begin{algorithm}[$t$-Sy$0$EC/$(t+1)$-Sy$0$ED/AU$0$ED recursive decoding algorithm for the codes in (\ref{eqEtX})]
\label{algrecdect0ecc}\rm
\\
{\bf Input}:
\begin{itemize}
\item[1)]
The (received) word $Y_{t}\el\iint_{2}^{l(Y_{t})}\sse\iint_{2}^{*}$ having length $l(Y_{t})\el\inat$.
\end{itemize}
{\bf Output}:
\begin{itemize}
\item[1)] A word $\tilde{E}_{t}=\mathcal{E}_{t}(\tilde{X}_{t})\el\iint_{2}^{n_{t}}$, for some $\tilde{X}_{t}\el\iint_{2}^{k}$ (the word $\tilde{E}_{t}$ represents the estimate of the sent codeword $E_{t}=\mathcal{E}_{t}(X)\el\iint_{2}^{n_{t}}$ associated with the information word $X=X_{t}\el\iint_{2}^{k}$); and,
\item[2)] a signal $cor_{t}\el\{0,1\}$. Recall that $cor=1$ means the received word has been corrected.
\end{itemize}
Execute the following steps.\\
{\bf Base case step}: If $t=t_{b}$ then compute the codeword estimate $\tilde{E}\el\iint_{2}^{n_{t}}$ and signal $cor_{t}\el\{0,1\}$ by executing any $t$-Sy$0$EC/$(t+1)$-Sy$0$ED/AU$0$ED algorithm for the code defined by $\mathcal{E}_{t_{b}}^{Base}$ with input word $Y_{t}$ and \textbf{exit}.\\
{\bf Recursive case step}: Otherwise, if $t>t_{b}$ then assume $E_{t}=\mathcal{E}_{t}(X_{t})=X\mu_{t}\,E_{t-1}\el\iint_{2}^{n_{t}}$ was sent and
$$
Y_{t}\css{def}{=}Z\nu_{t}\,Y_{t-1}\el\iint_{2}^{*},
$$
was received, where $Z\nu_{t}\el\iint_{2}^{*}1$ is the (unknown length) erroneous versions of $X\mu_{t}\el\iint_{2}^{k+t}1\sse\iint_{2}^{k+t+1}$.\\
Execute the following steps.
\\
{\bf S1}: check for more than $t$ unidirectional $0$-errors in $Y_{t}$: compute
$$
\delta\css{def}{=}l(Y_{t})-l(E_{t})=l(Y_{t})-n_{t}\el\iint
$$
and afterwards, if $\delta\nel[-t,t]$ then\\
{\bf S1.1}: set $\tilde{E}_{t}=\mathcal{E}_{t}(\tilde{X}_{t})\el\iint_{2}^{n_{t}}$, for some $\tilde{X}_{t}\el\iint_{2}^{k}$, set $cor_{t}=0$ and \textbf{exit};
\\
otherwise, if $\delta\el[-t,t]$ then\\
{\bf S1.2}: execute the following steps.
\\
{\bf S2}: extract $Z$ and $Y_{t-1}$ from the received sequence $Y_{t}$: execute the following steps.
\\
{\bf S2.1}: compute
$$
\tau_{-}
\css{def}{=}
\iia\frac{t-\delta}{2}\iic
\quad
\mbox{and}
\quad
\tau_{+}
\css{def}{=}t-\tau_{-}.
$$
Note that $0\leq\tau_{-},\tau_{+}\leq t$ (because, here $|\delta|\leq t$) and $\tau_{-}+\tau_{+}=t$.
\\
{\bf S2.2} Let
$$
i\css{def}{=}i(\delta)\css{def}{=}k+\tau_{+}+1\el[k+1,k+t+1]\sse\inat
$$
and execute $Extract(Y_{t},\mu_{t},t,i)$ in Algorithm \ref{algextract}. Let $Z\el\iint_{2}^{*}$ and $Y_{t-1}\el\iint_{2}^{*}$ be the prefix and suffix, respectively, of $Y_{t}$ which are output from $Extract(Y_{t},\mu_{t},t,i)$.
\\
{\bf S3}: Recursively decode $Y_{t-1}\el\iint_{2}^{*}$: compute
$$
E_{t-1}'=\mathcal{E}_{t-1}(X_{t-1}')=X_{t-1}'\mu_{t-1}\ldots\el\iint_{2}^{n_{t-1}}
$$
and $cor_{t-1}\el\{0,1\}$.
\\
{\bf S4}: Compute $X_{t}'$, estimate of $X_{t}$, as follows. Let $w\css{def}{=}w(Z)$. If $cor_{t-1}=0$ then\\
{\bf S4.1}: set $X_{t}'=Z$;\\
otherwise, if $cor_{t-1}=1$ then\\
{\bf S4.2}: set $X_{t}'=V^{-1}(W)\el\iint_{2}^{k}$, where $W\el\iint_{m}^{w+1}$ is the output word from the general $t$-SyEC/$(t+1)$-SyED/AUED decoding Algorithm \ref{alggentsyECdUEDcwc} with input word $Y\css{def}{=}V(Z)\el\iint_{m}^{w+1}$ and input code
\setlength{\jot}{2pt}
\begin{alignat*}{1}
\mathcal{A}&\css{def}{=}\mathcal{A}_{z^{t+1},\tilde{\sigma}}(\iint_{m},w+1,k-w)\sse\\
&\sse\mathcal{S}(\iint_{m},w+1,k-w),\quad\mbox{with $w=w(Z)=w(X)$};
\end{alignat*}
where, $\tilde{\sigma}(z)\el\left\{1+\mathbb{F}[z]/(z^{t+1})\right\}$ is computed from $X_{t-1}'$ (see the definition (\ref{eqEtX})).
\\
{\bf S5}: Using the definition (\ref{eqEtX}), compute $\mathcal{E}_{t}(X_{t}')$, encoding of $X_{t}'$, as the estimate of $E_{t}$ as follows. If $X_{t}'$ is in the domain $\iint_{2}^{k}$ of $\mathcal{E}_{t}$ (i.~e., $l(X_{t}')=k$) then\\
{\bf S5.1}: set $E_{t}'=\mathcal{E}_{t}(X_{t}')$;\\
otherwise, if $X_{t}'\nel\iint_{2}^{k}$ (i.~e., $l(X_{t}')\neq k$) then\\
{\bf S5.2}: set $E_{t}'=\mathcal{E}_{t}(X_{t}')$, for some $X_{t}'\el\iint_{2}^{k}$.
\\
{\bf S6}: If $\did(E_{t}',Y_{t})\leq t$ then
\\
{\bf S6.1}: set $\tilde{E}_{t}=E_{t}'$, $\tilde{X}_{t}=X_{t}'$, $cor_{t}=1$ and \textbf{exit}.
\\
If, otherwise, $\did(E_{t}',Y_{t})>t$ then\\
{\bf S6.2}: set $\tilde{E}_{t}=\mathcal{E}_{t}(\tilde{X}_{t})\el\iint_{2}^{n_{t}}$, for some $\tilde{X}_{t}\el\iint_{2}^{k}$, set $cor_{t}=0$ and \textbf{exit}.
\end{algorithm}

The following theorem holds.
\begin{theorem}[correctness of Algorithm \ref{algrecdect0ecc}]
\label{thmcorralg0eccrec}
Let $t,k\el\inat$ and
$$
\mathcal{C}
\css{def}{=}
\mathcal{C}_{t}
\css{def}{=}
\{\mathcal{E}_{t}(X):\;X\el\iint_{2}^{k}\},
$$
be the code defined by (\ref{eqEtX}) where $\mathcal{C}_{t_{b}}$ is any $t_{b}$-Sy$0$EC/$(t_{b}+1)$-Sy$0$ED/AU$0$ED code (such as the ones given in Subsection \ref{subseclimit}, Subsection \ref{subsecreeds}, or the identity function when $t_{b}=0$). Then
\begin{itemize}
\item[1)] the minimum $0$-deletion/insertion distance of the code is $\did(\mathcal{C})>2t$;
\item[2)] $\mathcal{C}$ is a $t$-Sy$0$EC/$(t+1)$-Sy$0$ED/AU$0$ED code; and,
\item[3)] Algorithm \ref{algrecdect0ecc} is a correct $t$-Sy$0$EC/$(t+1)$-Sy$0$ED/AU$0$ED decoding algorithm.
\end{itemize}
\end{theorem}
\begin{IEEEproof}
Let $\mathcal{C}$ be defined with a marker $\mu_{t}$ immune by at most $t$ $0$-insertions as the one given in (\ref{eqMtas}). In first place, we prove that, when $t>t_{b}$, if $\did(E_{t},Y_{t})\leq t$ then $Extract(Y_{t},\mu_{t},t,k+\tau_{+}+1)$ in Step S2.2 of Algorithm \ref{algrecdect0ecc} is sucessful according to Definition~\ref{defsucckeybit}. Let
$$
\did(E_{t},Y_{t})\css{def}{=}e+f<\infty
$$
be the number of $0$-errors between the sent codeword $E_{t}$ and received word $Y_{t}$; where $e,f\el\inat$ are the number of $0$-deletions and $0$-insertions, respectively. Also let $e',f'\el\inat$ be the number of $0$-deletions and $0$-insertions, respectively, between $X\mu_{t}$ and $Z\nu_{t}$. Since $Z\nu_{t}$ is the erroneous version of $X\mu_{t}$, it follows that $Z$ is the erroneous version of $X$ and $w(X)=w(Z)$. So, as in Theorem \ref{thmconcatprop1},
\begin{alignat*}{1}
e=\,&|V(E_{t})\dotmin V(Y_{t})|=\\
&|V(X\mu_{t})\dotmin V(Z\nu_{t})|+|V(E_{t-1})\dotmin V(Y_{t-1})|\geq\\
&|V(X\mu_{t})\dotmin V(Z\nu_{t})|=e';
\end{alignat*}
and, analogously, $f\geq f'$. Now, from this and the definition of $\delta$ in Step S1, it follows,
\setlength{\jot}{2pt}
\begin{alignat*}{1}
&f+e=\did(E_{t},Y_{t});\quad\mbox{whereas},\\
&f-e=l(Y_{t})-l(E_{t})=\delta;
\end{alignat*}
and so,
\setlength{\jot}{4pt}
\begin{alignat*}{1}
f'\leq f=\,&\frac{\did(E_{t},Y_{t})+\delta}{2}\quad\mbox{and}\\
e'\leq e=&\frac{\did(E_{t},Y_{t})-\delta}{2}.
\end{alignat*}
From above and the definition of $\tau_{-}$ and $\tau_{+}$ in Step S2.1, if we assume $\did(E_{t},Y_{t})\leq t$ then
\setlength{\jot}{4pt}
\begin{alignat*}{1}
f'\leq f\leq&\iia\frac{t+\delta}{2}\iic\leq t-t_{-}=t_{+}\quad\mbox{and}\\
e'\leq e\leq&\iia\frac{t-\delta}{2}\iic=t_{-}.
\end{alignat*}
Since $Extract(Y,\mu_{t},t,k+\tau_{+}+1)$ defined in Algorithm~\ref{algextract} is successful for $(t_{-},t_{+})$-$0$EC immunity, the above relations imply that $Extract(Y_{t},\mu_{t},t,k+\tau_{+}+1)$ in Step S2.2 of Algorithm \ref{algrecdect0ecc} makes the correct parsing; whenever $\did(E_{t},Y_{t})\leq t$.

At this point, let us first prove that Algorithm \ref{algrecdect0ecc} is a correct $t$-Sy$0$EC/$(t+1)$-Sy$0$ED/AU$0$ED decoding algorithm when no $0$-deletions occur during transmission (i.~e., $e=0$). This particular case of 3) will imply 1) and 2) because of Theorem \ref{thtdi0ECCequidec2}. Now, in all the cases, statement 3) follows by induction. In fact, recall Definition \ref{deftSy0EC/t+1Sy0ED/AU0ED} with $d=t+1$, consider Algorithm \ref{algrecdect0ecc} and assume, as base case of an inductive argument, that Algorithm \ref{algrecdect0ecc} is correct for $t=t_{b}$. In any case, if $cor=cor_{t}=1$ then Step S6.1 of Algorithm \ref{algrecdect0ecc} is executed. This implies that steps S2--S6 are executed and so $|\delta|\leq t$. So, if the $0$-errors are of unidirectional type (i.~e., either $f=0$ or $e=0$) and $cor=cor_{t}=1$ (as in case C1 and C2 of Definition \ref{deftSy0EC/t+1Sy0ED/AU0ED}) then the number of $0$-errors is
$$
\did(Y_{t},E_{t})=f+e=|f-e|=|\delta|\leq t.
$$
because either $f=0$ or $e=0$. In this way, condition C3) of Definition \ref{deftSy0EC/t+1Sy0ED/AU0ED} will imply C1) and C2). Now, with regard to condition C3), assume $\did(Y_{t},E_{t})\leq t+1$, $cor=cor_{t}=1$ and no $0$-deletions during transmission (i.~e., $e=0$). In this case, $f=\did(Y_{t},E_{t})=|\delta|\leq t$ because $e=0$ and steps S2--S6 are executed (being $cor=cor_{t}=1$). This imples that $Extract(Y_{t},\mu_{t},t,k+\tau_{+}+1)$ in Step S2.2 is succesful and correctly parses $Y_{t}=Z\nu_{t}\,Y_{t-1}$ as $(Z,Y_{t-1})$ where $w(Z)=w(X)$. Hence, $\did(Y_{t},E_{t})\leq t$, Theorem \ref{thmconcatprop1} and $E_{t}=X\mu_{t}\,E_{t-1}$ imply
\setlength{\jot}{0pt}
\begin{alignat}{1}
\did(Z,X)+\,&\did(Y_{t-1},E_{t-1})=\label{eqdidzxye}\\
&\did(Y_{t},E_{t})=|\delta|\leq t<\infty.\notag
\end{alignat} 
Now, if $cor_{t-1}=0$ in step S3 then $\did(Y_{t-1},E_{t-1})\geq t$ because, from the inductive hypotesis, condition C4) of Definition \ref{deftSy0EC/t+1Sy0ED/AU0ED} holds true for the $(t-1)$-Sy$0$EC/$t$-Sy$0$ED/AU$0$ED code $\mathcal{C}_{t-1}$. So, from (\ref{eqdidzxye}), $\did(Y_{t-1},E_{t-1})=t$ and $\did(Z,X)=0$. This implies $X_{t}'=Z=X$ (because Step S4.1 is executed), that $E_{t}'=E_{t}$ (i.~e., all the errors in $E_{t}$ are corrected because $X_{t}'=X\el\iint_{2}^{k}$) in step S5 and that step S6.1 is executed (because $\did(Y_{t},E_{t}')=\did(Y_{t},E_{t})\leq t$) in such a way that $\tilde{E}_{t}=E_{t}'=E_{t}$ and $cor_{t}=1$. If instead $cor_{t-1}=1$ in step S3 then $\did(Y_{t-1},E_{t-1})\leq t$ (because of (\ref{eqdidzxye})) and $cor_{t-1}=1$. Thus, Step S4.2 is executed with the correct value of
$$
\tilde{\sigma}=\sigma_{\hat{V}_{X_{t}}}\bmod{z^{t+1}}
$$
because, from the inductive hypotesis, condition C3) of Definition \ref{deftSy0EC/t+1Sy0ED/AU0ED} holds true for $\mathcal{C}_{t-1}$. Hence, in Step S4.2, $X_{t}'=X$ because the errors in $Z$ are corrected using Algorithm \ref{alggentsyECdUEDcwc}. Also in this case, step S6.1 is executed so that $\tilde{E}_{t}=E_{t}$ and $cor_{t}=1$. In this way, if $\did(Y_{t},E_{t})\leq t+1$ and $cor=cor_{t}=1$ then $\tilde{E}_{t}=E_{t}$; and so, C1), C2) and C3) of Definition \ref{deftSy0EC/t+1Sy0ED/AU0ED} hold for $\mathcal{C}_{t}$. Indeed, with regard to C4), if $\did(Y_{t},E_{t})\leq t$ then $Extract(Y_{t},\mu_{t},t,k+\tau_{+}+1)$ in Step S2.2 is sucessful and makes the correct parsing. Now, only the following three cases are possible.
\begin{itemize}
\item[C1:\hglue -4pt] $\did(Y_{t-1},E_{t-1})=t$ (i.~e., $Y_{t-1}$ contains exactly $t$ errors) and $cor_{t-1}=0$. In this case, as above, from (\ref{eqdidzxye}), $\did(Z,X)=0$, and so, $X_{t}'=Z=X$.
\item[C2:\hglue -4pt] $\did(Y_{t-1},E_{t-1})=t$ and $cor_{t-1}=1$. Here, as above, $X_{t}'=X$ because, from the inductive hypotesis, condition C3) of Definition \ref{deftSy0EC/t+1Sy0ED/AU0ED} holds true for the $(t-1)$-Sy$0$EC/$t$-Sy$0$ED/AU$0$ED code $\mathcal{C}_{t-1}$.
\item[C3:\hglue -4pt] $\did(Y_{t-1},E_{t-1})<t$ (and $Z$ may contain some errors). In this case, $cor_{t-1}=1$ because, from the inductive hypotesis, condition C4) of Definition \ref{deftSy0EC/t+1Sy0ED/AU0ED} holds true for $\mathcal{C}_{t-1}$. Hence, in this case, $\did(Y_{t-1},E_{t-1})\leq t$ and $cor_{t-1}=1$ holds. So, from the inductive hypotesis, condition C3) of Definition \ref{deftSy0EC/t+1Sy0ED/AU0ED} holds true for $\mathcal{C}_{t-1}$, and this condition implies $X_{t}'=X$.
\end{itemize}
Thus, in any case, if $\did(Y_{t},E_{t})\leq t$ then $X_{t}'=X$ and, as above, $\tilde{E}_{t}=E_{t}'=E_{t}$ and $cor_{t}=1$; that is, C4) is also true when no $0$-deletion occurs during the transmission. In this way, 1) and 2) are true. In order to prove 3) in the general case, assume $e,f\el[0,t]$, $f+e=\did(Y_{t},E_{t})\leq t+1$ and $cor=cor_{t}=1$. Since $cor_{t}=1$, step S6.1 of Algorithm \ref{algrecdect0ecc} is executed, and so, there exists an $E_{t}'\el\mathcal{C}$ computed in step S5 (estimate of the sent codeword $E_{t}\el\mathcal{C}$) such that $\did(E_{t}',Y_{t})\leq t$. This and $\did(Y_{t},E_{t})\leq t+1$ imply
\setlength{\jot}{0pt}
\begin{alignat*}{1}
\did(E_{t}',E_{t})\leq\,&\did(E_{t}',Y_{t})+\did(Y_{t},E_{t})\leq\\
&t+(t+1)=2t+1;
\end{alignat*}
and so, from 1) and Theorem \ref{thtdi0ECCequidec2}, $\tilde{E}_{t}=E_{t}'=E_{t}$. Thus, condition C3) of Definition \ref{deftSy0EC/t+1Sy0ED/AU0ED} holds in the general case. With regard to C4), if $\did(Y_{t},E_{t})\leq t$ then, as (\ref{eqdidzxye}),
$$
\did(Z,X)+\did(Y_{t-1},E_{t-1})=\did(Y_{t},E_{t})\leq t,
$$
and so, C4) follows exactly as the above case of no $0$-deletions where, really, we did not use the hypothesis $e=0$.
\end{IEEEproof}

\section{Redundancy Analysis}
\label{secredun}
In (\ref{eqEtX}), given $t,k\css{def}{=}k_{t+1}\el\inat$, let the field $\mathbb{F}=\mathbb{F}(k,t)\css{def}{=}\mathbb{F}_{k}$, where $\mathbb{F}_{k}$ is the smallest field, $\mathbb{K}$, whose cardinality is $|\mathbb{K}|>k$, when $t>1$, and $\mathbb{F}_{k}=(\iint_{k+1},+\bmod{(k+1)})$ when $t=1$. In this way, from Bertrand's Postulate it follows
\begin{equation}
\label{eqredana1}
k<|\mathbb{F}_{k}|<2k.
\end{equation}
From (\ref{eqEtX}) and (\ref{eqktdef}), the length of the recursive code design of Section \ref{secsyste_defi} satisfies the following recurrence relation,
\setlength{\jot}{0pt}
\begin{alignat}{1}
n_{t}(k)=k+t+1+n_{t-1}\left(k_{t}\right)\label{eqntkrec}
\end{alignat}
with
$$
k_{t}\css{def}{=}\min\{k,\isa t\log_{2}|\mathbb{F}_{k}|\isc\}\leq\isa t\log_{2}|\mathbb{F}_{k}|\isc.
$$
In this case, from $k_{t}\leq\isa t\log_{2}|\mathbb{F}_{k}|\isc$ and (\ref{eqredana1}), the length of the code design is bounded as
\setlength{\jot}{0pt}
\begin{alignat}{1}
k+t\leq n_{t}(k)=\,&k+t+1+n_{t-1}\left(k_{t}\right)\leq\label{eqntkrec2}\\
&k+t+1+n_{t-1}\left(\isa t\log_{2}(2k)\isc\right).\notag
\end{alignat}
First, let $n(t,k)=n^{\mbox{\scriptsize\ref{subsecreeds}}}(t,k)$ be the code length (\ref{eqredBCD_A}) of the $t$-Sy$0$EC/$(t+1)$-Sy$0$ED/AU$0$ED code with $k\el\inat$ information bits of Subsection \ref{subsecreeds}. From (\ref{eqntkrec2}), it is readily seen that, using only one recursive step and this code as base code,
\setlength{\jot}{4pt}
\begin{alignat}{1}
n_{t}\css{def}{=}n_{t}(k)\leq&\,k+t+1+n_{t-1}\left(\isa t\log_{2}(2k)\isc\right)=\label{eqntsb}\\
&k+t+1+n^{\mbox{\scriptsize\ref{subsecreeds}}}\left(t-1,\isa t\log_{2}(2k)\isc\right).\notag
\end{alignat}
Now, the asymptotic upper bound of Theorem \ref{thmubnsmin} applied to
$$
n^{\mbox{\scriptsize\ref{subsecreeds}}}\css{def}{=}n^{\mbox{\scriptsize\ref{subsecreeds}}}\left(t-1,\isa t\log_{2}(2k)\isc\right),
$$
can be used because
\setlength{\jot}{4pt}
\begin{alignat}{1}
\isa t\log_{2}(2k)\isc=\,&\isa t\log_{2}2+t\log_{2}k\isc=\label{eqeqntsb35}\\
&\isa t\log_{2}k\isc+t=\Theta(t\log k),\notag\\
b=&\,b\left(t-1,\isa t\log_{2}(2k)\isc\right)=\notag\\
&\isa\log_{2}(\isa t\log_{2}(2k)\isc+t-1)\isc=\notag\\
&\isa\log_{2}(\isa t\log_{2}(k)\isc+2t-1)\isc=\notag\\
&\Theta\left(\log(t\log k)\right),\notag\\
\log_{2}b=&\,\Theta\left(\log^{(2)}(t\log k)\right),\label{eqeqntsb36}\\
\tau=&\,\tau\left(t-1,\isa t\log_{2}(2k)\isc\right)=\notag\\
&\isa\sqrt{\frac{(t-1)b}{2\isa \isa t\log_{2}(2k)\isc/b\isc\log_{2}b}}\blackbox{7mm}{0mm}{0mm}\isc=\notag\\
&\Theta\left(\frac{b}{\sqrt{\log(k)\log^{(2)}(t\log k)}}\right);\notag
\end{alignat}
and so,
$$
\frac{\tau}{b}=\Theta\left(\frac{1}{\sqrt{\log(k)\log^{(2)}(t\log k)}}\right)=o(1)\leq \log_{e}\sqrt{2}
$$
when $\max\{t,k\}$ increases. So, from the asymptotic upper bound of Theorem \ref{thmubnsmin} applied to $n^{\mbox{\scriptsize\ref{subsecreeds}}}\css{def}{=}n^{\mbox{\scriptsize\ref{subsecreeds}}}\left(t-1,\isa t\log_{2}(2k)\isc\right)$, it follows that, if $\max\{t,k\}$ is large then $\max\{t-1,\isa t\log_{2}(2k)\isc\}$ is large, and then
\setlength{\jot}{4pt}
\begin{alignat}{1}
n^{\mbox{\scriptsize\ref{subsecreeds}}}\lesssim\,&\isa t\log_{2}(2k)\isc+\label{eqntsb2}\\
&\sqrt{8(t-1)\isa t\log_{2}(2k)\isc\log_{2}b}+2(t-1)\log_{2}b;\notag
\end{alignat}
and so, from (\ref{eqntsb2}), (\ref{eqeqntsb35}) and (\ref{eqeqntsb36}), it follows,
\setlength{\jot}{4pt}
\begin{alignat}{1}
n^{\mbox{\scriptsize\ref{subsecreeds}}}\lesssim\;&t\log_{2}k+O(t)+\label{eqeqntsb37}\\
&O\left(t\sqrt{\log k\log^{(2)}(t\log k)}\right)+\notag\\
&O\left(t\log^{(2)}(t\log k)\right)=\notag\\
&\,t\log_{2}k+O\left(t\log^{(2)}(t\log k)\sqrt{\log k}\right);\notag
\end{alignat}
and hence, from (\ref{eqntsb}), the redundancy is
\setlength{\jot}{2pt}
\begin{alignat}{1}
r_{t}\css{def}{=}r_{t}(k)\css{def}{=}\,&n_{t}(k)-k\leq\label{eqntasyO}\\
&t\log_{2}k+O\left(t\log^{(2)}(t\log k)\sqrt{\log k}\right),\notag
\end{alignat}
proving the leftmost relation in (\ref{eqsysreccdlenbound}).

Now, if $\max\{t,k\}\leq n_{t}$ is large then, from (\ref{eqntasyO}) and the leftmost relation in (\ref{eqntkrec2}), it follows,
\setlength{\jot}{6pt}
\begin{alignat*}{1}
\frac{r_{t}-t\log_{2}n_{t}}{t\log_{2}n_{t}}\leq\,&\frac{r_{t}-t\log_{2}k}{t\log_{2}n_{t}}\leq\\
&O\left(\frac{\cancel{\,t\,}\log^{(2)}(t\log k)\sqrt{\log k}}{\cancel{\,t\,}\log n_{t}}\right)\leq\\
&O\left(\frac{\log^{(2)}(t\log k)\sqrt{\log k}}{\sqrt{\log n_{t}}\sqrt{\log n_{t}}}\right)\leq\\
&O\left(\frac{\log^{(2)}(t\log k)\cancel{\sqrt{\log k}}}{\sqrt{\log n_{t}}\cancel{\sqrt{\log k}}}\right)\leq\\
&O\left(\frac{\log^{(2)}[(t+k)\log (k+t)]}{\sqrt{\log(k+t)}}\right)=o(1);
\end{alignat*}
and so,
\begin{equation}
\label{eqrtasyopt}
r_{t}=t\log_{2}k+o(t\log_{2}n_{t})\leq t\log_{2}n_{t}+o(t\log_{2}n_{t}).
\end{equation}
The last relation and Theorem \ref{thoptrel} imply that if $t=2^{o(\log n)}$ (say, $t=2^{\sqrt{\log_{2}k}}$) then the family of codes presented here are asymptotically optimal according to Definition \ref{defaoc}. Actually, note that (\ref{eqrtasyopt}) sets a stronger results because
$$
r_{t}(k)=t\log_{2}k+o(t\log_{2}n_{t}(k)).
$$
The above redundancy is actually achieved with the base code design of Subsection \ref{subsecreeds} for (see Theorem \ref{thmubnsmin})
\setlength{\jot}{6pt}
\begin{alignat*}{1}
\tau\css{def}{=}\,&\isa\sqrt{\frac{(t-1)b}{2\isa k_{t}/b\isc\log_{2}b}}\;\isc\simeq\sqrt{\frac{(t-1)}{2k_{t}\log_{2}b}}\,b,\\
k_{t}\css{def}{=}&\min\{k,\isa t\log_{2}|\mathbb{F}_{k}|\isc\},\;\mbox{and}\\
b\css{def}{=}&\isa\log_{2}(k_{t}+t-1)\isc\leq\isa\log_{2}(t\log_{2}k+2t)\isc.
\end{alignat*}
We note that, if $\tau=1$ then the balanced encoding $\beta_{\mathcal{I},\tau}1$ in (\ref{eqbetadef}) can be implemented with the efficient methods given in \cite{KNU86,PEL15,TAL98,TAL99}. If $\tau>1$ the encoding can be efficiently implemented with a table look-up to store $2^{b}\leq t\log_{2}k+2t+1$ codewords of length $b+\tau\log_{2}b+O(1)$. We just want to mention that choosing the same value of $b$ and $\tau=1$ in this base code design, and letting one recursive step, a more clever analysis shows that
\setlength{\jot}{6pt}
\begin{alignat}{1}
r_{t}(k)\lesssim t\log_{2}k+\,&t\log_{2}t+\left(\frac{\log_{2}b+3}{2b}\right)t\log_{2}k+\label{eqntsbwith_A}\\
&O\left(t\log_{2}^{(2)}k+t\log_{2}^{(2)}t\right).\notag
\end{alignat}
Note that $b$ increases as $\max\{t,k\}$ increases.

However it is possible to use any efficient $t$-Sy$0$EC/$(t+1)$-Sy$0$ED/AU$0$ED code as base code, going on recurring at least once, and stopping when it is most convenient in terms of redundancy. In this case, with the codes defined in this paper, the length can be computed exactly as
\setlength{\jot}{4pt}
\begin{alignat}{1}
&n_{t}(k)=k+t+1+\label{eqlentab}\\
&\;\min\!\left\{
n_{t-1}\left(k_{t}\right),
n^{\mbox{\scriptsize\ref{subseclimit}}}(t-1,k_{t}),
n^{\mbox{\scriptsize\ref{subsecreeds}}}(t-1,k_{t}),
n^{\mbox{\scriptsize\ref{subsecrepet}}}(t-1,k_{t}),
n^{\mbox{\scriptsize\ref{subsecdisti}}}(t-1,k_{t})
\right\}\!,\notag\\
&\;\;\mbox{with}\;k_{t}=\min\!\left\{k,\isa t\log_{2}|\mathbb{F}_{k}|\isc\right\}
\;\mbox{and}\;n_{0}(k)=k.\notag
\end{alignat}
where $n^{\mbox{\scriptsize\ref{subseclimit}}}(t,k)$, $n^{\mbox{\scriptsize\ref{subsecreeds}}}(t,k)$, $n^{\mbox{\scriptsize\ref{subsecrepet}}}(t,k)$ and $n^{\mbox{\scriptsize\ref{subsecdisti}}}(t,k)$ are the code lengths of the smallest $t$-Sy$0$EC/$(t+1)$-Sy$0$ED/AU$0$ED codes obtained in Subsection \ref{subseclimit}, Subsection \ref{subsecreeds}, Subsection \ref{subsecrepet} and Subsection \ref{subsecdisti} with $2^{k}$ codewords, respectively. The redundancy coming from (\ref{eqlentab}) has been computed together with the redundancies of the systematic codes in Subsection \ref{subsecrepet} and Subsection \ref{subsecdisti} for some values of $k$ and $t$. For some values of $k$ and $t$, Table \ref{tabV} shows the best redundancy among these three choices:
\begin{equation}
\label{eqredTABLEV}
r(t,k)
\css{def}{=}
\min\!\left\{n_{t}(k),n^{\mbox{\scriptsize\ref{subsecrepet}}}(t,k),n^{\mbox{\scriptsize\ref{subsecdisti}}}(t,k)
\right\}
-k.
\end{equation}
In Table \ref{tabV}, for the value of $k$ given in the first column and the value of $t$ given in the first row, if the entry is $r_{t_{b},T,k_{t_{b}}}^{n_{t_{b}},b,\tau}$ then
\begin{itemize}
\item
$r\css{def}{=}r(t,k)\el\inat$ is the number of redundant bits defined in (\ref{eqredTABLEV}) for the $t$-Sy$0$EC code design with $k$ information bits (so that the length of the code design is $n=k+r$);
\item
$t_{b}\el\inat$ is the error correcting capability of the base code (so that the number of recurring steps of the code design is $t-t_{b}$);
\item
$T\el\{I,W,R,M,S\}$ is the base code type used. The label $I$ indicates the identity code of length $n_{0}(k)=k$, $W$ the systematic distinct weight code in Subsection \ref{subsecdisti}, $R$ the systematic repetition code in Subsection \ref{subsecrepet}, $M$ the limited magnitude based code in Subsection \ref{subseclimit} and $S$ the Reed-Solomon based code in Subsection \ref{subsecreeds};
\item
$k_{t_{b}}\el\inat$ is the number of information bits of the base code;
\item
$n_{t_{b}}\el\inat$ is the base code length;
\item
if $T=M$ then the base code is obtained from the design in Subsection \ref{subseclimit} as follows. The $k$ bit information word is split into $\iia k/32\iic$ bytes of length $s=32$ bit plus a last byte of length $s=(k\bmod{32})\leq32$ bits (i.~e., for a total of $\iia k/32\iic+(k\bmod{32})$ bytes), and then each byte is encoded with the smallest $t$-Sy$0$EC/$(t+1)$-Sy$0$ED/AU$0$ED code in Subsection \ref{subseclimit} containing $2^{s}$ codewords. Then these byte encodings are concatenated by putting a $t+1$ bit long synchronizing sequence between two consecutive byte codewords. In the table, $b=\iia k/32\iic\el\inat$ and $\tau=k\bmod{32}\el\inat$.
\item
if $T=S$ then the base code is the one given in Subsection \ref{subsecreeds} so that $b\el\inat$ and $\tau\el\inat$ are the byte length, $b_{min}$, and error correcting capability, $\tau_{min}$, of the encoding $\beta_{\iint_{2}^{k_{t_{b}}},\tau}1$ in (\ref{eqbetadef}), respectively, which minimize the base code length $n(t_{b},k_{t_{b}},\tau,b)$ in (\ref{eqexactlengEbaseA}). In this scheme the Reed Solomon code designated distance is $\iia t/\tau\iic+1$.\end{itemize}
From the table, it can be noticed that for small values of $t/k$ the number of recursive steps is essentially $t$; whereas, for large values of $t/k$ the number of recursive steps is close to $0$ and $r/t$ start approaching $\log_{2}k$ (for example, for $k=t=256$, $r/t=2326/256=9.085\ldots\approx8=\log_{2}(256)$). We mention that there may be other base code designs that improve the redundacy of these recursive $t$-Sy$0$EC/$(t+1)$-Sy$0$ED/AU$0$ED codes. In this respect, the values reported in Table \ref{tabV} should be considered as upper-bounds on what can be practically achieved.
\begin{table*}[!htp]
\vglue -6mm
\tiny
\renewcommand{\tabcolsep}{0.94pt}
\renewcommand{\arraystretch}{1.273}
\caption{\vglue 0mm Systematic code design parameters for some values of $k$ and $t$.}
\begin{center}
\vglue -2mm
\begin{tabular}{|r||r|r|r|r|r|r|r|r|r|r|r|r|r|}
\hline
$k\backslash t$ & $1$& $2$& $3$& $4$& $5$& $6$& $7$& $8$& $16$& $32$& $64$& $128$& $256$ \\\hline\hline
$1$ & $0_{1,W,1}^{1}$ & $0_{2,W,1}^{1}$ & $0_{3,W,1}^{1}$ & $0_{4,W,1}^{1}$ & $0_{5,W,1}^{1}$ & $0_{6,W,1}^{1}$ & $0_{7,W,1}^{1}$ & $0_{8,W,1}^{1}$ & $0_{16,W,1}^{1}$ & $0_{32,W,1}^{1}$ & $0_{64,W,1}^{1}$ & $0_{128,W,1}^{1}$ & $0_{256,W,1}^{1}$ \\\hline
$2$ & $1_{1,W,2}^{3}$ & $1_{2,W,2}^{3}$ & $1_{3,W,2}^{3}$ & $1_{4,W,2}^{3}$ & $1_{5,W,2}^{3}$ & $1_{6,W,2}^{3}$ & $1_{7,W,2}^{3}$ & $1_{8,W,2}^{3}$ & $1_{16,W,2}^{3}$ & $1_{32,W,2}^{3}$ & $1_{64,W,2}^{3}$ & $1_{128,W,2}^{3}$ & $1_{256,W,2}^{3}$ \\\hline
$3$ & $3_{1,R,3}^{6}$ & $4_{2,W,3}^{7}$ & $4_{3,W,3}^{7}$ & $4_{4,W,3}^{7}$ & $4_{5,W,3}^{7}$ & $4_{6,W,3}^{7}$ & $4_{7,W,3}^{7}$ & $4_{8,W,3}^{7}$ & $4_{16,W,3}^{7}$ & $4_{32,W,3}^{7}$ & $4_{64,W,3}^{7}$ & $4_{128,W,3}^{7}$ & $4_{256,W,3}^{7}$ \\\hline
$4$ & $4_{1,R,4}^{8}$ & $8_{2,R,4}^{12}$ & $11_{3,W,4}^{15}$ & $11_{4,W,4}^{15}$ & $11_{5,W,4}^{15}$ & $11_{6,W,4}^{15}$ & $11_{7,W,4}^{15}$ & $11_{8,W,4}^{15}$ & $11_{16,W,4}^{15}$ & $11_{32,W,4}^{15}$ & $11_{64,W,4}^{15}$ & $11_{128,W,4}^{15}$ & $11_{256,W,4}^{15}$ \\\hline
$5$ & $5_{1,R,5}^{10}$ & $10_{2,R,5}^{15}$ & $13_{2,M,5}^{9,0,5}$ & $15_{3,M,5}^{10,0,5}$ & $17_{4,M,5}^{11,0,5}$ & $19_{5,M,5}^{12,0,5}$ & $21_{6,M,5}^{13,0,5}$ & $23_{7,M,5}^{14,0,5}$ & $26_{16,W,5}^{31}$ & $26_{32,W,5}^{31}$ & $26_{64,W,5}^{31}$ & $26_{128,W,5}^{31}$ & $26_{256,W,5}^{31}$ \\\hline
$6$ & $5_{0,I,3}^{3}$ & $12_{2,R,6}^{18}$ & $15_{2,M,6}^{11,0,6}$ & $17_{3,M,6}^{12,0,6}$ & $20_{4,M,6}^{14,0,6}$ & $22_{5,M,6}^{15,0,6}$ & $24_{6,M,6}^{16,0,6}$ & $26_{7,M,6}^{17,0,6}$ & $42_{15,M,6}^{25,0,6}$ & $57_{32,W,6}^{63}$ & $57_{64,W,6}^{63}$ & $57_{128,W,6}^{63}$ & $57_{256,W,6}^{63}$ \\\hline
$7$ & $5_{0,I,3}^{3}$ & $12_{1,M,6}^{9,0,6}$ & $16_{2,M,7}^{12,0,7}$ & $19_{3,M,7}^{14,0,7}$ & $22_{4,M,7}^{16,0,7}$ & $25_{5,M,7}^{18,0,7}$ & $27_{6,M,7}^{19,0,7}$ & $30_{7,M,7}^{21,0,7}$ & $47_{15,M,7}^{30,0,7}$ & $78_{31,M,7}^{45,0,7}$ & $120_{64,W,7}^{127}$ & $120_{128,W,7}^{127}$ & $120_{256,W,7}^{127}$ \\\hline
$8$ & $6_{0,I,4}^{4}$ & $13_{1,M,7}^{10,0,7}$ & $18_{2,M,8}^{14,0,8}$ & $22_{3,M,8}^{17,0,8}$ & $25_{4,M,8}^{19,0,8}$ & $28_{5,M,8}^{21,0,8}$ & $30_{6,M,8}^{22,0,8}$ & $33_{7,M,8}^{24,0,8}$ & $53_{15,M,8}^{36,0,8}$ & $85_{31,M,8}^{52,0,8}$ & $148_{63,M,8}^{83,0,8}$ & $247_{128,W,8}^{255}$ & $247_{256,W,8}^{255}$ \\\hline
$9$ & $6_{0,I,4}^{4}$ & $13_{1,M,7}^{10,0,7}$ & $20_{2,M,9}^{16,0,9}$ & $24_{3,M,9}^{19,0,9}$ & $27_{4,M,9}^{21,0,9}$ & $30_{5,M,9}^{23,0,9}$ & $33_{6,M,9}^{25,0,9}$ & $36_{7,M,9}^{27,0,9}$ & $58_{15,M,9}^{41,0,9}$ & $95_{31,M,9}^{62,0,9}$ & $158_{63,M,9}^{93,0,9}$ & $284_{127,M,9}^{155,0,9}$ & $502_{256,W,9}^{511}$ \\\hline
$10$ & $6_{0,I,4}^{4}$ & $13_{1,M,7}^{10,0,7}$ & $22_{2,M,10}^{18,0,10}$ & $26_{3,M,10}^{21,0,10}$ & $30_{4,M,10}^{24,0,10}$ & $33_{5,M,10}^{26,0,10}$ & $36_{6,M,10}^{28,0,10}$ & $40_{7,M,10}^{31,0,10}$ & $63_{15,M,10}^{46,0,10}$ & $106_{31,M,10}^{73,0,10}$ & $172_{63,M,10}^{107,0,10}$ & $298_{127,M,10}^{169,0,10}$ & $551_{255,M,10}^{294,0,10}$ \\\hline
$11$ & $6_{0,I,4}^{4}$ & $15_{1,M,8}^{12,0,8}$ & $24_{2,M,11}^{20,0,11}$ & $28_{3,M,11}^{23,0,11}$ & $32_{4,M,11}^{26,0,11}$ & $36_{5,M,11}^{29,0,11}$ & $40_{6,M,11}^{32,0,11}$ & $43_{7,M,11}^{34,0,11}$ & $69_{15,M,11}^{52,0,11}$ & $113_{31,M,11}^{80,0,11}$ & $191_{63,M,11}^{126,0,11}$ & $318_{127,M,11}^{189,0,11}$ & $572_{255,M,11}^{315,0,11}$ \\\hline
$12$ & $6_{0,I,4}^{4}$ & $15_{1,M,8}^{12,0,8}$ & $26_{2,M,12}^{22,0,12}$ & $30_{3,M,12}^{25,0,12}$ & $35_{4,M,12}^{29,0,12}$ & $39_{5,M,12}^{32,0,12}$ & $43_{6,M,12}^{35,0,12}$ & $46_{7,M,12}^{37,0,12}$ & $74_{15,M,12}^{57,0,12}$ & $121_{31,M,12}^{88,0,12}$ & $209_{63,M,12}^{144,0,12}$ & $345_{127,M,12}^{216,0,12}$ & $600_{255,M,12}^{343,0,12}$ \\\hline
$13$ & $6_{0,I,4}^{4}$ & $15_{1,M,8}^{12,0,8}$ & $26_{2,M,12}^{22,0,12}$ & $32_{3,M,13}^{27,0,13}$ & $37_{4,M,13}^{31,0,13}$ & $41_{5,M,13}^{34,0,13}$ & $46_{6,M,13}^{38,0,13}$ & $50_{7,M,13}^{41,0,13}$ & $79_{15,M,13}^{62,0,13}$ & $130_{31,M,13}^{97,0,13}$ & $221_{63,M,13}^{156,0,13}$ & $383_{127,M,13}^{254,0,13}$ & $638_{255,M,13}^{381,0,13}$ \\\hline
$14$ & $6_{0,I,4}^{4}$ & $15_{1,M,8}^{12,0,8}$ & $26_{2,M,12}^{22,0,12}$ & $34_{3,M,14}^{29,0,14}$ & $39_{4,M,14}^{33,0,14}$ & $44_{5,M,14}^{37,0,14}$ & $49_{6,M,14}^{41,0,14}$ & $53_{7,M,14}^{44,0,14}$ & $84_{15,M,14}^{67,0,14}$ & $139_{31,M,14}^{106,0,14}$ & $233_{63,M,14}^{168,0,14}$ & $413_{127,M,14}^{284,0,14}$ & $692_{255,M,14}^{435,0,14}$ \\\hline
$15$ & $6_{0,I,4}^{4}$ & $15_{1,M,8}^{12,0,8}$ & $26_{2,M,12}^{22,0,12}$ & $37_{3,M,15}^{32,0,15}$ & $42_{4,M,15}^{36,0,15}$ & $47_{5,M,15}^{40,0,15}$ & $52_{6,M,15}^{44,0,15}$ & $56_{7,M,15}^{47,0,15}$ & $90_{15,M,15}^{73,0,15}$ & $148_{31,M,15}^{115,0,15}$ & $246_{63,M,15}^{181,0,15}$ & $431_{127,M,15}^{302,0,15}$ & $767_{255,M,15}^{510,0,15}$ \\\hline
$16$ & $7_{0,I,5}^{5}$ & $16_{1,M,9}^{13,0,9}$ & $27_{2,M,13}^{23,0,13}$ & $39_{3,M,16}^{34,0,16}$ & $44_{4,M,16}^{38,0,16}$ & $50_{5,M,16}^{43,0,16}$ & $55_{6,M,16}^{47,0,16}$ & $60_{7,M,16}^{51,0,16}$ & $95_{15,M,16}^{78,0,16}$ & $156_{31,M,16}^{123,0,16}$ & $262_{63,M,16}^{197,0,16}$ & $450_{127,M,16}^{321,0,16}$ & $817_{255,M,16}^{560,0,16}$ \\\hline
$18$ & $7_{0,I,5}^{5}$ & $16_{1,M,9}^{13,0,9}$ & $27_{2,M,13}^{23,0,13}$ & $41_{3,M,17}^{36,0,17}$ & $49_{4,M,18}^{43,0,18}$ & $55_{5,M,18}^{48,0,18}$ & $61_{6,M,18}^{53,0,18}$ & $66_{7,M,18}^{57,0,18}$ & $106_{15,M,18}^{89,0,18}$ & $173_{31,M,18}^{140,0,18}$ & $293_{63,M,18}^{228,0,18}$ & $496_{127,M,18}^{367,0,18}$ & $874_{255,M,18}^{617,0,18}$ \\\hline
$20$ & $7_{0,I,5}^{5}$ & $18_{1,M,10}^{15,0,10}$ & $29_{2,M,14}^{25,0,14}$ & $45_{3,M,19}^{40,0,19}$ & $54_{4,M,20}^{48,0,20}$ & $61_{5,M,20}^{54,0,20}$ & $67_{6,M,20}^{59,0,20}$ & $73_{7,M,20}^{64,0,20}$ & $116_{15,M,20}^{99,0,20}$ & $191_{31,M,20}^{158,0,20}$ & $320_{63,M,20}^{255,0,20}$ & $556_{127,M,20}^{427,0,20}$ & $947_{255,M,20}^{690,0,20}$ \\\hline
$22$ & $7_{0,I,5}^{5}$ & $18_{1,M,10}^{15,0,10}$ & $29_{2,M,14}^{25,0,14}$ & $45_{3,M,19}^{40,0,19}$ & $59_{4,M,22}^{53,0,22}$ & $66_{5,M,22}^{59,0,22}$ & $73_{6,M,22}^{65,0,22}$ & $80_{7,M,22}^{71,0,22}$ & $127_{15,M,22}^{110,0,22}$ & $208_{31,M,22}^{175,0,22}$ & $350_{63,M,22}^{285,0,22}$ & $600_{127,M,22}^{471,0,22}$ & $1057_{255,M,22}^{800,0,22}$ \\\hline
$24$ & $7_{0,I,5}^{5}$ & $18_{1,M,10}^{15,0,10}$ & $29_{2,M,14}^{25,0,14}$ & $45_{3,M,19}^{40,0,19}$ & $64_{4,M,24}^{58,0,24}$ & $72_{5,M,24}^{65,0,24}$ & $79_{6,M,24}^{71,0,24}$ & $86_{7,M,24}^{77,0,24}$ & $137_{15,M,24}^{120,0,24}$ & $225_{31,M,24}^{192,0,24}$ & $379_{63,M,24}^{314,0,24}$ & $647_{127,M,24}^{518,0,24}$ & $1141_{255,M,24}^{884,0,24}$ \\\hline
$26$ & $7_{0,I,5}^{5}$ & $18_{1,M,10}^{15,0,10}$ & $31_{2,M,15}^{27,0,15}$ & $47_{3,M,20}^{42,0,20}$ & $64_{4,M,24}^{58,0,24}$ & $77_{5,M,26}^{70,0,26}$ & $85_{6,M,26}^{77,0,26}$ & $93_{7,M,26}^{84,0,26}$ & $148_{15,M,26}^{131,0,26}$ & $243_{31,M,26}^{210,0,26}$ & $407_{63,M,26}^{342,0,26}$ & $702_{127,M,26}^{573,0,26}$ & $1212_{255,M,26}^{955,0,26}$ \\\hline
$28$ & $7_{0,I,5}^{5}$ & $18_{1,M,10}^{15,0,10}$ & $31_{2,M,15}^{27,0,15}$ & $47_{3,M,20}^{42,0,20}$ & $67_{4,M,25}^{61,0,25}$ & $83_{5,M,28}^{76,0,28}$ & $91_{6,M,28}^{83,0,28}$ & $100_{7,M,28}^{91,0,28}$ & $157_{15,S,28}^{140,4,2}$ & $253_{31,S,28}^{220,4,2}$ & $437_{63,M,28}^{372,0,28}$ & $749_{127,M,28}^{620,0,28}$ & $1300_{255,M,28}^{1043,0,28}$ \\\hline
$30$ & $7_{0,I,5}^{5}$ & $18_{1,M,10}^{15,0,10}$ & $31_{2,M,15}^{27,0,15}$ & $47_{3,M,20}^{42,0,20}$ & $67_{4,M,25}^{61,0,25}$ & $88_{5,M,30}^{81,0,30}$ & $97_{6,M,30}^{89,0,30}$ & $105_{7,S,30}^{96,6,2}$ & $160_{15,S,30}^{143,5,2}$ & $257_{31,S,30}^{224,5,3}$ & $443_{63,S,30}^{378,5,3}$ & $796_{127,M,30}^{667,0,30}$ & $1399_{255,M,30}^{1142,0,30}$ \\\hline
$31$ & $7_{0,I,5}^{5}$ & $18_{1,M,10}^{15,0,10}$ & $31_{2,M,15}^{27,0,15}$ & $47_{3,M,20}^{42,0,20}$ & $67_{4,M,25}^{61,0,25}$ & $88_{5,M,30}^{81,0,30}$ & $100_{6,M,31}^{92,0,31}$ & $108_{7,S,31}^{99,7,2}$ & $164_{15,S,31}^{147,5,2}$ & $261_{31,S,31}^{228,4,2}$ & $449_{63,S,31}^{384,5,3}$ & $822_{127,M,31}^{693,0,31}$ & $1439_{255,M,31}^{1182,0,31}$ \\\hline
$32$ & $8_{0,I,6}^{6}$ & $19_{1,M,11}^{16,0,11}$ & $33_{2,M,16}^{29,0,16}$ & $50_{3,M,21}^{45,0,21}$ & $71_{4,M,27}^{65,0,27}$ & $94_{5,M,32}^{87,1,0}$ & $104_{6,M,32}^{96,1,0}$ & $110_{7,S,32}^{101,7,2}$ & $166_{15,S,32}^{149,5,2}$ & $263_{31,S,32}^{230,4,2}$ & $453_{63,S,32}^{388,5,3}$ & $849_{127,M,32}^{720,1,0}$ & $1477_{255,M,32}^{1220,1,0}$ \\\hline
$36$ & $8_{0,I,6}^{6}$ & $19_{1,M,11}^{16,0,11}$ & $33_{2,M,16}^{29,0,16}$ & $50_{3,M,21}^{45,0,21}$ & $71_{4,M,27}^{65,0,27}$ & $94_{5,M,32}^{87,1,0}$ & $113_{6,S,36}^{105,4,1}$ & $117_{7,S,36}^{108,6,2}$ & $173_{15,S,36}^{156,6,2}$ & $273_{31,S,36}^{240,4,2}$ & $463_{63,S,36}^{398,5,3}$ & $863_{127,S,36}^{734,5,3}$ & $1682_{255,S,36}^{1425,6,5}$ \\\hline
$40$ & $8_{0,I,6}^{6}$ & $19_{1,M,11}^{16,0,11}$ & $35_{2,M,17}^{31,0,17}$ & $52_{3,M,22}^{47,0,22}$ & $71_{4,M,27}^{65,0,27}$ & $96_{5,S,33}^{89,7,2}$ & $118_{6,S,38}^{110,4,1}$ & $124_{7,S,40}^{115,7,2}$ & $182_{15,S,40}^{165,5,2}$ & $283_{31,S,40}^{250,4,2}$ & $471_{63,S,40}^{406,5,3}$ & $871_{127,S,40}^{742,5,3}$ & $1703_{255,S,40}^{1446,7,4}$ \\\hline
$44$ & $8_{0,I,6}^{6}$ & $20_{1,M,12}^{17,0,12}$ & $35_{2,M,17}^{31,0,17}$ & $54_{3,M,23}^{49,0,23}$ & $74_{4,M,28}^{68,0,28}$ & $97_{5,S,34}^{90,7,2}$ & $119_{6,S,39}^{111,4,1}$ & $132_{7,S,44}^{123,7,2}$ & $191_{15,S,44}^{174,6,2}$ & $296_{31,S,44}^{263,5,2}$ & $484_{63,S,44}^{419,5,3}$ & $884_{127,S,44}^{755,5,3}$ & $1720_{255,S,44}^{1463,7,4}$ \\\hline
$48$ & $8_{0,I,6}^{6}$ & $20_{1,M,12}^{17,0,12}$ & $35_{2,M,17}^{31,0,17}$ & $54_{3,M,23}^{49,0,23}$ & $76_{4,M,29}^{70,0,29}$ & $97_{5,S,34}^{90,7,2}$ & $120_{6,S,40}^{112,4,1}$ & $134_{7,S,45}^{125,7,2}$ & $197_{15,S,48}^{180,6,2}$ & $305_{31,S,48}^{272,5,2}$ & $497_{63,S,48}^{432,5,3}$ & $897_{127,S,48}^{768,5,3}$ & $1726_{255,S,48}^{1469,7,4}$ \\\hline
$52$ & $8_{0,I,6}^{6}$ & $20_{1,M,12}^{17,0,12}$ & $36_{2,M,18}^{32,0,18}$ & $54_{3,M,23}^{49,0,23}$ & $76_{4,M,29}^{70,0,29}$ & $98_{5,S,35}^{91,7,2}$ & $123_{6,S,41}^{115,4,1}$ & $136_{7,S,46}^{127,7,2}$ & $207_{15,S,52}^{190,6,2}$ & $314_{31,S,52}^{281,5,2}$ & $509_{63,S,52}^{444,5,3}$ & $909_{127,S,52}^{780,5,3}$ & $1743_{255,S,52}^{1486,7,4}$ \\\hline
$56$ & $8_{0,I,6}^{6}$ & $20_{1,M,12}^{17,0,12}$ & $36_{2,M,18}^{32,0,18}$ & $56_{3,M,24}^{51,0,24}$ & $79_{4,M,30}^{73,0,30}$ & $102_{5,S,36}^{95,7,2}$ & $125_{6,S,42}^{117,4,1}$ & $138_{7,S,48}^{129,7,2}$ & $212_{15,S,56}^{195,7,2}$ & $323_{31,S,56}^{290,5,2}$ & $540_{63,S,56}^{475,5,3}$ & $943_{127,S,56}^{814,6,4}$ & $1748_{255,S,56}^{1491,7,4}$ \\\hline
$60$ & $8_{0,I,6}^{6}$ & $20_{1,M,12}^{17,0,12}$ & $36_{2,M,18}^{32,0,18}$ & $56_{3,M,24}^{51,0,24}$ & $79_{4,M,30}^{73,0,30}$ & $102_{5,S,36}^{95,7,2}$ & $125_{6,S,42}^{117,4,1}$ & $138_{7,S,48}^{129,7,2}$ & $221_{15,S,60}^{204,6,2}$ & $330_{31,S,60}^{297,5,2}$ & $548_{63,S,60}^{483,5,3}$ & $949_{127,S,60}^{820,6,4}$ & $1765_{255,S,60}^{1508,7,4}$ \\\hline
$63$ & $8_{0,I,6}^{6}$ & $20_{1,M,12}^{17,0,12}$ & $36_{2,M,18}^{32,0,18}$ & $56_{3,M,24}^{51,0,24}$ & $79_{4,M,30}^{73,0,30}$ & $102_{5,S,36}^{95,7,2}$ & $125_{6,S,42}^{117,4,1}$ & $138_{7,S,48}^{129,7,2}$ & $225_{15,S,63}^{208,7,2}$ & $338_{31,S,63}^{305,5,2}$ & $560_{63,S,63}^{495,5,3}$ & $965_{127,S,63}^{836,6,4}$ & $1769_{255,S,63}^{1512,7,4}$ \\\hline
$64$ & $9_{0,I,7}^{7}$ & $22_{0,I,4}^{4}$ & $38_{2,M,19}^{34,0,19}$ & $58_{3,M,25}^{53,0,25}$ & $81_{4,M,31}^{75,0,31}$ & $104_{5,S,37}^{97,7,2}$ & $126_{6,S,43}^{118,4,1}$ & $139_{7,S,49}^{130,7,2}$ & $229_{15,S,64}^{212,7,2}$ & $340_{31,S,64}^{307,5,2}$ & $561_{63,S,64}^{496,5,3}$ & $966_{127,S,64}^{837,6,4}$ & $1779_{255,S,64}^{1522,7,4}$ \\\hline
$127$ & $9_{0,I,7}^{7}$ & $23_{0,I,4}^{4}$ & $42_{2,M,21}^{38,0,21}$ & $64_{2,M,15}^{27,0,15}$ & $91_{3,M,21}^{45,0,21}$ & $111_{5,S,42}^{104,7,2}$ & $135_{6,S,49}^{127,5,1}$ & $152_{7,S,56}^{143,7,2}$ & $316_{15,S,112}^{299,7,2}$ & $466_{31,S,127}^{433,7,2}$ & $693_{63,S,127}^{628,6,2}$ & $1143_{127,S,127}^{1014,6,3}$ & $1968_{255,S,127}^{1711,7,4}$ \\\hline
$128$ & $10_{0,I,8}^{8}$ & $24_{0,I,4}^{4}$ & $44_{1,M,10}^{15,0,10}$ & $65_{2,M,15}^{27,0,15}$ & $92_{3,M,21}^{45,0,21}$ & $115_{5,S,43}^{108,7,2}$ & $136_{6,S,50}^{128,5,1}$ & $156_{7,S,57}^{147,7,2}$ & $320_{15,S,113}^{303,7,2}$ & $468_{31,S,128}^{435,7,2}$ & $695_{63,S,128}^{630,6,2}$ & $1147_{127,S,128}^{1018,6,3}$ & $1972_{255,S,128}^{1715,7,4}$ \\\hline
$255$ & $10_{0,I,8}^{8}$ & $26_{0,I,5}^{5}$ & $46_{1,M,10}^{15,0,10}$ & $70_{2,M,16}^{29,0,16}$ & $98_{3,M,22}^{47,0,22}$ & $123_{5,S,48}^{116,7,2}$ & $147_{6,S,56}^{139,5,1}$ & $168_{7,S,64}^{159,5,1}$ & $346_{15,S,128}^{329,6,1}$ & $704_{31,S,255}^{671,7,2}$ & $944_{63,S,255}^{879,7,2}$ & $1424_{127,S,255}^{1295,7,2}$ & $2326_{255,S,255}^{2069,7,3}$ \\\hline
$256$ & $11_{0,I,9}^{9}$ & $27_{0,I,5}^{5}$ & $47_{1,M,10}^{15,0,10}$ & $71_{2,M,16}^{29,0,16}$ & $99_{3,M,22}^{47,0,22}$ & $124_{5,S,49}^{117,7,2}$ & $149_{6,S,57}^{141,5,1}$ & $169_{7,S,65}^{160,5,1}$ & $347_{15,S,129}^{330,6,1}$ & $706_{31,S,256}^{673,7,2}$ & $946_{63,S,256}^{881,7,2}$ & $1426_{127,S,256}^{1297,7,2}$ & $2327_{255,S,256}^{2070,7,3}$ \\\hline
$511$ & $11_{0,I,9}^{9}$ & $28_{0,I,5}^{5}$ & $49_{1,M,10}^{15,0,10}$ & $74_{2,M,16}^{29,0,16}$ & $105_{3,M,23}^{49,0,23}$ & $133_{5,S,54}^{126,6,1}$ & $158_{6,S,63}^{150,5,1}$ & $180_{7,S,72}^{171,6,1}$ & $368_{15,S,144}^{351,6,1}$ & $765_{31,S,288}^{732,7,2}$ & $1417_{63,S,511}^{1352,7,2}$ & $1960_{127,S,511}^{1831,7,2}$ & $2938_{255,S,511}^{2681,8,3}$ \\\hline
$512$ & $12_{0,I,10}^{10}$ & $29_{0,I,5}^{5}$ & $50_{1,M,10}^{15,0,10}$ & $77_{2,M,17}^{31,0,17}$ & $106_{3,M,23}^{49,0,23}$ & $135_{5,S,55}^{128,5,1}$ & $159_{6,S,64}^{151,5,1}$ & $183_{7,S,73}^{174,5,1}$ & $371_{15,S,145}^{354,6,1}$ & $767_{31,S,289}^{734,7,2}$ & $1421_{63,S,512}^{1356,7,2}$ & $1964_{127,S,512}^{1835,7,2}$ & $2939_{255,S,512}^{2682,8,3}$ \\\hline
$2^{10}$ & $13_{0,I,11}^{11}$ & $31_{0,I,5}^{5}$ & $53_{1,M,10}^{15,0,10}$ & $81_{2,M,17}^{31,0,17}$ & $111_{3,M,23}^{49,0,23}$ & $145_{5,S,61}^{138,6,1}$ & $169_{6,S,71}^{161,6,1}$ & $195_{7,S,81}^{186,6,1}$ & $394_{15,S,161}^{377,6,1}$ & $825_{31,S,321}^{792,7,2}$ & $1661_{63,S,641}^{1596,7,2}$ & $2915_{127,S,1024}^{2786,7,2}$ & $4066_{255,S,1024}^{3809,7,2}$ \\\hline
$2^{11}$ & $14_{0,I,12}^{12}$ & $33_{0,I,5}^{5}$ & $57_{1,M,11}^{16,0,11}$ & $85_{2,M,17}^{31,0,17}$ & $118_{3,M,24}^{51,0,24}$ & $154_{5,S,67}^{147,6,1}$ & $179_{6,S,78}^{171,6,1}$ & $206_{7,S,89}^{197,6,1}$ & $419_{15,S,177}^{402,6,1}$ & $873_{31,S,353}^{840,6,1}$ & $1772_{63,S,705}^{1707,12,2}$ & $3560_{127,S,1409}^{3431,12,2}$ & $5897_{255,S,2048}^{5640,11,2}$ \\\hline
$2^{12}$ & $15_{0,I,13}^{13}$ & $35_{0,I,5}^{5}$ & $60_{1,M,11}^{16,0,11}$ & $90_{2,M,18}^{32,0,18}$ & $123_{3,M,24}^{51,0,24}$ & $163_{5,S,73}^{156,6,1}$ & $191_{6,S,85}^{183,6,1}$ & $219_{7,S,97}^{210,6,1}$ & $443_{15,S,193}^{426,6,1}$ & $922_{31,S,385}^{889,6,1}$ & $1874_{63,S,769}^{1809,12,2}$ & $3757_{127,S,1537}^{3628,13,2}$ & $7528_{255,S,3073}^{7271,13,2}$ \\\hline
$2^{13}$ & $16_{0,I,14}^{14}$ & $37_{0,I,5}^{5}$ & $63_{1,M,11}^{16,0,11}$ & $94_{2,M,18}^{32,0,18}$ & $130_{3,M,25}^{53,0,25}$ & $170_{4,M,32}^{78,1,0}$ & $202_{6,S,92}^{194,6,1}$ & $231_{7,S,105}^{222,6,1}$ & $466_{15,S,209}^{449,6,1}$ & $970_{31,S,417}^{937,6,1}$ & $1976_{63,S,833}^{1911,12,2}$ & $3953_{127,S,1665}^{3824,13,2}$ & $7919_{255,S,3329}^{7662,14,2}$ \\\hline
$2^{14}$ & $17_{0,I,15}^{15}$ & $39_{0,I,5}^{5}$ & $67_{1,M,12}^{17,0,12}$ & $98_{2,M,18}^{32,0,18}$ & $135_{3,M,25}^{53,0,25}$ & $181_{3,M,21}^{45,0,21}$ & $212_{6,S,99}^{204,6,1}$ & $242_{7,S,113}^{233,6,1}$ & $491_{15,S,225}^{474,6,1}$ & $1017_{31,S,449}^{984,6,1}$ & $2076_{63,S,897}^{2011,12,2}$ & $4148_{127,S,1793}^{4019,13,2}$ & $8304_{255,S,3585}^{8047,14,2}$ \\\hline
$2^{15}$ & $18_{0,I,16}^{16}$ & $41_{0,I,5}^{5}$ & $70_{1,M,12}^{17,0,12}$ & $102_{2,M,18}^{32,0,18}$ & $142_{3,M,26}^{55,0,26}$ & $187_{3,M,21}^{45,0,21}$ & $222_{6,S,106}^{214,6,1}$ & $255_{7,S,121}^{246,6,1}$ & $515_{15,S,241}^{498,6,1}$ & $1066_{31,S,481}^{1033,6,1}$ & $2178_{63,S,961}^{2113,12,2}$ & $4346_{127,S,1921}^{4217,13,2}$ & $8689_{255,S,3841}^{8432,14,2}$ \\\hline
$2^{16}$ & $19_{0,I,17}^{17}$ & $44_{0,I,6}^{6}$ & $73_{1,M,12}^{17,0,12}$ & $108_{2,M,19}^{34,0,19}$ & $147_{3,M,26}^{55,0,26}$ & $194_{3,M,21}^{45,0,21}$ & $232_{5,S,42}^{104,7,2}$ & $267_{7,S,129}^{258,6,1}$ & $538_{15,S,257}^{521,6,1}$ & $1114_{31,S,513}^{1081,6,1}$ & $2263_{63,S,1025}^{2198,13,2}$ & $4529_{127,S,2049}^{4400,14,2}$ & $9072_{255,S,4097}^{8815,14,2}$ \\\hline
$2^{17}$ & $20_{0,I,18}^{18}$ & $46_{0,I,6}^{6}$ & $76_{1,M,12}^{17,0,12}$ & $112_{2,M,19}^{34,0,19}$ & $152_{3,M,26}^{55,0,26}$ & $200_{3,M,21}^{45,0,21}$ & $239_{5,S,42}^{104,7,2}$ & $278_{7,S,137}^{269,6,1}$ & $563_{15,S,273}^{546,6,1}$ & $1161_{31,S,545}^{1128,6,1}$ & $2362_{63,S,1089}^{2297,13,2}$ & $4720_{127,S,2177}^{4591,14,2}$ & $9454_{255,S,4353}^{9197,14,2}$ \\\hline
$2^{18}$ & $21_{0,I,19}^{19}$ & $48_{0,I,6}^{6}$ & $79_{1,M,12}^{17,0,12}$ & $116_{2,M,19}^{34,0,19}$ & $159_{3,M,27}^{57,0,27}$ & $207_{3,M,21}^{45,0,21}$ & $246_{5,S,42}^{104,7,2}$ & $291_{7,S,145}^{282,6,1}$ & $587_{15,S,289}^{570,6,1}$ & $1234_{31,S,577}^{1201,12,2}$ & $2461_{63,S,1153}^{2396,13,2}$ & $4912_{127,S,2305}^{4783,14,2}$ & $9841_{255,S,4609}^{9584,14,2}$ \\\hline
$2^{19}$ & $22_{0,I,20}^{20}$ & $50_{0,I,6}^{6}$ & $82_{1,M,12}^{17,0,12}$ & $120_{2,M,19}^{34,0,19}$ & $164_{3,M,27}^{57,0,27}$ & $213_{3,M,21}^{45,0,21}$ & $257_{5,S,43}^{108,7,2}$ & $303_{6,S,52}^{133,4,1}$ & $610_{15,S,305}^{593,6,1}$ & $1284_{31,S,609}^{1251,12,2}$ & $2560_{63,S,1217}^{2495,13,2}$ & $5103_{127,S,2433}^{4974,14,2}$ & $10224_{255,S,4865}^{9967,14,2}$ \\\hline
$2^{20}$ & $23_{0,I,21}^{21}$ & $52_{0,I,6}^{6}$ & $85_{1,M,12}^{17,0,12}$ & $126_{2,M,20}^{36,0,20}$ & $169_{3,M,27}^{57,0,27}$ & $219_{3,M,21}^{45,0,21}$ & $266_{5,S,44}^{110,7,2}$ & $311_{6,S,52}^{133,4,1}$ & $650_{15,S,321}^{633,6,1}$ & $1336_{31,S,641}^{1303,6,1}$ & $2658_{63,S,1281}^{2593,13,2}$ & $5294_{127,S,2561}^{5165,14,2}$ & $10607_{255,S,5121}^{10350,14,2}$ \\\hline
$2^{21}$ & $24_{0,I,22}^{22}$ & $54_{0,I,6}^{6}$ & $90_{0,I,4}^{4}$ & $130_{2,M,20}^{36,0,20}$ & $174_{3,M,27}^{57,0,27}$ & $225_{3,M,21}^{45,0,21}$ & $273_{5,S,44}^{110,7,2}$ & $320_{6,S,53}^{134,5,1}$ & $674_{15,S,337}^{657,6,1}$ & $1385_{31,S,673}^{1352,6,1}$ & $2757_{63,S,1345}^{2692,13,2}$ & $5488_{127,S,2689}^{5359,14,2}$ & $10992_{255,S,5377}^{10735,14,2}$ \\\hline
$2^{22}$ & $25_{0,I,23}^{23}$ & $56_{0,I,6}^{6}$ & $93_{0,I,4}^{4}$ & $134_{2,M,20}^{36,0,20}$ & $181_{2,M,15}^{27,0,15}$ & $232_{3,M,21}^{45,0,21}$ & $280_{5,S,44}^{110,7,2}$ & $328_{6,S,53}^{134,5,1}$ & $697_{15,S,353}^{680,6,1}$ & $1433_{31,S,705}^{1400,6,1}$ & $2856_{63,S,1409}^{2791,13,2}$ & $5681_{127,S,2817}^{5552,14,2}$ & $11377_{255,S,5633}^{11120,14,2}$ \\\hline
$2^{23}$ & $26_{0,I,24}^{24}$ & $58_{0,I,6}^{6}$ & $96_{0,I,4}^{4}$ & $138_{2,M,20}^{36,0,20}$ & $186_{2,M,15}^{27,0,15}$ & $241_{3,M,22}^{47,0,22}$ & $289_{5,S,45}^{112,5,1}$ & $337_{6,S,54}^{135,5,1}$ & $722_{15,S,369}^{705,6,1}$ & $1480_{31,S,737}^{1447,6,1}$ & $2955_{63,S,1473}^{2890,13,2}$ & $5873_{127,S,2945}^{5744,14,2}$ & $11760_{255,S,5889}^{11503,14,2}$ \\\hline
$2^{24}$ & $27_{0,I,25}^{25}$ & $60_{0,I,6}^{6}$ & $99_{0,I,4}^{4}$ & $142_{2,M,20}^{36,0,20}$ & $191_{2,M,15}^{27,0,15}$ & $247_{3,M,22}^{47,0,22}$ & $296_{5,S,45}^{112,5,1}$ & $345_{6,S,54}^{135,5,1}$ & $740_{14,S,130}^{322,6,1}$ & $1529_{31,S,769}^{1496,6,1}$ & $3053_{63,S,1537}^{2988,13,2}$ & $6064_{127,S,3073}^{5935,14,2}$ & $12142_{255,S,6145}^{11885,14,2}$ \\\hline
$2^{25}$ & $28_{0,I,26}^{26}$ & $62_{0,I,6}^{6}$ & $102_{0,I,4}^{4}$ & $148_{2,M,21}^{38,0,21}$ & $196_{2,M,15}^{27,0,15}$ & $253_{3,M,22}^{47,0,22}$ & $303_{5,S,45}^{112,5,1}$ & $354_{6,S,55}^{136,5,1}$ & $757_{14,S,131}^{323,6,1}$ & $1577_{31,S,801}^{1544,6,1}$ & $3151_{63,S,1601}^{3086,13,2}$ & $6256_{127,S,3201}^{6127,14,2}$ & $12529_{255,S,6401}^{12272,14,2}$ \\\hline
$2^{26}$ & $29_{0,I,27}^{27}$ & $64_{0,I,6}^{6}$ & $105_{0,I,4}^{4}$ & $152_{2,M,21}^{38,0,21}$ & $202_{2,M,15}^{27,0,15}$ & $259_{3,M,22}^{47,0,22}$ & $312_{5,S,46}^{114,7,2}$ & $362_{6,S,55}^{136,5,1}$ & $773_{14,S,131}^{323,6,1}$ & $1624_{31,S,833}^{1591,6,1}$ & $3249_{63,S,1665}^{3184,13,2}$ & $6447_{127,S,3329}^{6318,14,2}$ & $12912_{255,S,6657}^{12655,14,2}$ \\\hline
$2^{27}$ & $30_{0,I,28}^{28}$ & $66_{0,I,6}^{6}$ & $108_{0,I,4}^{4}$ & $156_{2,M,21}^{38,0,21}$ & $207_{2,M,15}^{27,0,15}$ & $265_{3,M,22}^{47,0,22}$ & $319_{5,S,46}^{114,7,2}$ & $370_{6,S,55}^{136,5,1}$ & $790_{14,S,132}^{324,6,1}$ & $1673_{31,S,865}^{1640,6,1}$ & $3345_{63,S,1729}^{3280,13,2}$ & $6638_{127,S,3457}^{6509,14,2}$ & $13295_{255,S,6913}^{13038,14,2}$ \\\hline
$2^{28}$ & $31_{0,I,29}^{29}$ & $68_{0,I,6}^{6}$ & $111_{0,I,4}^{4}$ & $160_{2,M,21}^{38,0,21}$ & $212_{2,M,15}^{27,0,15}$ & $272_{3,M,22}^{47,0,22}$ & $326_{5,S,46}^{114,7,2}$ & $378_{6,S,55}^{136,5,1}$ & $809_{14,S,133}^{327,6,1}$ & $1720_{30,S,305}^{758,6,1}$ & $3444_{63,S,1793}^{3379,13,2}$ & $6832_{127,S,3585}^{6703,14,2}$ & $13680_{255,S,7169}^{13423,14,2}$ \\\hline
\end{tabular}
\end{center}
\label{tabV}
\end{table*}

\section{Concluding remarks}
\label{secconcl}
Some theory and efficient design of binary block codes capable of controlling the deletions and/or insertions of the symbol ``$0$'' (i.~e., the $0$-errors) are given. It is shown that the design of codes for insertion and/or deletion of zeros is equivalent to the design of the $L_{1}$ metric error control codes. Some efficient asymptotically optimal systematic codes for correcting these errors are described and their encoding and decoding methods are also explained. 

Please note that based on the theory developed in this paper and \cite{BOS82,WEB92,TAL08,TAL10,TAL10b,TAL11a,TAL12a,TAL13,TAL18,TAL18b,TAL23}, whenever it is possible to define an isometry from the metric space which characterizes a given coding problem to the $L_{1}$ metric (as the mapping $V$ in (\ref{eqfunV})), any information on codes for the $L_{1}$ metric reflects in the analogous information for that coding problem. In particular, the sticky channel error control problem \cite{DOL10,TAL10b,MAH17} can be reduced to the $L_{1}$ metric error control problem through the isometry given by the composition of the Gray mapping and the $V$ mapping. Also, lower bounds, upper bounds, code designs and decoding algorithms can be given for $t$-Sy$0$EC codes which satisfy the $RLL(d,k)$ constraint \cite{LEV93,PAL12} by using $L_{1}$ error control codes over $\iint_{m}$, with $m\el\inat\cup\{\infty\}$. This is because the set of all $RLL(d,k)$ binary words of length $n$ and weight $w$ with the $\did$ metric can be put in bijection with $\left(\iint_{k-d+1}^{w+1}\!,d_{L_{1}}^{sy}\right)$ through the following isometry
$$
0^{v_{1}}10^{v_{2}}1\ldots0^{v_{w}}10^{v_{w+1}}\quad\leftrightarrow\quad(v_{1}-d,v_{2}-d,\ldots,v_{w}-d).
$$
Likewise, the bit-shift coding problem described in \cite{LEV93,KOV19} can be solved with the following isometry from the appropriate metric space $\left(\mathcal{S}(\iint_{2},n,w),d_{bit-shift}\right)$ into the metric space $\left(\iint_{n}^{w}\!,d_{L_{1}}^{sy}\right)$,
$$
0^{v_{1}}10^{v_{2}}1\ldots0^{v_{w}}10^{v_{w+1}}\quad\leftrightarrow\quad(v_{1},v_{1}+v_{2}+1,\ldots,v_{1}+v_{2}+\ldots+v_{w}+w-1)
$$
which associates any binary word with its support.

In addition, if we restrict the number of errors within the bucket of zeros to be at most $l\el\inat$, then the zero error capacity codes for the $L_1$ metric described in \cite{TAL18} can be used to design error correcting codes for this scenarios.

\section*{Acknowledgments}
This work is supported by the NSF grant CCF-2006571.

\section*{Appendix}
\begin{IEEEproof}[Proof of Theorem \ref{thmconcatprop}]
If $w(X_{1})\neq w(Y_{1})$ then, from Theorem \ref{thmisodidl1}, the righthand side value of (\ref{concsubadd}) is $\infty$, and therefore (\ref{concsubadd}) is true. Now, assume $w(X_{1})=w(Y_{1})$ and prove (\ref{concadd}). If $w(X_{1})=w(Y_{1})$ and $\did(X_{1}X_{2},Y_{1}Y_{2})=\infty$ then
\setlength{\jot}{2pt}
\begin{align*}
\did(X_{1}X_{2},Y_{1}Y_{2})=\infty\;&\imp\;w(X_{1}X_{2})\neq w(Y_{1}Y_{2})\;\imp\\
w(X_{2})\neq w(Y_{2})\;&\imp\;\did(X_{2},Y_{2})=\infty;
\end{align*}
and so, (\ref{concadd}) is true with equality because $|Q|\el\inat$. At this point, assume $w_{1}\css{def}{=}w(X_{1})=w(Y_{1})$ and $\did(X_{1}X_{2},Y_{1}Y_{2})<\infty$. In this case, from (\ref{eqdidugdl0}),
\setlength{\jot}{2pt}
\begin{align*}
w_{1}+w(X_{2})=\;&w(X_{1})+w(X_{2})=w(X_{1}X_{2})=\\
&w(Y_{1}Y_{2})=w(Y_{1})+w(Y_{2})=w_{1}+w(Y_{2}),
\end{align*}
and so, $w_{2}\css{def}{=}w(X_{2})=w(Y_{2})$. In particular, from (\ref{eqdidugdl0}) and (\ref{eqdistdef}),
\setlength{\jot}{4pt}
\begin{alignat*}{1}
\did&(X_{1}X_{2},Y_{1}Y_{2})=d_{L_{1}}^{sy}(V(X_{1}X_{2}),V(Y_{1}Y_{2}))=\\
&\sum_{i=1}^{w_{1}+w_{2}+1}|v_{i}(X_{1}X_{2})-v_{i}(Y_{1}Y_{2})|=\\
&\sum_{i=1}^{w_{1}}|v_{i}(X_{1}X_{2})-v_{i}(Y_{1}Y_{2})|+\\
&|v_{w_{1}+1}(X_{1}X_{2})-v_{w_{1}+1}(Y_{1}Y_{2})|+\\
&\sum_{i=2}^{w_{2}+1}|v_{w_{1}+i}(X_{1}X_{2})-v_{w_{1}+i}(Y_{1}Y_{2})|=\\
&\sum_{i=1}^{w_{1}+1}|v_{i}(X_{1})-v_{i}(Y_{1})|-|v_{w_{1}+1}(X_{1})-v_{w_{1}+1}(Y_{1})|+\\
&|v_{w_{1}+1}(X_{1}X_{2})-v_{w_{1}+1}(Y_{1}Y_{2})|-\\
&|v_{1}(X_{2})-v_{1}(Y_{2})|+\sum_{i=1}^{w_{2}+1}|v_{i}(X_{2})-v_{i}(Y_{2})|=\\
&d_{L_{1}}^{sy}(V(X_{1}),V(Y_{1}))-|v_{w_{1}+1}(X_{1})-v_{w_{1}+1}(Y_{1})|+\\
&|v_{w_{1}+1}(X_{1}X_{2})-v_{w_{1}+1}(Y_{1}Y_{2})|-\\
&|v_{1}(X_{2})-v_{1}(Y_{2})|+d_{L_{1}}^{sy}(V(X_{2}),V(Y_{2}))=\\
&\did(X_{1},Y_{1})+\did(X_{2},Y_{2})+\\
&|v_{w_{1}+1}(X_{1}X_{2})-v_{w_{1}+1}(Y_{1}Y_{2})|-\\
&(|v_{w_{1}+1}(X_{1})-v_{w_{1}+1}(Y_{1})|+|v_{1}(X_{2})-v_{1}(Y_{2})|)=\\
&\did(X_{1},Y_{1})+\did(X_{2},Y_{2})+\\
&|v_{w_{1}+1}(X_{1})+v_{1}(X_{2})-(v_{w_{1}+1}(Y_{1})+v_{1}(Y_{2}))|-\\
&(|v_{w_{1}+1}(X_{1})-v_{w_{1}+1}(Y_{1})|+|v_{1}(X_{2})-v_{1}(Y_{2})|) =\\
&\did(X_{1},Y_{1})+\did(X_{2},Y_{2})-Q.
\end{alignat*}
Actually, note that $Q$ is non negative because of the absolute value triangle inequality.
\end{IEEEproof}


\begin{thebibliography}{99}
\baselineskip 5mm
\bibitem{ABD12}
K. A. S. Abdel-Ghaffar, F. Paluncic, H. C. Ferreira and W. A. Clarke, ``On Helberg's Generalization of the Levenshtein Code for Multiple Deletion/Insertion Error Correction'', \emph{IEEE Transactions on Information Theory}, vol. 58, pp. 1804--1808, March 2012.
\bibitem{ABR18}
M. Abroshan, R. Venkataramanan and A. Guill\'en i F\`abregas, ``Coding for Segmented Edit Channels'', \emph{IEEE Transactions on Information Theory}, vol. 64, pp. 3086--3098, April 2018.
\bibitem{BER45}
``Bertrand's postulate'' at ``\url{https://en.wikipedia.org/wiki/Bertrand's_postulate}''.
\bibitem{BLA93}
M. Blaum, \emph{Codes for Detecting and Correcting Unidirectional Errors}. IEEE Computer Society Press, Washington, DC, USA, 1993.
\bibitem{BOR82}
J. M. Borden, ``Optimal asymmetric error detecting codes'', \emph{Information and Control}, vol. 53, n. 1--2, pp. 66--73, 1982.
\bibitem{BOS82}
B. Bose and T. R. N. Rao, ``Theory of undirectional error correcting/detecting codes'', \emph{IEEE Transactions on Computers}, vol. 31, pp. 521--530, June 1982.
\bibitem{TAL18}
B. Bose, N. Elarief and L. G. Tallini, ``On Codes Achieving Zero Error Capacities in Limited Magnitude Error Channels'',  \emph{IEEE Transactions on Information Theory}, vol. 64, pp. 257--273, Jan. 2018.
\bibitem{CON79}
S. D.Constantin and T. R. N. Rao, ``On the theory of binary asymmetric error correcting codes'', \emph{Information and Control}, vol. 40, pp. 20--36, Jan. 1979.
\bibitem{DOL10}
L. Dolecek and V. Anantharam, ``Repetition error correcting sets: Explicit constructions and prefixing methods'', \emph{SIAM Journal on Discrete Mathematics}, vol. 23, no. 4, pp. 2120--2146, 2010.
\bibitem{FAZ15}
A. Fazeli, A. Vardy and E. Yaakobi, ``Generalized Sphere Packing Bound'', \emph{IEEE Transactions on Information Theory}, vol. 61, pp. 2313--2334, May 2015.
\bibitem{FER97}
H. C. Ferreira, W. A. Clarke, A. S. J. Helberg, K. A. S. Abdel-Ghaffar, and A. J. Han Vinck, ``Insertion/Deletion Correction with Spectral Nulls'', \emph{IEEE Transactions on Information Theory}, vol. 43, pp. 722--732, March 1997.
\bibitem{GUR21}
V. Guruswami and J. H\aa stad. ``Explicit two-deletion codes with redundancy matching the existential bound'', \emph{Proceedings of the 32nd Annual ACM-SIAM Symposium on Discrete Algorithms}, pp. 21--32, Jan. 2021.
\bibitem{HEL02}
A. S. J. Helberg and H. C. Ferreira, ``On Multiple Insertion/Deletion Correcting Codes'', \emph{IEEE Transactions on Information Theory}, vol. 48, pp. 305--308, Jan. 2002. 
\bibitem{JAI17}
S. Jain, F. Farnoud, M. Schwartz, and J. Bruck, ``Duplication-Correcting Codes for Data Storage in the DNA of Living Organisms'', \emph{IEEE Transactions on Information Theory}, vol. 63, pp. 4996--5010, Aug. 2017.
\bibitem{KAU65}
W. Kautz, ``Fibonacci codes for synchronization control'', \emph{IEEE Transactions on Information Theory}, vol. 11, pp. 284--292, April 1965.
\bibitem{KNU86}
D. E. Knuth, ``Efficient Balanced Codes'', \emph{IEEE Transactions on Information Theory}, vol. 32, pp. 51--53, Jan. 1986.
\bibitem{KOV18}
M. Kova\u{c}evi\'{c} and V. Y. F. Tan, ``Asymptotically Optimal Codes Correcting Fixed-Length Duplication Errors in DNA Storage Systems'', \emph{IEEE Communications Letters}, vol. 22, pp. 2194--2197, Nov. 2018.
\bibitem{KOV19}
M. Kova\u{c}evi\'{c}, ``Runlength-Limited Sequences and Shift-Correcting Codes: Asymptotic Analysis'', \emph{IEEE Transactions on Information Theory}, vol. 65, pp. 4804--4814, Aug. 2019.
\bibitem{KLO95}
T. Kl\o ve, ``Error correction codes for the asymmetric channel'', \emph{Report, Dept. of Informatics, University of Bergen}, 1981. (Updated in 1995.)
\bibitem{KUL14}
A. A. Kulkarni, ``Insertion and deletion errors with a forbidden symbol'', \emph{2014 IEEE ITW}, pp. 596--600, Nov. 2014.
\bibitem{KUL13}
A. A. Kulkarni and N. Kiyavash, ``Non-asymptotic Upper Bounds for Deletion Correcting Codes'', \emph{IEEE Transactions on Information Theory}, vol 59,  pp 5115--5130, Aug. 2013.
\bibitem{LEV65}
V. I. Levenshtein, ``Binary codes with correction for deletions and insertions of the symbol $1$'', \emph{Probl. Peredachi Inf.}, vol. 1, n. 1, pp. 12--25, 1965 (in Russian). An english translation can be found in, ``Binary codes capable of correcting spurious insertions and deletions of ones'', \emph{Problems of Information Transmission}, vol. 1, pp. 8--17, 1965.
\bibitem{LEV66}
V. I. Levenshtein, ``Binary codes capable of correcting deletions, insertions and reversals'', \emph{Sov. Phys. Dokl.}, vol. 10, no. 8, pp. 707--710, 1966.
\bibitem{LEV93}
V. I. Levenshtein and A. J. H. Vinck, ``Perfect $(d,k)$-codes capable of correcting single peak-shifts'', \emph{IEEE Transactions on Information Theory}, vol. 39, pp. 656--662, March 1993.
\bibitem{LIR15}
Y. Liron and M. Langberg, ``A Characterization of the Number of Subsequences Obtained via the Deletion Channel'', \emph{IEEE Transactions on Information Theory}, vol. 61, pp. 2300--2312, May 2015.
\bibitem{MAH17}
H. Mahdavifar and A. Vardy, ``Asymptotically optimal sticky-insertion-correcting codes with efficient encoding and decoding'', \emph{2017 IEEE ISIT}, pp. 2683--2687, June 2017.
\bibitem{PAL12}
F. Palun\u{c}i\'{c}, K. A. S. Abdel-Ghaffar, H. C. Ferreira, and W. A. Clarke, ``A Multiple Insertion/Deletion Correcting Code for Run-Length Limited Sequences'', \emph{IEEE Transactions on Information Theory}, vol. 58, pp. 1809--1824, March 2012.
\bibitem{PEL15}
D. Pelusi, S. Elmougy, L. G. Tallini and B. Bose ``$m$-ary Balanced Codes With Parallel Decoding'', \emph{IEEE Transactions on Information Theory}, vol. 61, pp. 3251--3264, June 2015.
\bibitem{PEZ12}
L. Pezza, L. G. Tallini, B. Bose, ``Variable Length Unordered Codes'', \emph{IEEE Transactions on Information Theory}, vol. 58, pp. 548--569, Feb. 2012.
\bibitem{SEL62}
F. Sellers, ``Bit loss and gain correction code'', \emph{IRE Transactions on Information Theory}, vol. 8, pp. 35--38, Jan. 1962.
\bibitem{SIM19}
J. Sima and J. Bruck, ``Optimal $k$-Deletion Correcting Codes'', \emph{2019 IEEE ISIT}, pp. 847--851, July 2019.
\bibitem{SIMA20a}
J. Sima, N. Raviv and J. Bruck, ``Two Deletion Correcting Codes From Indicator Vectors'', \emph{IEEE Transactions on Information Theory}, vol. 66, pp. 2375--2391, April 2020.
\bibitem{SIMA20b}
J. Sima, R. Gabrys and J. Bruck, ``Optimal Systematic $t$-Deletion Correcting Codes'', \emph{2020 IEEE ISIT}, pp. 769--774, June 2020.
\bibitem{SIMA20c}
J. Sima, R. Gabrys and J. Bruck, ``Optimal Codes for the $q$-ary Deletion Channel'', \emph{2020 IEEE ISIT}, pp. 740--745, June 2020.
\bibitem{SLO00}
N. J. A. Sloane, ``On single-deletion-correcting codes'', in \emph{Codes and Designs}, Ohio State University (Ray-Chaudhuri Festschrift), pp. 273--291, 2000. Online: \url{https://arxiv.org/abs/math/0207197}.
\bibitem{TAL98}
L. G. Tallini and B. Bose, ``Design of Balanced and Constant Weight Codes for {VLSI} Systems'', \emph{IEEE Transactions on Computers}, vol. 47, pp. 556--572, May 1998.
\bibitem{TAL99}
L. G. Tallini, U. Vaccaro, ``Efficient $m$-ary balanced codes'', \emph{Discrete Applied Mathematics}, vol. 92, n. 1, pp. 17--56, March 1999.
\bibitem{TAL08}
L. G. Tallini and B. Bose, ``On a new class of error control codes and symmetric functions'', \emph{2008 IEEE ISIT}, pp. 980--984, July 2008.
\bibitem{TAL10}
L. G. Tallini and B. Bose, ``On decoding some error control codes using the elementary symmetric functions''. In {\it Trends in Incidence and Galois Geometries: a Tribute to Giuseppe Tallini - Quaderni di Matematica}, F. Mazzocca, N. Melone and D. Olanda Ed.. vol. 19, p. 265-297, Caserta, Dipartimento di Matematica, Seconda Universit\`{a} di Napoli, 2010.
\bibitem{TAL10b}
L. G. Tallini, N. Elarief and B. Bose, ``On efficient repetition error correcting codes'', \emph{2010 IEEE ISIT}, pp. 1012--1016, June 2010.
\bibitem{TAL11a}
L. G. Tallini and B. Bose, ``On $L_{1}$-distance error control codes'', \emph{2011 IEEE ISIT}, pp. 1026--1030, July/Aug. 2011.
\bibitem{TAL12a}
L. G. Tallini, B. Bose, ``On symmetric $L_{1}$ distance error control codes and elementary symmetric functions'', \emph{2012 IEEE ISIT}, pp. 741--745, July 2012.
\bibitem{TAL13}
L. G. Tallini and B. Bose, ``On $L_{1}$ metric asymmetric/unidirectional error control codes, constrained weight codes and $\sigma$-codes'', \emph{2013 IEEE ISIT}, pp. 694--698, July 2013.
\bibitem{TAL18b}
L. G. Tallini and B. Bose, ``On Some New $\iint_{m}$ Linear Codes Based on Elementary Symmetric Functions'', \emph{2018 IEEE ISIT}, pp. 1665--1669, June 2018.
\bibitem{TAL19}
L. G. Tallini, N. Alqwaifly and B. Bose, ``On Deletions and Insertions of the Symbol ``$0$'' and Asymmetric/Unidirectional Error Control Codes'', \emph{2019 IEEE ISIT}, pp. 2384--2388, July 2019.
\bibitem{TAL22}
L. G. Tallini, N. Alqwaifly and B. Bose, ``Zero Deletion/Insertion Codes and Zero Error Capacity'', \emph{2022 IEEE ISIT}, pp. 986--991, July 2022.
\bibitem{TAL23}
L. G. Tallini, N. Alqwaifly and B. Bose, ``Deletions and Insertions of the Symbol ``$0$'' and Asymmetric/Unidirectional Error Control Codes for the $L_{1}$ Metric'', \emph{IEEE Transactions on Information Theory}, vol. 69, pp. 86--106, Jan. 2023.
\bibitem{WEB92}
J. H. Weber, C. de Vroedt, D. E. Boekee, ``Necessary and sufficient conditions on block codes correcting/detecting errors of various types'', \emph{IEEE Transactions on Computers}, vol. 41, pp. 1189--1193, Sept. 1992.
\end{thebibliography}
\end{document}